\documentclass[12pt]{article}

\input epsf
\usepackage{amsmath,amsfonts,epsfig,color,latexsym}
\usepackage{amssymb}
\usepackage[english, USenglish]{babel}
\usepackage{epsfig}
\usepackage{a4wide}
\usepackage{rotating}
\usepackage{tikz}
\usepackage{pdflscape}
\usetikzlibrary{decorations.markings}

\usepackage{amsmath}
\usepackage{bm}
\usepackage{float}
\restylefloat{table}
\restylefloat{figure}
\usepackage{shuffle}
\allowdisplaybreaks

\csname @addtoreset\endcsname{equation}{section}

\newcommand{\be}{\begin{equation}}
\newcommand{\ee}{\end{equation}}
\newcommand{\bea}{\begin{eqnarray}}
\newcommand{\eaa}{\end{eqnarray}}

\newcommand{\ap}{\alpha'}

\newcommand{\bt}[1]{{\bar t}}

\csname @addtoreset\endcsname{equation}{section}
\makeatletter
\renewcommand{\@fnsymbol}[1]{\@alph{#1}}
\makeatother
\newcommand{\comment}[1]{}

\def\Nc{{\cal N}}
\def\fc#1#2{{\frac{#1}{#2}}}
\newcommand{\req}[1]{(\ref{#1})}

\def\Oc{{\cal O}}
\def\Ic{{\cal I}}\def\Jc{{\cal J}}
\def\h{\fc{1}{2}}

\def\eps{\epsilon}
\def\Li{{\cal L}i}

\newcommand{\eea}{\end{eqnarray}}
\def\lf{\left}
\def\ri{\right}
\def\z{\zeta}
\def\ra{{\rightarrow}}
\def\lra{{\longrightarrow}}

\newcommand\floor[1]{\lfloor#1\rfloor}

\def\Tr{{\rm Tr}}

\date{}
\graphicspath{{figures/}}

\usepackage{shuffle}

\author{\\[-1cm]\large S.~Hohenegger$^{\text{a}}$\ \  \ 
and\ \ \ S.~Stieberger$^{\text{b}}$\\[0.45cm]}
\title{\begin{flushright}{\vspace{-2.25cm}\normalsize MPP--2017--001}\end{flushright}
\date{}
\vspace{1cm}
\bf{Monodromy Relations in \\ Higher--Loop String Amplitudes\\[0.5cm]}}
\dateUSenglish

\date{}
\graphicspath{{figures/}}

\begin{document}
\begin{titlepage}
\maketitle
\begin{center}
\renewcommand{\thefootnote}{\fnsymbol{footnote}}\vspace{-0.75cm}
${}^{\footnotemark[1]}$ Universit\'e de Lyon\\
UMR 5822, CNRS/IN2P3, Institut de Physique Nucl\'eaire de Lyon\\ 4 rue Enrico Fermi, 69622 Villeurbanne Cedex, France\\[0.2cm]
Fields, Gravity \& Strings, CTPU\\
Institute for Basic Sciences, Daejeon 34047, \rm Korea\\[0.2cm]
Email: {\tt s.hohenegger@ipnl.in2p3.fr}\\[0.5cm]
${}^{\footnotemark[2]}$ Max--Planck--Institut f\"ur Physik\\
Werner--Heisenberg--Institut, 80805 M\"unchen, Germany\\[0.2cm]
Email: {\tt stephan.stieberger@mpp.mpg.de}\\[1.25cm]
\end{center}
\begin{abstract}
\noindent\baselineskip18pt
New monodromy relations of loop amplitudes are derived in open string theory. 
We particularly study $N$--point (planar and non--planar) one--loop amplitudes described by a world--sheet cylinder and derive a set of relations between subamplitudes of different color orderings. 
Various consistency checks are performed by matching
$\alpha'$--expansions of planar and non--planar amplitudes involving elliptic iterated integrals
 with the resulting periods giving rise to two sets of  multiple elliptic zeta values.
 The latter refer to the two  homology cycles on the once--punctured complex elliptic curve 
 and the monodromy equations provide relations between these two sets of  multiple elliptic zeta values. 
 Furthermore, our monodromy relations involve new objects for which we present a tentative interpretation in terms of open string scattering amplitudes in the presence of a non--trivial gauge field flux.
Finally, we provide an outlook on how to generalize the new monodromy relations to the non--oriented case and beyond the one--loop level. Comparing a subset of our results with recent findings in the literature we find therein several serious issues related to the structure and significance of monodromy phases and 
 the relevance of missed contributions from contour integrations.

\end{abstract}
\thispagestyle{empty}
\end{titlepage}
\setcounter{tocdepth}{2}
\tableofcontents
\section{Introduction}

The study of perturbative scattering amplitudes in string theory has been a very active field of research in recent years. In particular, a better understanding of the properties, structure and symmetries of string amplitudes has shed light on similar quantities  in gauge and gravity theories. In fact, many results for scattering amplitudes in $\Nc=4$ supersymmetric Yang Mills (SYM) theory were first derived in string theory \cite{Gross:1970eg,Green:1982sw}. Studying the geometric properties of 
the string world--sheet proves to be very fruitful along these directions.
Important relations between objects in field theory such as the Bern--Carrasco--Johansson (BCJ) or Kawai--Lewellen--Tye (KLT) relations have been derived from string theory. Indeed, these relations follow from monodromy properties of the tree--level string world--sheet 
\cite{Kawai:1985xq,Stieberger:2009hq,BjerrumBohr:2009rd}.
Non--trivial monodromies  are related to vertex operator positions on the string world--sheet and 
branch cuts originating from the universal Koba--Nielsen factor accounting for the plane wave correlator
in the string amplitudes. Studying the resulting
 monodromy contours provides interesting mathematical identities and physical constraints.

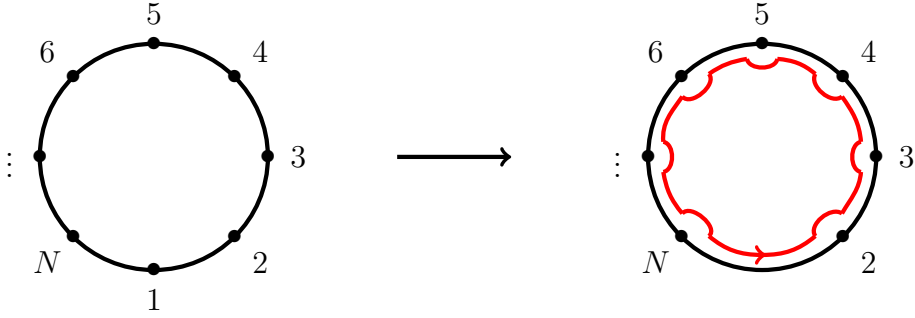
\begin{figure}[H]
\begin{center}
\begin{tikzpicture}
\draw[ultra thick] (0,0) ellipse (1.5cm and 1.5cm);
\node at (0,-1.5) {$\bullet$};
\node at (0,-1.9) {$1$};
\node at (-1.5,0) {$\bullet$};
\node at (-1.9,0) {$\vdots$};
\node at (1.5,0) {$\bullet$};
\node at (1.9,0) {$3$};
\node at (0,1.5) {$\bullet$};
\node at (0,1.9) {$5$};
\node at (1.06,1.06) {$\bullet$};
\node at (1.4,1.4) {$4$};
\node at (1.06,-1.06) {$\bullet$};
\node at (1.4,-1.4) {$2$};
\node at (-1.06,1.06) {$\bullet$};
\node at (-1.4,1.4) {$6$};
\node at (-1.06,-1.06) {$\bullet$};
\node at (-1.4,-1.4) {$N$};
\draw[ultra thick,->] (3.2,0) -- (4.7,0); 
\draw[ultra thick] (8,0) ellipse (1.5cm and 1.5cm);
\node at (6.5,0) {$\bullet$};
\node at (6.1,0) {$\vdots$};
\node at (9.5,0) {$\bullet$};
\node at (9.9,0) {$3$};
\node at (8,1.5) {$\bullet$};
\node at (8,1.9) {$5$};
\node at (9.06,1.06) {$\bullet$};
\node at (9.4,1.4) {$4$};
\node at (9.06,-1.06) {$\bullet$};
\node at (9.4,-1.4) {$2$};
\node at (6.94,1.06) {$\bullet$};
\node at (6.6,1.4) {$6$};
\node at (6.94,-1.06) {$\bullet$};
\node at (6.6,-1.4) {$N$};
\draw[ultra thick,red] (7.3,-1.05) to [out=320,in=180] (8,-1.3) to [out=0,in=220] (8.7,-1.05);
\draw[ultra thick,red] (8.7,-1.05) to [out=125,in=145] (9.05,-0.75);
\draw[ultra thick,red] (9.05,-0.75) to [out=60,in=265] (9.3,-0.2);
\draw[ultra thick,red] (9.3,-0.2) to [out=170,in=190] (9.3,0.2);
\draw[ultra thick,red] (9.05,0.8) to [out=305,in=95] (9.3,0.2);
\draw[ultra thick,red] (8.7,1.1) to [out=240,in=220] (9.05,0.8);
\draw[ultra thick,red] (8.2,1.3) to [out=355,in=140] (8.7,1.1);
\draw[ultra thick,red] (7.8,1.3) to [out=190,in=35] (7.3,1.1);
\draw[ultra thick,red] (6.95,0.8) to [out=235,in=95] (6.7,0.2);
\draw[ultra thick,red] (6.95,-0.75) to [out=130,in=280] (6.7,-0.2);
\draw[ultra thick,red] (8.2,1.3) to [out=260,in=280] (7.8,1.3);
\draw[ultra thick,red] (6.95,0.8) to [out=310,in=310] (7.3,1.1);
\draw[ultra thick,red] (6.7,0.2) to [out=350,in=10] (6.7,-0.2);
\draw[ultra thick,red] (7.3,-1.05) to [out=40,in=50] (6.95,-0.75);
\draw[ultra thick, red, ->] (7.95,-1.3)-- (8.05,-1.3);
\end{tikzpicture}
\end{center}
\caption{\sl Left: Color-ordered $N$-point open string amplitude on the disk. Right: Contour Prescription for obtaining the monodromy relation for $N$-point open string disk amplitudes.}
\label{Fig:ScatteringDisk}
\end{figure}

More concretely, we start with the color decomposition of the tree--level $N$--point open string 
amplitude ${\frak A}^{(0)}_N$  of $N$ adjoint gluons \cite{Mafra:2011nv}. The latter can be
decomposed w.r.t. the color ordered subamplitudes as
\be\label{TreeColor}
{\frak A}^{(0)}_N=g^{N-2}\ \sum_{\sigma\in S_{N-1}} 
\Tr(T^{a_1}T^{a_{\sigma(2)}}\ldots T^{a_{\sigma(N)}})\ A^{(0)}(1,\sigma(2),\ldots,\sigma(N))\ ,
\ee
with the gauge coupling constant $g$ and the Chan--Paton factors $T^a$ being generators of  the $U(N_c)\supset SU(N_c)$ color group. The single trace
structure is invariant under cyclic transformations. As a consequence the first entry can be fixed and the sum is over all non--cyclic permutations $S_{N-1}=S_N\slash {\bf Z}_N$. In the following we shall consider 
the color--ordered open string $N$--point tree--level subamplitude $A^{(0)}(1,2,\ldots,N)$. The latter is described by a world--sheet disk which contains $N$ ordered vertex operator insertions at the boundary of the disk (cf. the left part of Fig.~\ref{Fig:ScatteringDisk}).
For a given color--ordering all vertices are cyclically ordered, i.e. their relative ordering
is kept fixed. In order to obtain relations between different  orderings, we consider moving one vertex
insertion (e.g. the vertex labelled by $1$) along the contour shown in red in the right part of
Figure~\ref{Fig:ScatteringDisk}. As indicated, this contour is chosen in a specific way to avoid
the branch points arising when the vertex 1 collides with any of the other vertex operator insertions on
the boundary. Thus, since the integrand of the subamplitude 
$A^{(0)}(1,2,\ldots,N)$ is a holomorphic function, which is integrated along a closed contour and encloses
no poles\footnote{Here we do not consider additional closed string bulk insertions. 
For a generalization of the discussion including closed string insertions, see \cite{Stieberger:2016lng}.},
the resulting integral vanishes.

\noindent
 This amounts to the following tree--level identity \cite{Stieberger:2009hq,BjerrumBohr:2009rd}
\begin{align}
A^{(0)}(1,2,\ldots,N)+e^{i\pi s_{12}}A^{(0)}(2,1,\ldots,N)+e^{i\pi s_{1,23}}\ A^{(0)}(2,3,1,4,\ldots,N)&\nonumber\\
+\ldots+e^{i\pi s_{1,2\ldots N-1}}\ A^{(0)}(2,3,\ldots,N-1,1,N)&=0\ , \label{TreeAmpRel}
\end{align}
where
\be\label{Mandel}
s_{ij}=\ap\ (k_i+k_j)^2=2\alpha'k_i\cdot k_j\ ,\ \ \ 
s_{i,j\ldots l}=2\ap k_i(k_j+\ldots+k_l)\ \ \ ,\ \ \ \sum_{i=1}^Nk_i=0\ ,
\ee
with $k_j$ the momentum of the $j$--th vertex insertion. Furthermore, the phase factors stem from 
encircling the vertex positions to avoid the branch points at the boundary. It is important to notice that (\ref{TreeAmpRel})
is valid to all orders in $\alpha'$. On the other hand, at leading order 
in $\ap$ we gain identities for field theory: the real part of the field theory limit of \req{TreeAmpRel}
provides the Kleiss--Kuijf relations (photon decoupling relations) \cite{Kleiss:1988ne}, while its imaginary part  yields the BCJ relations \cite{Bern:2008qj}.
At any rate, the relation \req{TreeAmpRel} and permutations thereof allow to
express any tree--level open string amplitude of a given color ordering in terms of a basis of $(N-3)!$ subamplitudes \cite{Stieberger:2009hq,BjerrumBohr:2009rd}.
As a consequence, the decomposition \req{TreeColor} boils down to a sum over a basis of $(N-3)!$ 
subamplitudes.

It is an obvious task to try to generalize the above tree--level results \req{TreeAmpRel} 
to higher--loop open string amplitudes and extract similar relations for field theory. At higher loops oriented and unoriented string 
world--sheets with boundaries appear. 
The latter are quotients of compact higher genus Riemann surface
without boundaries. The presence of boundaries hampers using the virtue of periodicity properties of correlation functions. 
Although the general strategy of considering closed monodromy circles on the 
world--sheet remains the same as in the tree--level case, new and unexpected phase factors and additional boundary terms will show up in the monodromy relations. 
Furthermore, monodromy relations have to be discussed separately for oriented and unoriented surfaces.
In this work we shall elucidate these issues and derive higher loop analogs of \req{TreeAmpRel}.

A first step is to consider the worldÐ-sheet cylinder describing 
planar and non--planar oriented one--loop string amplitudes.
Similarly to the tree--level case \req{TreeColor}, oriented open string one--loop amplitudes 
${\frak A}^{(1)}_N$  of $N$ adjoint gluons can be decomposed w.r.t. the color ordering as
\begin{align}
{\frak A}^{(1)}_N&=g^N\ \Big\{\ N_c\sum_{\sigma\in S_{N-1}} 
\Tr(T^{a_{\sigma(1)}}\ldots T^{a_{\sigma(N)}})\ A^{(1)}(\sigma(1),\ldots,\sigma(N))\nonumber\\
&+\sum_{c=2}^{\floor{\tfrac{N}{2}}+1}\sum_{\sigma\in S_N\slash S_{N;c}} \Tr(T^{a_{\sigma(1)}}\ldots T^{a_{\sigma(c-1)}})\
\Tr(T^{a_{\sigma(c)}}\ldots T^{a_{\sigma(N)}})\label{Fullfledged}\\
&\times A^{(1)}(\sigma(1),\ldots,\sigma(c-1)|\sigma(c),\ldots,\sigma(N))\ \Big\}\ ,\nonumber
\end{align}
with 
$\Tr(1)=N_c$ and  $S_{N;c}$ describing the subset of permutations $S_N$, which leave the double trace
structure invariant. The corresponding open string world--sheet is described by a cylinder with
$c-1$ external states distributed along one boundary and $N-c+1$  along the other boundary.
Similar results on  color decomposition \req{Fullfledged} exist for the heterotic string 
\cite{Bern:1990ux}.
It is well known \cite{Mangano:1990by}, that in field theory enforcing $U(1)$ decoupling
allows for additional subamplitude relations, e.g.:
\be\label{FTmonoPl}
A^{(1)}_{YM}(1,2,\ldots,N)+A^{(1)}_{YM}(2,1,\ldots,N)+\ldots+A^{(1)}_{YM}(2,\ldots,N,1)=A^{(1)}_{YM}(1|2,\ldots,N)\ .
\ee
As a consequence in field theory all the subleading amplitudes with $c>1$ can be expressed in terms of the
leading ones $c=1$ \cite{Bern:1994zx}
\be\label{FTCOLOR}
A_{YM}^{(1)}(1,\ldots,c-1|c,c+1,\ldots,N)=(-1)^{c-1}\ \sum_{\sigma\in COP(\alpha\cup\beta)}
  A_{YM}^{(1)}(\sigma(1),\ldots,\sigma(N))\ ,\ \ \ c>1\ ,
\ee
with the sets $\alpha=\{c-1,c-2,\ldots,2,1\}$, $\beta=\{c,c+1,\ldots,N-1,N\}$ and
$COP(\alpha\cup\beta))$ the set of all permutations of $\{1,2,\ldots,N\}$ with $N$ held fixed
that preserve the cyclic ordering of the elements $\alpha_i$ and $\beta_j$ within each set $\alpha$ and 
$\beta$, while allowing for all possible relative orderings of the $\alpha_i$ w.r.t. $\beta_j$. 
E.g. for $N\!=\!4$ eq. \req{FTCOLOR} gives:
\begin{align}
A_{YM}^{(1)}(1|2,3,4)&=-A_{YM}^{(1)}(1,2,3,4)-A_{YM}^{(1)}(2,1,3,4)-A_{YM}^{(1)}(2,3,1,4)\ ,\label{Dix1}\\
A_{YM}^{(1)}(1,2|3,4)&=A_{YM}^{(1)}(1,2,3,4)+A_{YM}^{(1)}(1,3,2,4)+A_{YM}^{(1)}(2,1,3,4)+\nonumber\\
&+A_{YM}^{(1)}(2,3,1,4)+A_{YM}^{(1)}(3,1,2,4)+A_{YM}^{(1)}(3,2,1,4)\ .\label{Dix2}
\end{align}
In lines of the tree--level situation encoded in the equations \req{TreeAmpRel} we shall find string monodromy relations for the one--loop subamplitudes $A^{(1)}$ involving both planar and non--planar amplitudes. Again, the SYM relations \req{FTmonoPl} and \req{FTCOLOR} should then be simple consequences of 
the more general string monodromy relations.

When generalizing the contour integral to open string one--loop amplitudes we encounter new objects that are different from the color--ordered amplitudes in a flat background. Indeed, they differ by crucial position dependent phase factors. In the case of the cylinder the latter appear when transporting one vertex from one boundary component to the other. Essentially, these new objects are related to the lack of periodicity and  make an interpretation of the monodromy relations from a physical perspective difficult. Here we shall give a  possible interpretation by showing that in the simplest example (for $N\!=\!4$), these new objects can be interpreted as four--point open string amplitudes in the presence of a non-trivial 
(space--time) gauge field flux. While a generalization to $N$--point functions is straight forward, we will leave the study of these new objects to future work \cite{Progress}.

Tree--level $N$--point open string amplitudes are described by iterated integrals on the moduli space of curves of genus zero  with $N-3$ ordered points. The corresponding periods yield multiple polylogarithms which in turn reduce to multiple zeta values (MZVs). For a recent account see \cite{Stieberger:2016xhs}.
As a consequence the relations \req{TreeAmpRel} give rise to identities between different iterated 
integrals which only differ by their integration path from zero to one. 
The elliptic analogues of the latter are iterated integrals on the configuration
space of $N-1$ marked points on a complex elliptic curve $E_\tau$ of genus one \cite{BL}.
The elliptic curve $E_\tau$ is an  analytic manifold ${\bf C}\slash{\bf Z}\tau\oplus{\bf Z}$ with complex structure $\tau\in{\bf C}$ satisfying $\Im(\tau)>0$.
For  certain paths of integrations (corresponding to the two canonical  cycles of the elliptic curve) the resulting periods are multiple elliptic zeta values (eMZVs) \cite{BE}.
As we shall see these objects will appear when expanding the planar and 
non--planar cylinder amplitude w.r.t. $\ap$. 
For   planar amplitudes the real line interval $[0,1]\in E_\tau$ describes the relevant integration path
and the corresponding elliptic iterated integrals are evaluated at real arguments yielding a specific system of  eMZVs.
On the other hand, for non--planar amplitudes the path from $0$ to $\tau$ has to be considered and a different system of elliptic iterated integrals appears yielding to an other  system of   eMZVs. Since our one--loop
monodromy relations mix planar and non--planar amplitudes these relations provide identities 
between these systems of eMZVs.

While we were preparing to publish our results, we became aware of the letter \cite{Tourkine:2016bak}, which also discusses monodromy relations in higher loop string amplitudes. However, after comparing a subset of our results with these findings, we noticed several discrepancies and serious issues related to the structure and significance of monodromy phases and 
 the relevance of omitted contributions from contour integrations.
While these issues could not be detected by the kind of checks performed in \cite{Tourkine:2016bak}, they become apparent by the more in--depth verification we have performed in sections~\ref{Sect:AlphaExpansion} and \ref{Sect:FieldTheoryLimit}, as we shall point out throughout the present work.

This work is organized as follows: In section~\ref{Sect:OneLoopAmps} we review the setup for one--loop open string scattering amplitudes, focusing particularly on the planar and non--planar annulus. In section~\ref{Sect:MonodromyOrientable} we discuss monodromy relations for orientable $N$--point one--loop amplitudes involving both planar and non--planar subamplitudes. In section~\ref{Sect:AlphaExpansion} we provide extensive checks of our monodromy relations by studying $\alpha'$--expansions of the string theory $N$--point amplitudes and explicitly calculating the ensuing elliptic iterated integrals. The latter give rise to two sets of elliptic multiple zeta values whose appearance and role in the monodromy relations will be elaborated. In section~\ref{Sect:FieldTheoryLimit} we discuss the field theory limit of the one--loop $N$--point amplitudes and show that the string monodromy equations give rise to relations involving subamplitudes in gauge theory. In section~\ref{Sect:NonOrientable} we generalize our results to non--orientable string amplitudes by discussing monodromy relations on the M\"obius strip. In section~\ref{Sect:NonCommutative} we provide a tentative interpretation of the new objects appearing in the planar string monodromy relations. In section~\ref{Sect:HigherLoop} we give an outlook on higher-loop monodromy relations and point out  several conceptual issues in the non--planar case. Finally, section~\ref{Sect:Conclusions} contains our concluding remarks. 
Several definitions and useful identities of Jacobi theta functions, as well as an list of explicit  results for iterated elliptic integrals along with further technical computations are relegated to three appendices.\\


\section{One--loop open string amplitudes}\label{Sect:OneLoopAmps}

In this section we discuss the general form of one--loop amplitudes for $N$ open string states. 
For flat backgrounds one--loop open string amplitudes  involve 
conformal field theory correlators on a Riemann surface of genus one. There are three distinct
topologies to consider for one-loop open string amplitudes: the planar cylinder diagram, which can conformally be mapped to an annulus with vertex operator insertion points on one boundary only, 
the non--planar cylinder, which allows for insertion points on both boundaries and the M\"obius strip.
In this section  we consider the case of orientable amplitudes on the cylinder (annulus), while in section~\ref{Sect:NonOrientable} we discuss the non--orientable case, specifically amplitudes on the M\"obius strip.

\subsection*{Amplitudes with cylinder world--sheet}

In open superstring theory the amplitude \req{Fullfledged} is derived by the use of conformal field theory techniques in flat $D$--dimensional space--time.  The relevant cylinder world--sheet diagram 
with $N$ vertex operator insertions requires computing the following correlator:
\begin{align}
{\frak A}_N^{(1)}=\sum_{N_1+N_2=N}\sum_{{\sigma_1\in S_{N_1-1}}\atop{\sigma_2\in S_{N_2-1}}}V^{-1}_{\text{CKG}}\,\left(\int_{\mathcal{J}_{\sigma_1}}\prod_{i=1}^{N_1}du_i\,\int_{\mathcal{J}_{\sigma_2}'}\prod_{j=1}^{N_2}dv_j\right)\left\langle\prod_{i=1}^{N_1}:V_o(u_i):\prod_{j=1}^{N_2}:V_o(v_j):\right\rangle\ .\label{OneLoopAmpGen}
\end{align}
The amplitude \req{OneLoopAmpGen} decomposes into a sum  over all configurations with $N_1$ vertex operators on one boundary of the cylinder and $N_2$ on the other one, with $N=N_1+N_2$. 
Furthermore, in \req{OneLoopAmpGen} the summations over $\sigma_{1,2}$ run over all $(N_1-1)!(N_2-1)!$ inequivalent cyclic orderings of the vertices on each boundary component, cf. the next figure. 
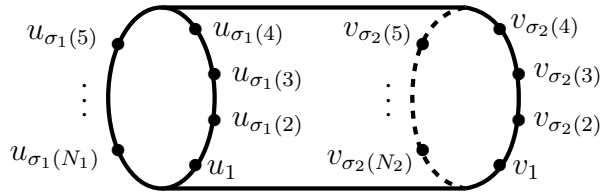
\begin{center}
\begin{tikzpicture}
\draw[ultra thick] (-1,0) ellipse (0.7cm and 1.2cm);
\draw[ultra thick,dashed] (3,1.2) to [out=190,in=90] (2.3,0) to [out=270,in=170] (3,-1.2);
\draw[ultra thick] (3,1.2) to [out=350,in=90] (3.7,0) to [out=270,in=10] (3,-1.2);
\draw[ultra thick] (-1,1.2) -- (3,1.2);
\draw[ultra thick] (-1,-1.2) -- (3,-1.2);
\node at (-0.55,-0.9) {$\bullet$};
\node at (-0.2,-0.95) {$u_1$};
\node at (-0.3,-0.3) {$\bullet$};
\node at (0.4,-0.35) {$u_{\sigma_1(2)}$};
\node at (-0.3,0.3) {$\bullet$};
\node at (0.4,0.25) {$u_{\sigma_1(3)}$};
\node at (-0.55,0.9) {$\bullet$};
\node at (0.15,0.85) {$u_{\sigma_1(4)}$};
\node at (-1.57,-0.7) {$\bullet$};
\node at (-2.4,-0.8) {$u_{\sigma_1(N_1)}$};
\node at (-1.57,0.7) {$\bullet$};
\node at (-2.3,0.8) {$u_{\sigma_1(5)}$};
\node at (3.45,-0.9) {$\bullet$};
\node at (3.8,-0.95) {$v_1$};
\node at (3.7,-0.3) {$\bullet$};
\node at (4.35,-0.35) {$v_{\sigma_2(2)}$};
\node at (3.7,0.3) {$\bullet$};
\node at (4.35,0.3) {$v_{\sigma_2(3)}$};
\node at (3.45,0.9) {$\bullet$};
\node at (4.05,0.95) {$v_{\sigma_2(4)}$};
\node at (2.44,-0.7) {$\bullet$};
\node at (1.75,-0.85) {$v_{\sigma_2(N_2)}$};
\node at (2.44,0.7) {$\bullet$};
\node at (1.85,0.85) {$v_{\sigma_2(5)}$};
\node[rotate=90] at (-2,0) {$\cdots$};
\node[rotate=90] at (2,0) {$\cdots$};
\end{tikzpicture}
\end{center}
\noindent
Subject to cyclic invariance the integrations ${\cal J}_{\sigma_k}$ on each boundary decompose into $(N_k-1)!$  regions according to the various orderings $\sigma_k$ of the $N_k$ vertex operator positions, respectively. For a suitable\footnote{Notice that the parameterization in (\ref{REGION}) simply indicates a particular (cyclic) ordering of the vertex insertions. To make the notation $u_i<u_j$ precise, in explicit computations we shall typically use the conformal Killing group to fix the position of one of the vertex operators. In this way, coordinates can be chosen to give a precise meaning to the notation in (\ref{REGION}).} parameterization we have:
\begin{align}
{\cal J}_{\sigma_1}&:=\left\{\ \bigcup_{i=1}^{N_1} u_i\ |\  u_1< u_{\sigma_1(2)}<\ldots u_{\sigma_1(N_1)}\ \right\}\ ,\nonumber\\
{\cal J}_{\sigma_2}'&:=\left\{\ \bigcup_{j=1}^{N_2}  v_j\ |\  v_1>v_{\sigma_2(2)}>\ldots v_{\sigma_2(N_2)}\ \right\}\ .\label{REGION}
\end{align}
Besides,  in (\ref{OneLoopAmpGen}) we divided by the volume $V_{\rm CKG}$ of the conformal Killing group. The latter accounts for the translation invariance on the genus one surface.
The vertex operator insertions in (\ref{OneLoopAmpGen}) are given by: 
\begin{align}
V_o(x;p,\xi)=T^a\ F(X,\psi,k,\xi)\,e^{ik_\mu X^\mu}(x)\ .\label{vertex}
\end{align}
The first factor represents the Chan--Paton factor $T^a$, which carries the gauge degrees of freedom.
In fact, each ordering $\sigma_{1,2}$ of the vertices gives rise to a group factor in the sum \req{Fullfledged}. 
The function $F$ specifies each vertex operator, with
 $X^\mu$ (with $\mu=0,\ldots, D-1$)  the world--sheet bosons and $\psi_\mu$ their fermionic superpartners. Furthermore, $x\in{\cal J}_{\sigma_1},{\cal J}_{\sigma_2}'$ is the world-sheet position of the vertex, $k_\mu$ its momentum and $\xi$ represent additional quantum numbers including polarizations. The third factor 
 of \req{vertex} represents the plane wave factor   
$e^{ik_\mu X^\mu}(x)$. In the sequel the latter will play a crucial role, while the specific form of $F$ will not be relevant.

The color--stripped subamplitude $A^{(1)}(1,\ldots,N_1|N_1+1,\ldots, N_2)$  has 
two--different (but equivalent) representations described by either the open-- or closed--string channel.

\subsubsection*{Open string channel}

In the open string channel, we parameterize the cylinder as the region $[0,\tfrac{1}{2}]\times [0,t]$ in the complex plane. 
Therefore, for the cylinder we have the one--loop modular parameter
\be\label{modop}
\tilde{\tau}=it\ ,
\ee
and the vertex operator positions are parameterized as:
\begin{align}\label{positionso}
&\tilde{z}_j=\left\{\begin{array}{lcl}ix_j & \text{if} & j=1,\ldots, N_1\,, \\ i x_j+\tfrac{1}{2} & \text{if} & j=N_1+1,\ldots, N\,,\end{array}\right.&&\text{with:} &&x_l\in[0,t]\hspace{0.2cm}\ ,\  l=1,\ldots,N\ .
\end{align}
The vertices are inserted along the boundary as depicted in the following figure.
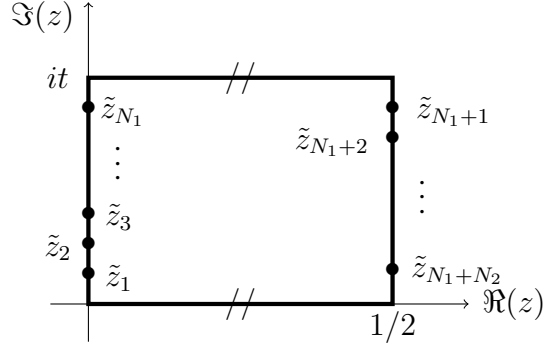
\begin{center}
\begin{tikzpicture}
\draw[->] (-0.5,0) -- (5,0);
\draw[->] (0,-0.5) -- (0,4);
\draw[ultra thick] (0,0) rectangle (4,3);
\node at (5.6,0) {$\Re (z)$};
\node at (-0.6,3.8) {$\Im (z)$};
\node at (2,0) {//};
\node at (2,3) {//};
\node at (-0.4,3) {$it$};
\node at (4,-0.3) {1/2};
\node[rotate=90] at (0.4,1.9) {$\cdots$};
\node at (0,0.4) {$\bullet$};
\node at (0.4,0.35) {$\tilde{z}_{1}$};
\node at (0,0.8) {$\bullet$};
\node at (-0.4,0.75) {$\tilde{z}_2$};
\node at (0,1.2) {$\bullet$};
\node at (0.4,1.15) {$\tilde{z}_3$};
\node at (0,2.6) {$\bullet$};
\node at (0.45,2.55) {$\tilde{z}_{N_1}$};
\node at (4,0.45) {$\bullet$};
\node at (4.85,0.45) {$\tilde{z}_{N_1+N_2}$};
\node at (4,2.2) {$\bullet$};
\node at (3.2,2.15) {$\tilde{z}_{N_1+2}$};
\node[rotate=90] at (4.4,1.45) {$\cdots$};
\node at (4,2.6) {$\bullet$};
\node at (4.8,2.55) {$\tilde{z}_{N_1+1}$};
\end{tikzpicture}
\end{center}

\noindent
and the integration regions \req{REGION} are given by:
\begin{align}\label{Regions}
{\cal J}_{\sigma}&=\lf\{\ \bigcup_{i=1}^{N_1}x_i\in E_{\tilde\tau}^{N_1}\ |\ 0\leq x_1\leq x_{\sigma(2)} \leq\ldots\leq x_{\sigma(N_1)} \leq t\ \ri\}\ ,\ \sigma\in S_{N_1-1}\ ,
\nonumber\\
{\cal J}'_{\sigma'}&=\lf\{\ \bigcup_{i=N_1+1}^{N_1+N_2}x_i\in E_{\tilde\tau}^{N_2}\ |\ 0\leq x_{N_1+N_2}\leq x_{\sigma'(N_1+N_2-1)} \leq\ldots\leq x_{\sigma'(N_1+1)} \leq t\ \ri\}\ ,\ \sigma'\in S_{N_2-1}\ .
\end{align}
The scalar Green function on the cylinder  (referring to  Neumann boundary conditions for the open string ends) is given by:
\be\label{Go}
\tilde{G}(\tilde{z},\tilde{\tau})=\ln\lf|\fc{\theta_1(\tilde{z},\tilde{\tau})}{\theta_1'(0,\tilde{\tau})}\ri|^2-\fc{2\pi}{\Im \tilde{\tau}}\ \Im(\tilde{z})^2\ .
\ee
With this notation, in $D$ space--time dimensions the color--ordered 
planar $N$--point open superstring amplitude assumes the generic form ($N_1=N, N_2=0$):
\begin{align}
A^{(1)}(1,\ldots,N)&=\delta(k_1+\ldots k_N)\ \int_0^\infty \fc{dt}{t^{1+\tfrac{D}{2}}}\ V_{\rm CKG}^{-1}\int_{\Jc_1}\prod_{i=1}^N d\tilde{z}_i\ P_N(\tilde{z}_1,\ldots,\tilde{z}_N,\tilde{\tau})\nonumber\\ 
&\times \exp\lf\{\tfrac{1}{2}\sum_{1\leq i<j\leq N}  s_{ij}\ G(\tilde{z}_{ji},\tilde{\tau})\ri\}\ ,\label{OpenStringChannelAmp}
\end{align}
with:
\be
\tilde{z}_{ji}:=\tilde{z}_j-\tilde{z}_i\ .
\ee
Above, we included the momentum--conserving  delta function and the kinematics invariants $s_{ij}$ given in \req{Mandel}.
Furthermore, $P_N$ comprising all the kinematical terms represents a function of modular weight $N+1-\tfrac{D}{2}$.
For $n:=N+1-\tfrac{D}{2}\geq 1$ the function $P_N$ is a polynomial in derivatives $\partial^k \tilde G,(\partial \tilde G)^k$ of the scalar Green function $\tilde G$ on the cylinder \req{Go}, and the Eisenstein series $G_k$ with $k\leq n$ such that the total modular weight is $n$, cf. \cite{Stieberger:2002wk,Broedel:2014vla}. 
Furthermore, $\theta_1(\tilde{z},\tilde{\tau})$ is (one of) the Jacobi theta functions, cf. appendix~\ref{Sect:JacobiTheta} for their precise definition in addition to useful identities and relations.

The  scalar Green function on the cylinder describing two positions on two distinct boundaries
is given by:
\be\label{GTo}
\tilde{G}_T(\tilde{z},\tilde{\tau})=\ln\lf|\fc{\theta_2(\tilde{z},\tilde{\tau})}{\theta_1'(0,\tilde{\tau})}\ri|^2-\fc{2\pi}{\Im \tilde{\tau}}\ \Im(\tilde{z})^2\ .
\ee
Note, that we have $\tilde{G}(\tilde{z}+\tfrac{1}{2},\tilde\tau)=\tilde{G}_T(\tilde{z},\tilde\tau)$ and $\tilde{G}_T(\tilde{z}+\tfrac{1}{2},\tilde\tau)=\tilde{G}(\tilde{z},\tilde\tau)$.
With \req{GTo} in $D$ space--time dimensions the non--planar $N$--point open superstring amplitude with $N_1$ open strings on one boundary
and $N_2$ on the second boundary assumes the generic form:  
\begin{align}
A^{(1)}(1,\ldots,N_1|N_1+1,\ldots,N_2)&=\delta(k_1+\ldots k_N)\ \int_0^\infty \fc{dt}{t^{1+\tfrac{D}{2}}}\ V_{\rm CKG}^{-1}\lf(\int_{\Jc_1}\prod_{i=1}^{N_1} d\tilde{z}_i\ 
\int_{\Jc_1'}\prod_{j=N_1+1}^{N} d\tilde{z}_j\ri) \nonumber\\ 
&\times \exp\lf\{\tfrac{1}{2}\sum_{1\leq i<j\leq N_1}  s_{ij}\ \tilde{G}(\tilde{z}_{ji},\tilde{\tau})+\tfrac{1}{2}\sum_{N_1+1\leq i<j\leq N}  s_{ij}\ \tilde{G}(\tilde{z}_{ij},\tilde{\tau})\ri\}\nonumber\\
&\times\exp\lf\{\tfrac{1}{2}\sum_{1\leq i\leq N_1\atop N_1+1\leq j\leq N}  s_{ij}\ \tilde{G}_T(\tilde{z}_{ji}-\tfrac{1}{2}),\tilde{\tau})\ri\}\ \ Q_N(\tilde{z}_1,\ldots,\tilde{z}_N,\tilde{\tau})\ .\label{GENERIC1}
\end{align}
Furthermore, $Q_N$ comprising all the kinematical terms is a function of modular weight $N+1-\tfrac{D}{2}$.
For $n:=N+1-\tfrac{D}{2}\geq 1$ the function  $Q_N$ can be expressed in terms of the objects  $\partial^k \tilde G_T,(\partial \tilde G_T)^k$, and $G_k$ such that the total modular weight is $n$.

Due to the presence of the non--holomorphic phase factors in (\ref{OpenStringChannelAmp}) and \req{GENERIC1} it is useful 
to transform the open string amplitudes into the dual closed string channel, cf. also the discussion 
in section \ref{Hongkong}.

\subsubsection*{Closed string channel}

In the closed string channel we work with the closed string modulus
\be\label{modcl}
\tau=il =\fc{i}{t}=-\fc{1}{\tilde \tau}\ ,
\ee
and the  vertex operator positions
\be\label{vertcl}
z=\fc{x}{t}=\fc{\tilde z}{\tilde \tau}\ ,
\ee
parameterized as:
\begin{align}\label{positionsc}
&z_j=\left\{\begin{array}{lcl}x_j & \text{if} & j=1,\ldots, N_1\,, \\ x_j-\tfrac{il}{2} & \text{if} & j=N_1+1,\ldots, N\,,\end{array}\right.&&\text{with:} &&x_l\in[0,1]\hspace{0.2cm}\ ,\  l=1,\ldots,N\ .
\end{align}
\noindent
Graphically, the vertices are arranged according to the following figure
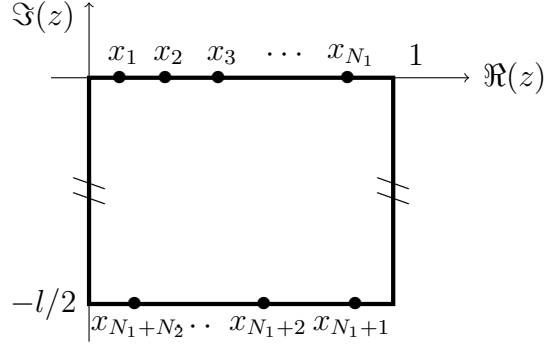
\begin{center}
\begin{tikzpicture}
\draw[->] (-0.5,3) -- (5,3);
\draw[->] (0,-0.5) -- (0,4);
\draw[ultra thick] (0,0) rectangle (4,3); 
\node at (5.6,3) {$\Re (z)$};
\node at (-0.6,3.8) {$\Im (z)$};
\node at (-0.6,0) {$-l/2$};
\node at (0.4,3) {$\bullet$};
\node at (0.45,3.3) {$x_1$};
\node at (1,3) {$\bullet$};
\node at (1.05,3.3) {$x_2$};
\node at (1.7,3) {$\bullet$};
\node at (1.75,3.3) {$x_3$};
\node at (2.6,3.3) {$\cdots$};
\node at (3.4,3) {$\bullet$};
\node at (3.45,3.3) {$x_{N_1}$};
\node at (4.3,3.3) {1};
\node at (3.5,0) {$\bullet$};
\node at (3.45,-0.3) {$x_{N_1+1}$};
\node at (2.3,0) {$\bullet$};
\node at (2.35,-0.3) {$x_{N_1+2}$};
\node at (1.4,-0.3) {$\cdots$};
\node at (0.6,0) {$\bullet$};
\node at (0.65,-0.3) {$x_{N_1+N_2}$};
\node[rotate=90] at (0,1.5) {//};
\node[rotate=90] at (4,1.5) {//};
\end{tikzpicture}
\end{center}

\noindent
and the integration regions \req{Regions} translate into:
\begin{align}\label{Regionsc}
{\cal I}_{\sigma}&=\lf\{\ \bigcup_{i=1}^{N_1}x_i\in E_\tau^{N_1}\ |\ 0\leq x_1\leq x_{\sigma(2)} \leq\ldots\leq x_{\sigma(N_1)} \leq 1\ \ri\}\ ,\ \sigma\in S_{N_1-1}\ ,
\nonumber\\
{\cal I}'_{\sigma'}&=\lf\{\ \bigcup_{i=N_1+1}^{N_1+N_2}x_i\in E_\tau^{N_2}\ |\ 0\leq x_{N_1+N_2}\leq x_{\sigma'(N_1+N_2-1)} \leq\ldots\leq x_{\sigma'(N_1+1)} \leq 1\ \ri\}\ ,\ \sigma'\in S_{N_2-1}\ .
\end{align}
Furthermore, the Green function \req{Go} in the closed string channel is represented by
\be\label{G}
G(z,\tau)=\ln\lf|\fc{\theta_1(z,\tau)}{\theta_1'(0,\tau)}\ri|^2\ ,
\ee
while for \req{GTo} we take:
\be\label{GT}
G_T(z,\tau)=\ln\lf|\fc{\theta_4(z,\tau)}{\theta_1'(0,\tau)}\ri|^2\ .
\ee
Contractions among insertions on the same boundary give rise to $\theta_1$--functions encoded in \req{G}, while contractions among vertices on different boundaries give rise to $\theta_4$--functions
encoded in \req{GT}.
Using the transformation properties (\ref{TrafoTheta}) and (\ref{TrafoEta}) we can express the planar amplitude (\ref{OpenStringChannelAmp}) in terms of the  quantities \req{modcl}--\req{GT} as follows:
\begin{align}
A^{(1)}(1,\ldots,N)&=\delta(k_1+\ldots k_N)\ \int_0^\infty dl\ 
V_{\rm CKG}^{-1} \int_{\Ic_1}\prod_{i=1}^N dz_i\ P_N(z_1,\ldots,z_N,\tau)\nonumber\\
&\times\exp\lf\{\tfrac{1}{2}\sum\limits_{1\leq i<j\leq N} s_{ij}\ G(z_{ji},\tau)\ri\}\label{ClosedStringChannelAmp}\\
&=\delta(k_1+\ldots k_N)\int_0^\infty  dl\ 
V_{\rm CKG}^{-1} \int_{\Ic_1}\prod_{i=1}^N dz_i\ P_N(z_1,\ldots,z_N,\tau)\hskip-0.2cm \prod_{1\leq i<j\leq N}\left|\fc{\theta_1\left(x_{ji},il\right)}{\theta'_1(0,il)}\right|^{s_{ij}}\hskip-0.4cm.\nonumber
\end{align}
Furthermore, in terms of the quantities \req{modcl}--\req{GT} the non--planar amplitude 
reads:
\begin{align}
A^{(1)}(1,\ldots,&N_1|N_1+1,\ldots,N)=\delta(k_1+\ldots k_N)\ \int dl\ 
V_{\rm CKG}^{-1}\ \left(\int_{\Ic_1}\prod_{j=1}^{N_1}dx_j\int_{\Ic_1'}\prod_{j=N_1+1}^{N}dx_j\right)\nonumber\\
&\times\exp\lf\{\tfrac{1}{2}\sum\limits_{1\leq i<j\leq N_1} s_{ij}\ G(z_{ji},\tau)+\tfrac{1}{2}\sum\limits_{N_1+1\leq i<j\leq N} s_{ij}\ G(z_{ij},\tau)\ri\}\nonumber\\ 
&\times\exp\lf\{\tfrac{1}{2}\sum\limits_{1\leq i\leq N_1\atop N_1+1\leq j\leq N} s_{ij}\ G_T(z_{ji}+\tfrac{il}{2},\tau)\ri\}\ Q_N(z_1,\ldots,z_N,\tau)\label{DefClosedChannelAmp}\\
&=\delta(k_1+\ldots k_N)\ \int dl\ 
V_{\rm CKG}^{-1}\ \left(\int_{\Ic_1}\prod_{j=1}^{N_1}dx_j\int_{\Ic_1'}\prod_{j=N_1+1}^{N}dx_j\right)\ Q_N(z_1,\ldots,z_N,\tau)\nonumber\\
&\times\prod_{1\leq i<j\leq N_1}\left|\fc{\theta_1\left(x_{ji},il\right)}{\theta'_1(0,il)}
\right|^{s_{ij}} \prod_{N_1+1\leq i<j\leq N}
\left|\fc{\theta_1\left(x_{ij},il\right)}{\theta'_1(0,il)}\right|^{s_{ij}}\ 
\prod_{1\leq i\leq N_1\atop N_1+1\leq j\leq N}
\left|\fc{\theta_4\left(x_{ji},il\right)}{\theta'_1(0,il)}\right|^{s_{ij}}.\nonumber 
\end{align}
In contrast to the open string channel the integrands of the expressions \req{ClosedStringChannelAmp} and \req{DefClosedChannelAmp} are holomorphic quantities and the virtue of holomorphicity can be applied
for considering contours on the elliptic curve.


\section{One--loop monodromy relations on the cylinder}\label{Sect:MonodromyOrientable}

In this section we shall discuss monodromy relations between a subset of the amplitudes introduced above. Actually, the relations hold at the level of
integrands before performing the $l$-- (or $t$)--integration, i.e. for generic 
$\tau$ (or $\tilde\tau$).
In particular, we may discuss their field theory limit 
$\Im\tilde\tau\ra\infty$, cf. section \ref{Sect:FieldTheoryLimit}.
Furthermore, the relations are not affected by any tadpole issues 
arising from the dilaton in the limit $\Im\tau\ra\infty$.
Our discussion will be rather generic, while specific checks in the case of four--point amplitudes are presented in the sequel. We also point out that we only use the monodromy properties of the vertex insertions on the world-sheet cylinder, which is in particular independent of the space--time 
dimension $D$, as well as the structure of the internal compact space.

\subsection{Planar amplitude relations}

\subsubsection{Planar four--point  amplitude relations}\label{Sect:4PtPlanar}
Before considering the general case of orientable $N$--point amplitudes on the cylinder, we first discuss monodromy relations in the four--point case involving amplitudes with all four points inserted on the same boundary.
Our strategy for finding relations between different color-orderings of planar four-point  amplitudes is to start with a setup in which all four vertex insertions are inserted on the upper boundary with $x_1<x_2<x_3<x_4$, corresponding to the planar amplitude $A^{(1)}(1,2,3,4)$. Next, while keeping the positions of the points $x_{2,3,4}$ on the upper boundary (with $x_2<x_3<x_4$) we integrate the position of $x_1$ along the following closed contour 
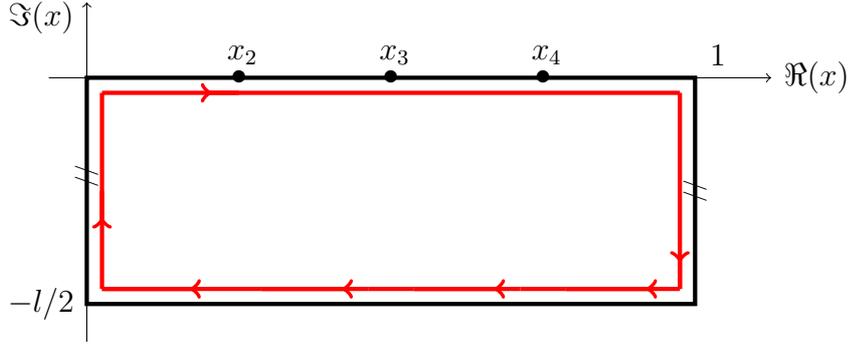
\begin{center}
\begin{tikzpicture}
\draw[->] (-0.5,3) -- (9,3);
\draw[->] (0,-0.5) -- (0,4);
\draw[ultra thick] (0,0) rectangle (8,3); 
\node at (9.6,3) {$\Re (x)$};
\node at (-0.6,3.8) {$\Im (x)$};
\node at (-0.6,0) {$-l/2$};
\node at (8.3,3.3) {1};
\node at (2,3) {$\bullet$};
\node at (2.05,3.3) {$x_2$};
\node at (4,3) {$\bullet$};
\node at (4.05,3.3) {$x_3$};
\node at (6,3) {$\bullet$};
\node at (6.05,3.3) {$x_4$};
\node[rotate=90,scale=0.8] at (0,1.7) {{\small //}};
\node[rotate=90,scale=0.8] at (8,1.5) {{\small //}};
\draw[red, ultra thick,yshift=0.2cm] (0.2,0) -- (1.7,0);
\draw[red, ultra thick,yshift=0.2cm] (2.3,0) -- (3.7,0);
\draw[red, ultra thick,yshift=0.2cm] (4.3,0) -- (5.7,0);
\draw[red, ultra thick,yshift=0.2cm] (6.3,0) -- (7.8,0);
%
\draw[red, ultra thick,yshift=0.2cm] (1.7,0) -- (2.3,0);
\draw[red, ultra thick,yshift=0.2cm] (3.7,0) -- (4.3,0);
\draw[red, ultra thick,yshift=0.2cm] (5.7,0) -- (6.3,0);
\draw[red, ultra thick,-<,yshift=0.2cm] (0.2,0) -- (1.5,0);
\draw[red, ultra thick,-<,yshift=0.2cm] (2.3,0) -- (3.5,0);
\draw[red, ultra thick,-<,yshift=0.2cm] (4.3,0) -- (5.5,0);
\draw[red, ultra thick,-<,yshift=0.2cm] (6.3,0) -- (7.5,0);
\draw[red, ultra thick,-<] (7.8,0.2) -- (7.8,0.7);
\draw[red, ultra thick,-<] (0.2,1.5) -- (0.2,1);
\draw[red, ultra thick,-<] (2,2.8) -- (1.5,2.8); 
\draw[red, ultra thick] (7.8,0.2) -- (7.8,2.8);
\draw[red, ultra thick] (0.2,0.2) -- (0.2,2.8);
\draw[red, ultra thick] (0.2,2.8) -- (7.8,2.8);
\end{tikzpicture}
\end{center}
Since $A^{(1)}(1,2,3,4)$ has no poles in the interior of this curve, using Cauchy's theorem, this contour integral vanishes, thus giving rise to relations between amplitudes with different orderings of the vertices. However, there are two important points
\begin{itemize}
\item Since $\theta_1(z,i\ell)=0$ for $z=0$, the contour integral on the upper boundary is ill-defined at the points $x_1=x_a$ for $a=2,3,4$. It is therefore necessary to provide a prescription how to avoid these singular points in the contour.
\item The vertical pieces of the contour correspond to a contribution in which the vertex at $x_1$ is not inserted on a boundary and thus cannot be interpreted in terms of an amplitude of the type (\ref{DefClosedChannelAmp}). Thus, we need to find a prescription for the contour integration, such that these two contributions mutually cancel.
\end{itemize}
In order to address both issues, we re-write the single contour integral introduced above as three separate contours according to the following picture 
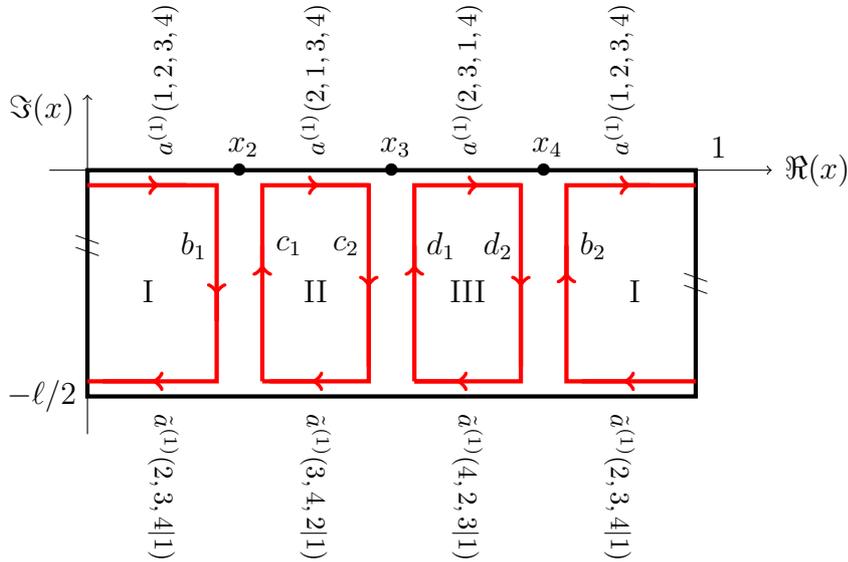
\begin{center}
\begin{tikzpicture}
\draw[->] (-0.5,3) -- (9,3);
\draw[->] (0,-0.5) -- (0,4);
\draw[ultra thick] (0,0) rectangle (8,3); 
\node at (9.6,3) {$\Re (x)$};
\node at (-0.6,3.8) {$\Im (x)$};
\node at (-0.6,0) {$-\ell/2$};
\node at (8.3,3.3) {1};
\node at (2,3) {$\bullet$};
\node at (2.05,3.3) {$x_2$};
\node at (4,3) {$\bullet$};
\node at (4.05,3.3) {$x_3$};
\node at (6,3) {$\bullet$};
\node at (6.05,3.3) {$x_4$};
\node[rotate=90,scale=0.8] at (0,2) {{\small //}};
\node[rotate=90,scale=0.8] at (8,1.5) {{\small //}};
\draw[red,ultra thick] (0,0.2) -- (1.7,0.2) -- (1.7,2.8) -- (0,2.8);
\draw[red, ultra thick,-<] (1.7,1) -- (1.7,1.5);
\draw[red, ultra thick,-<] (1.3,2.8) -- (0.8,2.8); 
\draw[red, ultra thick,-<,yshift=0.2cm] (0.2,0) -- (1,0);
\node[yshift=0.2cm] at (0.8,1.2) {I};
\draw[red,ultra thick] (8,0.2) -- (6.3,0.2) -- (6.3,2.8) -- (8,2.8);
\draw[red, ultra thick,-<] (6.5,0.2) -- (7.2,0.2);
\draw[red, ultra thick,-<] (6.3,2) -- (6.3,1.5);
\draw[red, ultra thick,-<] (7.4,2.8) -- (7,2.8);
\node[yshift=0.2cm] at (7.2,1.2) {I};
\node[yshift=0.2cm] at (1.4,1.8) {$b_1$};
\node[yshift=0.2cm] at (6.65,1.8) {$b_2$};
\node[yshift=0.2cm] at (3,1.2) {II};
\draw[red, ultra thick] (2.3,0.2) -- (3.7,0.2) -- (3.7,2.8) -- (2.3,2.8) -- (2.3,0.2);
\draw[red, ultra thick,-<] (2.4,0.2) -- (3.1,0.2);
\draw[red, ultra thick,-<] (3.7,0.4) -- (3.7,1.6);
\draw[red, ultra thick,-<] (3.7,2.8) -- (2.9,2.8);
\draw[red, ultra thick,-<] (2.3,2) -- (2.3,1.6);
\node[yshift=0.2cm] at (2.65,1.8) {$c_1$};
\node[yshift=0.2cm] at (3.4,1.8) {$c_2$};
\node[yshift=0.2cm] at (5,1.2) {III};
\draw[red, ultra thick] (4.3,0.2) -- (5.7,0.2) -- (5.7,2.8) -- (4.3,2.8) -- (4.3,0.2);
\draw[red, ultra thick,-<] (4.3,0.2) -- (5,0.2);
\draw[red, ultra thick,-<] (5.7,0.4) -- (5.7,1.6);
\draw[red, ultra thick,-<] (5.7,2.8) -- (5,2.8);
\draw[red, ultra thick,-<] (4.3,2) -- (4.3,1.6);
\node[yshift=0.2cm] at (4.65,1.8) {$d_1$};
\node[yshift=0.2cm] at (5.4,1.8) {$d_2$};
\node[rotate=90] at (1,4.2) {\footnotesize $a^{(1)}(1,2,3,4)$}; 
\node[rotate=90] at (3,4.2) {\footnotesize $a^{(1)}(2,1,3,4)$}; 
\node[rotate=90] at (5,4.2) {\footnotesize $a^{(1)}(2,3,1,4)$}; 
\node[rotate=90] at (7,4.2) {\footnotesize $a^{(1)}(1,2,3,4)$}; 
\node[rotate=270] at (1,-1.2) {\footnotesize $\tilde{a}^{(1)}(2,3,4|1)$}; 
\node[rotate=270] at (3,-1.2) {\footnotesize $\tilde{a}^{(1)}(3,4,2|1)$}; 
\node[rotate=270] at (5,-1.2) {\footnotesize $\tilde{a}^{(1)}(4,2,3|1)$}; 
\node[rotate=270] at (7,-1.2) {\footnotesize $\tilde{a}^{(1)}(2,3,4|1)$}; 
\end{tikzpicture}
\end{center}
Here, all three integrals I, II and III are well defined in the sense that they avoid the singular points $x_1=x_a$ (for $a=2,3,4$), however, at the price of having introduced new integral contributions in the bulk of cylinder, which we called $b_{1,2}$, $c_{1,2}$ and $d_{1,2}$. 

Focusing on the horizontal (boundary) contributions first, each of the three contours contains a planar amplitude contribution (with different orderings of the four vertex insertions) and a non-planar contribution with point $x_1$ inserted on the upper boundary. In order to describe these contributions we introduce 
\begin{align}
a^{(1)}(i_1,i_2,i_3,i_4)=V_{\rm CKG}^{-1}\ P_4\ \int_0^1 dx_{i_4}\int_0^{x_{i_4}}dx_{i_3}\int_0^{x_{i_3}}dx_{i_2}\int_0^{x_{i_2}}dx_{i_1}\,\prod_{1\leq a<b\leq 4}\left(\frac{\theta_1(x_{i_bi_a},i\ell)}{\theta'_1(0,i\ell)}\right)^{s_{i_ai_b}}.\label{DefPlanar4}
\end{align}
along with:
\begin{align}
\tilde{a}^{(1)}(i_1,i_2,i_3|j)&=V_{\rm CKG}^{-1}\ P_4\ \int_0^1 dx_{i_3}\int_0^{x_{i_3}}dx_{i_2}\int_0^{x_{i_2}}dx_{i_1}\int_0^{x_{i_1}}dx_j\ \exp\lf\{i\pi \sum_{l=1}^3s_{ji_l}x_{ji_l}\ri\}\ \nonumber\\
&\times \prod_{a=1}^3\left(\frac{\theta_4(x_{ji_a},il )}{\theta'_1(0,il )}\right)^{s_{ji_a}}\ \prod_{1\leq a<b\leq 3}\left(\frac{\theta_1(x_{i_bi_a},il )}{\theta'_1(0,il )}\right)^{s_{i_ai_b}}\ .\label{F4Integrand}
\end{align}
Recall, that the factor $P_4$ comprises all the kinematics for $N\!=\!4$ and the following discussion  
holds for any given kinematics.
Note that the expression 
\begin{align}
\delta(k_1+k_2+k_3+k_4)\int dl\ \left\{\tilde{a}^{(1)}(2,3,4|1)+\tilde{a}^{(1)}(3,4,2|1)+\tilde{a}^{(1)}(4,2,3|1)\right\}\,,
\end{align}
is different from the non--planar four--point amplitude $A^{(1)}(2,3,4|1)$ as defined in (\ref{DefClosedChannelAmp}), due to the position dependent phase factor $\exp\{i\pi \sum\limits_{j=2}^4s_{1j}x_{1j}\}$, which arises from moving the position $x_1$ to the upper boundary, cf. the shift identity (\ref{IdentityShiftTheta}) of the Jacobi--theta functions. 

Finally, considering  vertical contributions $b_{1}$, $c_{2}$ and $d_{2}$, they can be parametrised as
{\allowdisplaybreaks
\begin{align}
&b_1=V_{\rm CKG}^{-1}\ P_4\ \int_0^1 dx_4\int_0^{x_4} dx_3\int_0^{x_3}dx_2\int_0^{l /2} dr\prod_{1\leq a<b\leq 4}\left(\frac{\theta_1(x_{ab},il )}{\theta_1'(0,il )}\right)^{s_{ab}}\bigg|_{x_1=x_2+ir}\,,\nonumber\\
&c_2=e^{-i\pi s_{12}}\ V_{\rm CKG}^{-1}\ P_4\ \int_0^1 dx_4\int_0^{x_4} dx_3\int_0^{x_3}dx_2\int_0^{l /2} dr\prod_{1\leq a<b\leq 4}\left(\frac{\theta_1(x_{ab},il )}{\theta_1'(0,il )}\right)^{s_{ab}}\bigg|_{x_1=x_3+ir}\,,\nonumber\\
&d_2=e^{-i\pi (s_{12}+s_{13})}\ V_{\rm CKG}^{-1}\ P_4\ \int_0^1 dx_4\int_0^{x_4} dx_3\int_0^{x_3}dx_2\int_0^{l /2} dr\prod_{1\leq a<b\leq 4}\left(\frac{\theta_1(x_{ab},il )}{\theta_1'(0,il )}\right)^{s_{ab}}\bigg|_{x_1=x_4+ir}\ ,\label{VerticalPlanar}
\end{align}
}
while $c_{1}$, $d_{1}$ and $b_{2}$ are obtained through:
\begin{align}
&c_1=-e^{-i\pi s_{12}}\,b_1\,,&&d_1=-e^{-i\pi s_{13}}\,c_2\,,&&b_2=-e^{i\pi (s_{12}+s_{13})}\,d_2\,.\label{RelVertical4}
\end{align}
Putting all these contributions together, we find for the three contour integrals I, II and III:
\begin{align}
&\text{contour I:} &&a^{(1)}(1,2,3,4)-\tilde{a}^{(1)}(2,3,4|1)+b_1+b_2=0\ ,\nonumber\\
&\text{contour II:} &&a^{(1)}(2,1,3,4)-\tilde{a}^{(1)}(3,4,2|1)+c_1+c_2=0\ ,\nonumber\\
&\text{contour III:} &&a^{(1)}(2,3,1,4)-\tilde{a}^{(1)}(4,2,3|1)+d_1+d_2=0\ .\label{Eq3}
\end{align}
Due to (\ref{RelVertical4}), we can combine these equations such that the vertical contributions $b_{1,2}$, $c_{1,2}$ and $d_{1,2}$ cancel each other
\begin{align}
a^{(1)}(1,2,3,4)&+e^{i\pi s_{12}}\ a^{(1)}(2,1,3,4)+e^{i\pi (s_{12}+s_{13})}\ a^{(1)}(2,3,1,4)\nonumber\\
&=\tilde{a}^{(1)}(2,3,4|1)+e^{i\pi s_{12}}\ \tilde{a}^{(1)}(3,4,2|1)+e^{i\pi(s_{12}+s_{13})}\ \tilde{a}^{(1)}(4,2,3|1)\ ,\label{AmplitudeRelation}
\end{align}
thus leaving only a relation between the boundary contributions. 
Eq. (\ref{AmplitudeRelation}) is valid for any value of $l $. Therefore, integrating over the modulus $l$  we can (partially) interpret (\ref{AmplitudeRelation}) in terms of color--ordered amplitudes
\begin{align}
&A^{(1)}(1,2,3,4)+e^{i\pi s_{12}}\,A^{(1)}(2,1,3,4)+e^{i\pi(s_{12}+s_{13})}\,A^{(1)}(2,3,1,4)\nonumber\\
&\hspace{0.5cm}=\tilde{A}^{(1)}(2,3,4|1)+e^{i\pi s_{12}}\ \tilde{A}^{(1)}(3,4,2|1)+e^{i\pi(s_{12}+s_{13})}\ \tilde{A}^{(1)}(4,2,3|1)\ ,
\label{FourPointMonodromyRelation}
\end{align}
with:
\be\label{take}
\tilde{A}^{(1)}(i_1,i_2,i_3|j)=\delta(k_1+k_2+k_3+k_4)\,\int_0^\infty dl\  
\tilde{a}^{(1)}(i_1,i_2,i_3|j)\ .
\ee
Here, we have used the holomorphicity of (\ref{DefClosedChannelAmp}) along with the fact that on $\mathcal{I}_{1}$ the positions of all insertion points are ordered, to write the integrand of \emph{e.g.} $A^{(1)}(1,2,3,4)$ without absolute values. Notice furthermore, that due to the position dependent phase factor in (\ref{F4Integrand}), a priori the terms on the r.h.s. have no interpretation in terms of string 
four--point amplitudes. However, we refer the reader to section \ref{Sect:NonCommutative}.
\subsubsection[Planar $N$--point  amplitude relation]{Planar $\bm{N}$--point amplitude relation}\label{Sect:RelNonPlanarAmp4}

The discussion in subsection~\ref{Sect:4PtPlanar} points into an immediate generalization to $N\geq 4$ points. Indeed, starting with a configuration where the points $x_{2,\ldots,N}$ (with $x_2<x_3\ldots<x_N$) are inserted on one boundary, we can consider a closed contour integral of point $x_1$ which we can split into $N-1$ integrals according to the following figure
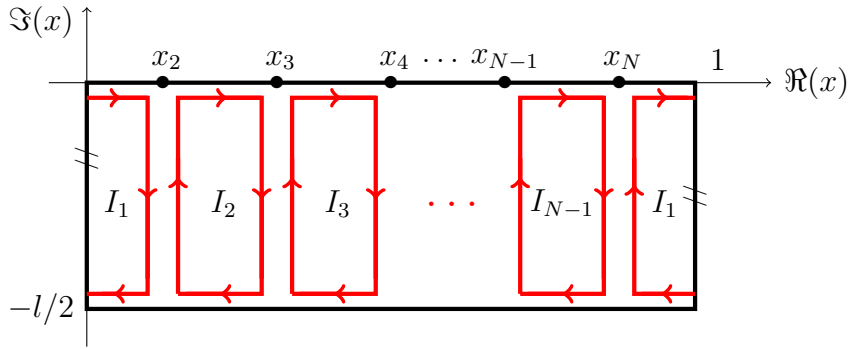
\begin{center}
\begin{tikzpicture}
\draw[->] (-0.5,3) -- (9,3);
\draw[->] (0,-0.5) -- (0,4);
\draw[ultra thick] (0,0) rectangle (8,3); 
\node at (9.6,3) {$\Re (x)$};
\node at (-0.6,3.8) {$\Im (x)$};
\node at (-0.6,0) {$-l /2$};
\node at (8.3,3.3) {1};
\node at (1,3) {$\bullet$};
\node at (1.05,3.3) {$x_2$};
\node at (2.5,3) {$\bullet$};
\node at (2.55,3.3) {$x_3$};
\node at (4,3) {$\bullet$};
\node at (4.05,3.3) {$x_4$};
\node at (4.65,3.3) {{$\ldots$}};
\node at (5.5,3) {$\bullet$};
\node at (5.5,3.3) {$x_{N-1}$};
\node at (7,3) {$\bullet$};
\node at (7.05,3.3) {$x_N$};
\node[rotate=90,scale=0.8] at (0,2) {{\small //}};
\node[rotate=90,scale=0.8] at (8,1.5) {{\small //}};
\draw[red,ultra thick] (0,0.2) -- (0.8,0.2) -- (0.8,2.8) -- (0,2.8);
\draw[red, ultra thick,-<] (0.8,0.4) -- (0.8,1.6);
\draw[red, ultra thick,-<] (0.8,2.8) -- (0.3,2.8); 
\draw[red, ultra thick,-<] (0.2,0.2) -- (0.5,0.2);
\node at (0.4,1.4) {$I_1$};
\draw[red,ultra thick] (8,0.2) -- (7.2,0.2) -- (7.2,2.8) -- (8,2.8);
\draw[red, ultra thick,-<] (7.3,0.2) -- (7.7,0.2);
\draw[red, ultra thick,-<] (7.2,2) -- (7.2,1.5);
\draw[red, ultra thick,-<] (7.9,2.8) -- (7.5,2.8);
\node at (7.6,1.4) {$I_1$};
\node at (1.8,1.4) {$I_2$};
\draw[red, ultra thick] (1.2,0.2) -- (2.3,0.2) -- (2.3,2.8) -- (1.2,2.8) -- (1.2,0.2);
\draw[red, ultra thick,-<] (1.3,0.2) -- (1.9,0.2);
\draw[red, ultra thick,-<] (2.3,0.4) -- (2.3,1.6);
\draw[red, ultra thick,-<] (2.3,2.8) -- (1.7,2.8);
\draw[red, ultra thick,-<] (1.2,2) -- (1.2,1.6);
\node[xshift=1.5cm] at (1.8,1.4) {$I_3$};
\draw[red, ultra thick, xshift=1.5cm] (1.2,0.2) -- (2.3,0.2) -- (2.3,2.8) -- (1.2,2.8) -- (1.2,0.2);
\draw[red, ultra thick,-<, xshift=1.5cm] (1.3,0.2) -- (1.9,0.2);
\draw[red, ultra thick,-<, xshift=1.5cm] (2.3,0.4) -- (2.3,1.6);
\draw[red, ultra thick,-<, xshift=1.5cm] (2.3,2.8) -- (1.7,2.8);
\draw[red, ultra thick,-<, xshift=1.5cm] (1.2,2) -- (1.2,1.6);
\node[xshift=4.5cm] at (1.75,1.4) {$I_{N-1}$};
\draw[red, ultra thick, xshift=4.5cm] (1.2,0.2) -- (2.3,0.2) -- (2.3,2.8) -- (1.2,2.8) -- (1.2,0.2);
\draw[red, ultra thick,-<, xshift=4.5cm] (1.3,0.2) -- (1.9,0.2);
\draw[red, ultra thick,-<, xshift=4.5cm] (2.3,0.4) -- (2.3,1.6);
\draw[red, ultra thick,-<, xshift=4.5cm] (2.3,2.8) -- (1.7,2.8);
\draw[red, ultra thick,-<, xshift=4.5cm] (1.2,2) -- (1.2,1.6);
\node[red, xshift=3cm] at (1.8,1.4) {{\Large $\ldots$}};
\end{tikzpicture}
\end{center}
To describe these contour integrations, we generalize the definitions (\ref{DefPlanar4}) and (\ref{F4Integrand})
\begin{align}
a^{(1)}(i_1,i_2,\ldots,i_N)=V_{\rm CKG}^{-1}\ \int_0^1 dx_{i_N}\int_0^{x_{i_N}}dx_{i_{N-1}}\ldots \int_0^{x_{i_2}}dx_{i_1}\ P_N\ \prod_{1\leq a<b\leq N}\left(\frac{\theta_1(x_{i_bi_a},il )}{\theta'_1(0,il )}\right)^{s_{i_ai_b}}\ ,\label{DefPlanarN}
\end{align}
along with:
\begin{align}
\tilde{a}^{(1)}(i_1,i_2,\ldots,i_{N-1}&|j)=V_{\rm CKG}^{-1}\ \int_0^1 dx_{i_{N-1}}\int_0^{x_{i_{N-1}}}dx_{i_{N-2}}\ldots \int_0^{x_{i_2}}dx_{i_1}\int_0^{x_{i_1}}dx_j\ P_N\nonumber\\
&\times \ \exp\lf\{i\pi \sum\limits_{l=1}^{N-1}s_{ji_l}x_{ji_l}\ri\}\ \prod_{a=1}^{N-1}\left(\frac{\theta_4(x_{ji_a},il )}{\theta'_1(0,il )}\right)^{s_{ji_a}}\ \prod_{1\leq a<b\leq N-1}\left(\frac{\theta_1(x_{i_bi_a},il )}{\theta'_1(0,il )}\right)^{s_{i_ai_b}}.\label{FNIntegrand}
\end{align}
Recall, that  $P_N$ is a modular function of weight $N-4$ and comprises all kinematical factors. In addition, for $N\geq 5$ the function $P_N$ depends also on the vertex operator positions $x_i$. Nonetheless, the following discussion  holds for any given kinematics and is unaffected by the specific form of $P_N$ since 
the latter does not imply branch cuts w.r.t. to the positions $x_i$. 
Similar to the four--point case, the $N-1$ conditions $I_a=0$ for $a=1,\ldots,N-1$ can again be combined in such a way that the vertical contributions cancel each other. We are thus left with only the horizontal contributions corresponding to the following relation between boundary 
integrals:
\begin{align}
&a^{(1)}(1,2,\ldots,N)+e^{i\pi s_{12}}\ a^{(1)}(2,1,\ldots,N)+\ldots+e^{i\pi s_{1,2\ldots N-1}}\ a^{(1)}(2,\ldots,N-1,1,N)\nonumber\\
&=\tilde{a}^{(1)}(2,\ldots,N|1)+e^{i\pi s_{12}}\ \tilde{a}^{(1)}(3,\ldots,N,2|1)+\ldots+e^{i\pi s_{1,2\ldots N-1}}\ \tilde{a}^{(1)}(N,2,\ldots,N-1|1)\ .\label{PlanarNIntegrandRel}
\end{align}
This relation is valid for any value of $\ell$ and thus, upon integrating over the world-sheet modulus, we find the following monodromy relation for the one-loop amplitudes 
\begin{align}
A^{(1)}&(1,2,\ldots,N)+e^{i\pi s_{12}}A^{(1)}(2,1,\ldots,N)+\ldots+e^{i\pi  s_{1,2\ldots N-1}}A^{(1)}(2,\ldots,N-1,1,N)\nonumber\\
&=\tilde{A}^{(1)}(2,\ldots,N|1)+e^{i\pi s_{12}}\ \tilde{A}^{(1)}(3,\ldots,N,2|1)+\ldots+e^{i\pi s_{1,2\ldots N-1}}\ \tilde{A}^{(1)}(N,2,\ldots,N-1|1)\ ,\label{AmplitudeNRelation}
\end{align}
with the $N$--point generalization of \req{take}:
\be\label{Take}
\tilde{A}^{(1)}(a_2,a_3,\ldots,a_N|a_1)=\delta(k_1+\ldots+k_N)\,\int_0^\infty dl\
\tilde{a}^{(1)}(a_2,a_3,\ldots,a_N|a_1)\ .
\ee

\subsection{Non--planar amplitude relations}\label{Sect:StrategyNonPlanar}

The starting point of the discussion in the previous section was a planar four--point amplitude, \emph{i.e.} a configuration where all four points were inserted on the same boundary. By integrating a single point along a closed contour on the cylinder we could relate this planar four--point amplitude to non--planar configurations. In this subsection we shall use the same strategy, but start from a non--planar configuration. We will comment on some of the (conceptual) difficulties arising in this case.

Specifically, we consider a configuration of four points, with $x_2$ inserted on the lower boundary and $x_{3,4}$ inserted on the upper boundary, while $x_1$ is integrated along (a) closed contour(s) on the cylinder. Following the same strategy as in the previous section we consider several different integrals in order to avoid (potential) singularities when $x_1=x_a$ for $a=2,3,4$. 
\begin{figure}[H]
\begin{center}
\begin{tikzpicture}
\draw[->] (-0.5,3) -- (9,3);
\draw[->] (0,-0.5) -- (0,4);
\draw[ultra thick] (0,0) rectangle (8,3); 
\node at (9.6,3) {$\Re (x)$};
\node at (-0.6,3.8) {$\Im (x)$};
\node at (-0.6,0) {$-l /2$};
\node at (8.3,3.3) {1};
\node at (2,3) {$\bullet$};
\node at (2.05,3.3) {$x_2$};
\node at (4,0) {$\bullet$};
\node at (4.05,-0.3) {$x_4$};
\node at (6,0) {$\bullet$};
\node at (6.05,-0.3) {$x_3$};
\node[rotate=90,scale=0.8] at (0,2) {{\small //}};
\node[rotate=90,scale=0.8] at (8,1.5) {{\small //}};
\draw[red,ultra thick] (0.2,0.2) -- (1.7,0.2) -- (1.7,2.8) -- (0.2,2.8) -- (0.2,0.2);
\draw[red, ultra thick,-<] (1.7,1) -- (1.7,1.5);
\draw[red, ultra thick,-<] (1.3,2.8) -- (0.8,2.8); 
\draw[red, ultra thick,-<] (0.2,1.5) -- (0.2,1);
\draw[red, ultra thick,-<,yshift=0.2cm] (0.2,0) -- (1,0);
\node[yshift=0.2cm] at (0.9,1.2) {I};
\draw[red,ultra thick] (7.8,0.2) -- (6.3,0.2) -- (6.3,2.8) -- (7.8,2.8) -- (7.8,0.2);
\draw[red, ultra thick,-<] (6.5,0.2) -- (7.2,0.2);
\draw[red, ultra thick,-<] (6.3,2) -- (6.3,1.5);
\draw[red, ultra thick,-<] (7.4,2.8) -- (7,2.8);
\draw[red, ultra thick,-<] (7.8,0.5) -- (7.8,1);
\node[yshift=0.2cm] at (7.1,1.2) {IV};
\node[yshift=0.2cm] at (0.6,1.8) {$\hat{b}^{(1)}_1$};
\node[yshift=0.2cm] at (1.35,1.8) {$\hat{b}^{(1)}_2$};
\node[yshift=0.2cm] at (6.7,1.8) {$\hat{e}^{(1)}_1$};
\node[yshift=0.2cm] at (7.45,1.8) {$\hat{e}^{(1)}_2$};
\node[yshift=0.2cm] at (3,1.2) {II};
\draw[red, ultra thick] (2.3,0.2) -- (3.7,0.2) -- (3.7,2.8) -- (2.3,2.8) -- (2.3,0.2);
\draw[red, ultra thick,-<] (2.4,0.2) -- (3.1,0.2);
\draw[red, ultra thick,-<] (3.7,0.4) -- (3.7,1.6);
\draw[red, ultra thick,-<] (3.7,2.8) -- (2.9,2.8);
\draw[red, ultra thick,-<] (2.3,2) -- (2.3,1.6);
\node[yshift=0.2cm] at (2.7,1.8) {$\hat{c}^{(1)}_1$};
\node[yshift=0.2cm] at (3.4,1.8) {$\hat{c}^{(1)}_2$};
\node[yshift=0.2cm] at (5,1.2) {III};
\draw[red, ultra thick] (4.3,0.2) -- (5.7,0.2) -- (5.7,2.8) -- (4.3,2.8) -- (4.3,0.2);
\draw[red, ultra thick,-<] (4.3,0.2) -- (5,0.2);
\draw[red, ultra thick,-<] (5.7,0.4) -- (5.7,1.6);
\draw[red, ultra thick,-<] (5.7,2.8) -- (5,2.8);
\draw[red, ultra thick,-<] (4.3,2) -- (4.3,1.6);
\node[yshift=0.2cm] at (4.7,1.8) {$\hat{d}^{(1)}_1$};
\node[yshift=0.2cm] at (5.4,1.8) {$\hat{d}^{(1)}_2$};
\node[rotate=90] at (1,4.15) {\footnotesize $a^{(1)}_{\text{I}}(1,2|3,4)$}; 
\node[rotate=90] at (3,4.2) {\footnotesize $a^{(1)}_{\text{II}}(1,2|3,4)$}; 
\node[rotate=90] at (5,4.2) {\footnotesize $a^{(1)}_{\text{III}}(1,2|,3,4)$}; 
\node[rotate=90] at (7,4.2) {\footnotesize $a^{(1)}_{\text{IV}}(1,2|3,4)$}; 
\node[rotate=270] at (1,-1.2) {\footnotesize $\hat{a}^{(1)}_{\text{I}}(2|3,4,1)$}; 
\node[rotate=270] at (3,-1.2) {\footnotesize $\hat{a}^{(1)}_{\text{II}}(2|3,4,1)$}; 
\node[rotate=270] at (5,-1.2) {\footnotesize $\hat{a}^{(1)}_{\text{III}}(2|3,1,4)$}; 
\node[rotate=270] at (7,-1.2) {\footnotesize $\hat{a}^{(1)}_{\text{IV}}(2|1,3,4)$}; 
\end{tikzpicture}
\end{center}
\end{figure}
\noindent 
For simplicity (and in order to keep the notation as light as possible), we consider the case\footnote{In order to make contact to actual string amplitudes, we also need to take into account $\Re(x_4)<\Re(x_2)<\Re(x_3)$ as well as $\Re(x_4)<\Re(x_3)<\Re(x_2)$, since there is no relative ordering between the points on the lower and upper boundary. See section~\ref{Sect:NonPlanarMonodromy4pt} for details of the explicit computations, where an explicit gauge for one of the insertions points will be chosen, to keep the discussion tractable.} $\Re(x_2)<\Re(x_4)<\Re(x_3)$. Our prescription is schematically shown in the previous figure.
 In order to explicitly describe the various contributions, we first need to properly define the integral regions for the four regions I, II, III, IV. Using the notation (\ref{Regionsc}), we define\footnote{Since this notation is very complex, we shall choose a particular gauge in section~\ref{Sect:AlphaExpansion} when performing explicit computations.}:
\begin{align}
\mathcal{I}_{\text{I}}&=\left\{\bigcup_{i=1}^4 x_i\in E_\tau ^4\ |\ x_1\leq x_2\leq x_4\leq x_3\right\}\ ,\\
\mathcal{I}_{R}&=\left\{\bigcup_{i=1}^4 x_i\in E_\tau ^4\ | \ x_2\leq x_{\pi_R(1)}\leq x_{\pi_R(4)}\leq x_{\pi_R(3)}\right\}\ ,
\end{align}
with  $\pi_{II}(1,4,3)=(1,4,3),\ \pi_{III}(1,4,3)=(4,1,3)$
and $\pi_{IV}(1,4,3)=(4,3,1)$ referring to the three regions $R=\{\text{II\,,III\,,IV}\}$, respectively. \begin{align}
a^{(1)}_\text{I}(1,2|3,4)&=V_{\text{CKG}}^{-1}\ Q_4\int_{\mathcal{I}_{\text{I}}}\prod_{j=1}^4dx_j &\left(\frac{\theta_1(x_{21},il)}{\theta'_1(0,il)}\right)^{s_{12}}\,\left(\frac{\theta_1(x_{34},il)}{\theta'_1(0,il)}\right)^{s_{34}}\ \prod_{1\leq i\leq 2\atop 3\leq j\leq  4}\left(\frac{\theta_4(x_{ji},il)}{\theta'_1(0,il)}\right)^{s_{ij}},\nonumber\\
a^{(1)}_R(1,2|3,4)&=V_{\text{CKG}}^{-1}\ Q_4\int_{\mathcal{I}_R}\prod_{j=1}^4dx_j &\left(\frac{\theta_1(x_{12},il)}{\theta'_1(0,il)}\right)^{s_{12}}\,\left(\frac{\theta_1(x_{34},il)}{\theta'_1(0,il)}\right)^{s_{34}}\ \prod_{1\leq i\leq 2\atop 3\leq j\leq  4}\left(\frac{\theta_4(x_{ji},il)}{\theta'_1(0,il)}\right)^{s_{ij}}\ ,\label{NonPlanar4Integrand}
\end{align}
and as well
\begin{align}
\hat{a}^{(1)}_{\text{I}}(2|3,&4,1)=V_{\text{CKG}}^{-1}\  Q_4\int_{\mathcal{I}_\text{I}}\prod_{j=1}^4dx_j \ \text{exp}\left\{i\pi \sum_{k=2}^4 s_{1k}x_{1k}\right\} \ \left(\frac{\theta_1(x_{34},il)}{\theta'_1(0,il)}\right)^{s_{34}}\nonumber\\
&\times \left(\frac{\theta_4(x_{21},il)}{\theta'_1(0,il)}\right)^{s_{12}}\left(\frac{\theta_1(x_{31},il)}{\theta'_1(0,il)}\right)^{s_{13}}\left(\frac{\theta_1(x_{41},il)}{\theta'_1(0,il)}\right)^{s_{14}}\ \prod_{i={3,4}}\left(\frac{\theta_4(x_{i2},il)}{\theta'_1(0,il)}\right)^{s_{2i}},\nonumber\\
\hat{a}^{(1)}_{R}&(2|\pi_R(3,4,1))=V_{\text{CKG}}^{-1}\ Q_4\int_{\mathcal{I}_R}\prod_{j=1}^4dx_j  \ \text{exp}\left\{i\pi \sum_{k=2}^4 s_{1k}x_{1k}\right\}\ \exp\{\tfrac{\pi i}{2}(-s+u+t)\}\ \nonumber \\
&\times \left(\frac{\theta_1(x_{34},il)}{\theta'_1(0,il)}\right)^{s_{34}}\left(\frac{\theta_4(x_{12},il)}{\theta'_1(0,il)}\right)^{s_{12}}\left(\frac{\theta_1(x_{31},il)}{\theta'_1(0,il)}\right)^{s_{13}}\left(\frac{\theta_1(x_{41},il)}{\theta'_1(0,il)}\right)^{s_{14}}\ \prod_{i={3,4}}\left(\frac{\theta_4(x_{i2},il)}{\theta'_1(0,il)}\right)^{s_{2i}}.
%
%
%
\end{align}
The vertical contributions $(\hat{b}_{1,2}^{(1)}\,,\hat{c}_{1,2}^{(1)}\,,\hat{d}_{1,2}^{(1)}\,,\hat{e}_{1,2}^{(1)})$ are defined in a similar fashion to (\ref{VerticalPlanar}) as integrals in the interior of the cylinder. We specifically  have
\begin{align}
\hat{b}^{(1)}_1=-V_{\text{CKG}}^{-1}\ \frac{\tau}{2}\ Q_4\int_{\mathcal{I_{\text{B}}}}\prod_{j=2}^4dx_j\int\limits_0^{-1}d\alpha\prod_{1\leq i\leq 2\atop 3\leq j\leq  4}\left(\frac{\theta_4(x_{ji},il)}{\theta'_1(0,il)}\right)^{s_{ij}}\ \left(\frac{\theta_1(x_{21},il)}{\theta'_1(0,il)}\right)^{s_{12}}\,\left(\frac{\theta_1(x_{34},il)}{\theta'_1(0,il)}\right)^{s_{34}}\bigg|_{x_1=\alpha\frac{\tau}{2}}.\label{BOUNDb}
\end{align}
where we introduced $\mathcal{I}_{\text{B}}=\left\{\bigcup\limits_{i=1}^3 x_i\in E_\tau ^3\ |\ x_2\leq x_4\leq x_3\right\}$, as well as:
\begin{align}
&\hat{b}^{(1)}_2=-e^{i\pi s_{12}}\,\hat{c}^{(1)}_1\,,&&\hat{c}^{(1)}_2=-\hat{d}^{(1)}_1\,,&&\hat{d}^{(1)}_2=-\hat{e}^{(1)}_1\,,&&\hat{e}^{(1)}_2=-\hat{b}^{(1)}_1\,.\label{PhaseRelsNonPlan}
\end{align}
Recall, that the factor $Q_4$ comprises all kinematical factors for $N\!=\!4$ and the following discussion  
holds for any given kinematics.
Since each of the four contour integrals I, II, III and IV contains no poles, we have from Cauchy's theorem:
\begin{align}
&\text{contour I:} &&a^{(1)}_{\text{I}}(1,2|3,4)+\hat{b}^{(1)}_2+\hat{a}^{(1)}_{\text{I}}(2|3,4,1)+\hat{b}^{(1)}_1=0\,,\nonumber\\
&\text{contour II:} &&a^{(1)}_{\text{II}}(1,2|3,4)+\hat{c}^{(1)}_2+\hat{a}^{(1)}_{\text{II}}(2|3,4,1)+\hat{c}^{(1)}_1=0\,,\nonumber\\
&\text{contour III:} &&a^{(1)}_{\text{III}}(1,2|3,4)+\hat{d}^{(1)}_2+\hat{a}^{(1)}_{\text{III}}(2|3,1,4)+\hat{d}^{(1)}_1=0\,,\nonumber\\
&\text{contour IV:} &&a^{(1)}_{\text{IV}}(1,2|3,4)+\hat{e}^{(1)}_2+\hat{a}^{(1)}_{\text{IV}}(2|1,3,4)+\hat{e}_1^{(1)}=0\,.\label{ContourRelsNonPlanar}
\end{align}
However, taking into account (\ref{PhaseRelsNonPlan}), there is no linear combination of the four equations (\ref{ContourRelsNonPlanar}) such that all bulk contributions $\hat{b}^{(1)}_{1,2}$, $\hat{c}^{(1)}_{1,2}$, $\hat{d}^{(1)}_{1,2}$ and $\hat{e}^{(1)}_{1,2}$ drop out. Thus, any relation between the 
(non--planar) boundary terms also involves contributions where $x_1$ is integrated in the interior of the cylinder whose physical interpretation we leave to \cite{Progress}. For instance, we can combine (\ref{ContourRelsNonPlanar}) in the following way
\begin{align}
&a^{(1)}_{\text{I}}(1,2|3,4)+\hat{a}^{(1)}_{\text{I}}(2|3,4,1)+e^{i\pi s_{12}}\left\{\ a^{(1)}_{\text{II}}(1,2|3,4)+a^{(1)}_{\text{III}}(1,2|3,4)+a^{(1)}_{\text{IV}}(1,2|3,4)\ \right\}\nonumber\\*
&+e^{i\pi s_{12}}\left\{\ \hat{a}^{(1)}_{\text{II}}(2|3,4,1)+\hat{a}^{(1)}_{\text{III}}(2|3,1,4)+\hat{a}^{(1)}_{\text{IV}}(2|1,3,4)\ \right\}+(1-e^{i\pi s_{12}})\ \hat{b}^{(1)}_1=0\ ,\label{PredictFormNonPlanar}
\end{align}
which only depends on the bulk contribution $\hat{b}^{(1)}_1$ (and $\hat{e}^{(1)}_2$). In section~\ref{Sect:AlphaExpansion} we shall work out explicitly some of the terms contributing to (\ref{PredictFormNonPlanar}) and show explicitly the necessity to include integrals in the bulk of the cylinder.

Another problem we shall comment on further in section~\ref{Sect:HigherLoop} is the issue of parameterization invariance of the cylinder. Notice, while the contributions $\hat{b}^{(1)}_2$, $\hat{c}^{(1)}_{1,2}$, $\hat{d}^{(1)}_{1,2}$ and $\hat{e}^{(1)}_1$ are parametrised by $\Re(x_1)=\Re(x_a)$ for $a=2,3,4$, the contributions $\hat{b}^{(1)}_1$ and $\hat{e}^{(1)}_2$ (appearing in \emph{e.g.} (\ref{PredictFormNonPlanar})) are related to the choice of parametrisation of the cylinder (\emph{i.e.} they are parametrised as $\Re(x_1)=0$ or $1$). A related issue, is the fact that the  terms $a^{(1)}_{\text{I}}(1,2|3,4)$ and $a^{(1)}_{\text{II}}(1,2|3,4)+a^{(1)}_{\text{III}}(1,2|3,4)+a^{(1)}_{\text{IV}}(1,2|3,4)$ contribute with a relative factor $e^{i\pi s_{12}}$. Indeed, the physical non-planar four-point amplitude (in our notation) is given by\footnote{\label{Comment1} Note, that factor of $2$ is required in order to reproduce the correct normalization for the field--theory limit, which takes into account a permutation of the labels $1$ and $2$ along the cylinder boundary, cf. also footnote~\ref{Comment2}.} 
\begin{align}
A^{(1)}(1,2|3,4)=2\ \delta(k_1+k_2+k_3+k_4)\int dl\ \lf\{\ a^{(1)}_{\text{I}}(1,2|3,4)+\ri.&a^{(1)}_{\text{II}}(1,2|3,4)+a^{(1)}_{\text{III}}(1,2|3,4)\nonumber\\
&\lf.+a^{(1)}_{\text{IV}}(1,2|3,4)\ \ri\}\ ,
\end{align}
in which the relative ordering of points $x_1$ and $x_2$ is indistinguishable. As we shall explain in section~\ref{Sect:HigherLoop}, the fact that (\ref{PredictFormNonPlanar}) (or other possible combinations of (\ref{ContourRelsNonPlanar})) distinguishes between different orderings of these points can again be related to a failure of the non--planar amplitude to be invariant under reparameterizations of the cylinder. A physical interpretation of (\ref{PredictFormNonPlanar}) is therefore not yet at hand and we refer the reader to \cite{Progress}.

At any rate, for the gauge choice $x_2\!=\!0$ eq. (\ref{PredictFormNonPlanar}) gives rise to the following monodromy relation
\be
\h\ A^{(1)}(2,1|3,4)+
\hat A^{(1)}(2|1,3,4)-(1-e^{-i\pi  s})\ B^{(1)}(2,1|3,4)=0\ ,\label{Mono2a}
\ee 
with:
\begin{align}
A^{(1)}(2,1|3,4)&=2\ \delta(k_1+k_2+k_3+k_4)\int dl\ \lf\{\ a^{(1)}_{\text{II}}(1,2|3,4)+a^{(1)}_{\text{III}}(1,2|3,4)+a^{(1)}_{\text{IV}}(1,2|3,4)\ \ri\},\nonumber\\
\hat A^{(1)}(2|1,3,4)&=\delta(k_1+k_2+k_3+k_4)\int dl\ \left\{\ \hat{a}^{(1)}_{\text{II}}(2|3,4,1)+\hat{a}^{(1)}_{\text{III}}(2|3,1,4)+\hat{a}^{(1)}_{\text{IV}}(2|1,3,4)\ \right\}\ ,\nonumber\\
\hat B^{(1)}(2,1|3,4)&=\delta(k_1+k_2+k_3+k_4)\int dl\ \hat{b}^{(1)}_1\ .
\end{align}
Finally, the $N$--point generalization of \req{Mono2a} immediately follows:
\be
\h\ A^{(1)}(2,1|3,\ldots,N)+\hat A^{(1)}(2|1,3,\ldots,N)-(1-e^{-i\pi  s})\ B^{(1)}(2,1|3,\ldots,N)=0\ .\label{MonoN}
\ee

\subsection{The structure of contour integrals and branch cuts}\label{Sect:BranchCuts}

The derivation of the $N$--point monodromy relations (\ref{AmplitudeNRelation}) and (\ref{PredictFormNonPlanar}) relies on a specific choice of the contour integration of the point $z_1$ along the cylinder. In this subsection we will explain this choice of contour through the structure of branch cuts of the integrand of the planar and non-planar $N$--point function in the complex plane. This will allow us to compare with similar results in the recent literature and particularly point out a number of discrepancies compared to~\cite{Tourkine:2016bak}.

\subsubsection{Planar $\bm{N}$--point amplitude}\label{Sect:PlanMonodromCuts}

We begin with the planar case and recall the particular contour integral chosen in section~\ref{Sect:RelNonPlanarAmp4} to derive the monodromy relation (\ref{AmplitudeNRelation}). To understand this choice of contour  from the structure of branch cuts associated with the  integrand of the closed string planar $N$--point amplitude, we treat for a given kinematics the integrand of $a^{(1)}(1,2,\ldots,N)$ (defined in (\ref{DefPlanarN}))
\begin{align}
I^{(1)}(1,2,\ldots,N)=\prod_{a=2}^{N}\left(\frac{\theta_1(x_{a1},il)}{\theta'_1(0,il)}\right)^{s_{a1}}\ \prod_{2\leq a<b\leq N}\left(\frac{\theta_1(x_{ba},il)}{\theta'_1(0,il)}\right)^{s_{ab}}\ ,
\end{align}
with $x_{a}\in {\bf R}$ fixed for $a=2,\ldots,N$ and 
\begin{align}
0<x_{2}<x_{3}<\ldots< x_{N}<1\,,
\end{align}
as a holomorphic function of the point $x_{1}$ in the (lower) complex (half-)plane. The function $I^{(1)}(1,2,\ldots,N)$ has branch points at $x_1=x_{a}+in l$ for $a=2,\ldots, N$ and $n\in{\bf Z}$. $I^{(1)}(1,2,\ldots,N)$ is thus a priori multivalued on the complex plane. In order to render it single-valued, we need to introduce a set of branch cuts, which we choose to be parallel to the imaginary axis. We remark that this choice is in general not unique and indeed other configurations are also possible. We stress, however, that the choice shown in the following figure can be imposed consistently at the level of the whole amplitude (\ref{Fullfledged}): in this way, monodromy relations derived for different colour orderings can be compared in a consistent fashion.\footnote{This is in fact imperative if one attempts to combine the former and attempt to solve them.}

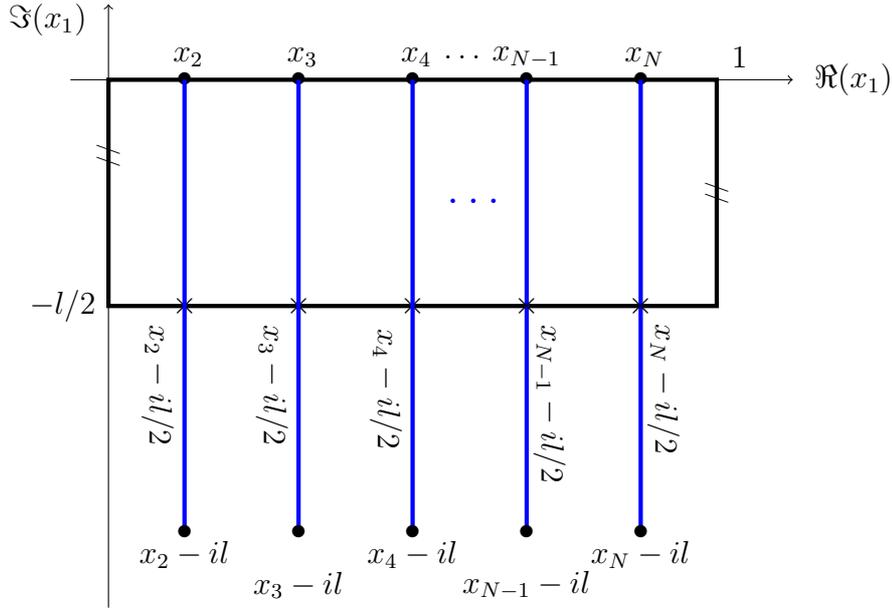
\begin{center}
\begin{tikzpicture}[decoration={
    markings,
    mark=at position 0.5 with {\arrow{>}}}]
\draw[->] (-0.5,3) -- (9,3);
\draw[->] (0,-4) -- (0,4);
\draw[ultra thick] (0,0) rectangle (8,3); 
\node at (9.8,3) {$\Re (x_1)$};
\node at (-0.8,3.8) {$\Im (x_1)$};
\node at (-0.6,0) {$-l /2$};
\node at (8.3,3.3) {1};
\node at (1,3) {$\bullet$};
\node at (1.05,3.3) {$x_2$};
\node at (2.5,3) {$\bullet$};
\node at (2.55,3.3) {$x_3$};
\node at (4,3) {$\bullet$};
\node at (4.05,3.3) {$x_4$};
\node at (4.65,3.3) {{$\ldots$}};
\node at (5.5,3) {$\bullet$};
\node at (5.5,3.3) {$x_{N-1}$};
\node at (7,3) {$\bullet$};
\node at (7.05,3.3) {$x_N$};
\node[rotate=90,scale=0.8] at (0,2) {{\small //}};
\node[rotate=90,scale=0.8] at (8,1.5) {{\small //}};
%
%
%
\node at (1,0) {$\times$};
\draw[ultra thick, blue] (1,3) -- (1,-3);
\node[rotate=270] at (0.65,-1.05) {$x_2-il/2$}; 
\node at (2.5,0) {$\times$};
\draw[ultra thick, blue] (2.5,3) -- (2.5,-3);
\node[rotate=270] at (2.15,-1.05) {$x_3-il/2$};
\node at (4,0) {$\times$};
\draw[ultra thick, blue] (4,3) -- (4,-3);
\node[rotate=270] at (3.65,-1.1) {$x_4-il/2$}; 
\node[blue, xshift=3cm] at (1.8,1.4) {{\Large $\ldots$}};
\draw[ultra thick, blue] (5.5,3) -- (5.5,-3);
\node at (5.5,0) {$\times$};
\node[rotate=270] at (5.8,-1.3) {$x_{N-1}-il/2$}; 
\draw[ultra thick, blue] (7,3) -- (7,-3);
\node at (7,0) {$\times$}; 
\node[rotate=270] at (7.3,-1.1) {$x_N-il/2$}; 
\node at (1,-3) {$\bullet$};
\node at (1,-3.3) {$x_2-il$};
\node at (2.5,-3) {$\bullet$};
\node at (2.5,-3.7) {$x_3-il$};
\node at (4,-3) {$\bullet$};
\node at (4,-3.3) {$x_4-il$};
\node at (5.5,-3) {$\bullet$};
\node at (5.5,-3.7) {$x_{N-1}-il$};
\node at (7,-3) {$\bullet$};
\node at (7,-3.3) {$x_N-il$};
\end{tikzpicture}
\end{center}

\noindent
The branch cuts shown in the above figure, intersect the second boundary (at $\Im(z)=-\tfrac{il}{2}$) of the cylinder at the 'mirror points' $x_{a}-\tfrac{il}{2}$ for $a=2,\ldots,N$. This means, however, when we integrate the point $x_1$ along a closed contour on the cylinder, we have to be careful when this contour approaches these mirror points, since we cannot simply cross the branch cuts. In fact, the contour integrals introduced in section~\ref{Sect:RelNonPlanarAmp4} are designed in such a way as to avoid the branch cuts entirely and only run in regions in which $I^{(1)}(1,i_1,\ldots,i_{N-1})$ is single--valued. Therefore, the integrated form $a^{(1)}(1,i_1,\ldots,i_{N-1})$, defined in (\ref{DefPlanarN}), is well defined. This can be seen in the following figure.
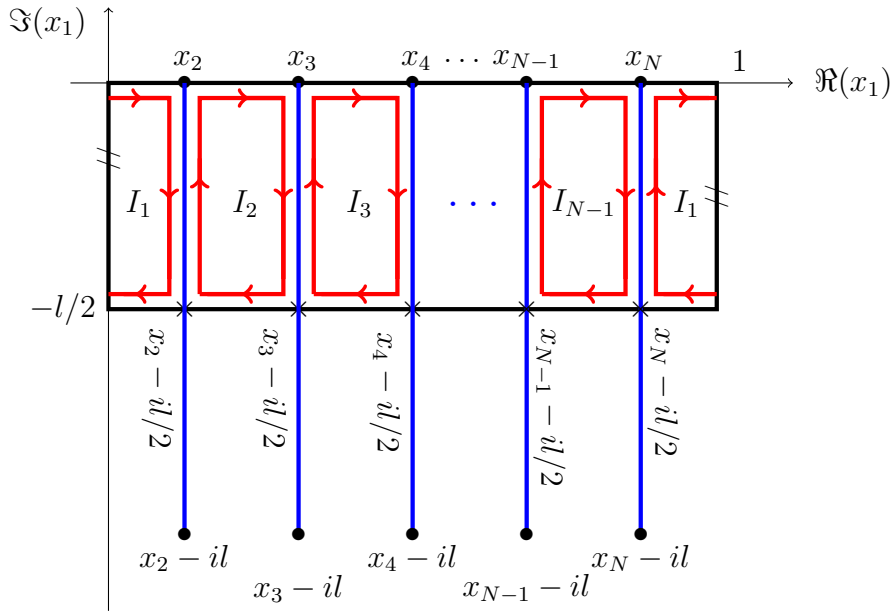
\begin{figure}[H]
\begin{center}
\begin{tikzpicture}[decoration={
    markings,
    mark=at position 0.5 with {\arrow{>}}}]
\draw[->] (-0.5,3) -- (9,3);
\draw[->] (0,-4) -- (0,4);
\draw[ultra thick] (0,0) rectangle (8,3); 
\node at (9.8,3) {$\Re (x_1)$};
\node at (-0.8,3.8) {$\Im (x_1)$};
\node at (-0.6,0) {$-l /2$};
\node at (8.3,3.3) {1};
\node at (1,3) {$\bullet$};
\node at (1.05,3.3) {$x_2$};
\node at (2.5,3) {$\bullet$};
\node at (2.55,3.3) {$x_3$};
\node at (4,3) {$\bullet$};
\node at (4.05,3.3) {$x_4$};
\node at (4.65,3.3) {{$\ldots$}};
\node at (5.5,3) {$\bullet$};
\node at (5.5,3.3) {$x_{N-1}$};
\node at (7,3) {$\bullet$};
\node at (7.05,3.3) {$x_N$};
\node[rotate=90,scale=0.8] at (0,2) {{\small //}};
\node[rotate=90,scale=0.8] at (8,1.5) {{\small //}};
%
%
%
\node at (1,0) {$\times$};
\draw[ultra thick, blue] (1,3) -- (1,-3);
\node[rotate=270] at (0.65,-1.05) {$x_2-il/2$}; 
\node at (2.5,0) {$\times$};
\draw[ultra thick, blue] (2.5,3) -- (2.5,-3);
\node[rotate=270] at (2.15,-1.05) {$x_3-il/2$};
\node at (4,0) {$\times$};
\draw[ultra thick, blue] (4,3) -- (4,-3);
\node[rotate=270] at (3.65,-1.1) {$x_4-il/2$}; 
\node[blue, xshift=3cm] at (1.8,1.4) {{\Large $\ldots$}};
\draw[ultra thick, blue] (5.5,3) -- (5.5,-3);
\node at (5.5,0) {$\times$};
\node[rotate=270] at (5.8,-1.3) {$x_{N-1}-il/2$}; 
\draw[ultra thick, blue] (7,3) -- (7,-3);
\node at (7,0) {$\times$}; 
\node[rotate=270] at (7.3,-1.1) {$x_N-il/2$}; 
\node at (1,-3) {$\bullet$};
\node at (1,-3.3) {$x_2-il$};
\node at (2.5,-3) {$\bullet$};
\node at (2.5,-3.7) {$x_3-il$};
\node at (4,-3) {$\bullet$};
\node at (4,-3.3) {$x_4-il$};
\node at (5.5,-3) {$\bullet$};
\node at (5.5,-3.7) {$x_{N-1}-il$};
\node at (7,-3) {$\bullet$};
\node at (7,-3.3) {$x_N-il$};%
\draw[red,ultra thick] (0,0.2) -- (0.8,0.2) -- (0.8,2.8) -- (0,2.8);
\draw[red, ultra thick,-<] (0.8,0.4) -- (0.8,1.6);
\draw[red, ultra thick,-<] (0.8,2.8) -- (0.3,2.8); 
\draw[red, ultra thick,-<] (0.2,0.2) -- (0.5,0.2);
\node at (0.4,1.4) {$I_1$};
\draw[red,ultra thick] (8,0.2) -- (7.2,0.2) -- (7.2,2.8) -- (8,2.8);
\draw[red, ultra thick,-<] (7.3,0.2) -- (7.7,0.2);
\draw[red, ultra thick,-<] (7.2,2) -- (7.2,1.5);
\draw[red, ultra thick,-<] (7.9,2.8) -- (7.5,2.8);
\node at (7.6,1.4) {$I_1$};
\node at (1.8,1.4) {$I_2$};
\draw[red, ultra thick] (1.2,0.2) -- (2.3,0.2) -- (2.3,2.8) -- (1.2,2.8) -- (1.2,0.2);
\draw[red, ultra thick,-<] (1.3,0.2) -- (1.9,0.2);
\draw[red, ultra thick,-<] (2.3,0.4) -- (2.3,1.6);
\draw[red, ultra thick,-<] (2.3,2.8) -- (1.7,2.8);
\draw[red, ultra thick,-<] (1.2,2) -- (1.2,1.6);
\node[xshift=1.5cm] at (1.8,1.4) {$I_3$};
\draw[red, ultra thick, xshift=1.5cm] (1.2,0.2) -- (2.3,0.2) -- (2.3,2.8) -- (1.2,2.8) -- (1.2,0.2);
\draw[red, ultra thick,-<, xshift=1.5cm] (1.3,0.2) -- (1.9,0.2);
\draw[red, ultra thick,-<, xshift=1.5cm] (2.3,0.4) -- (2.3,1.6);
\draw[red, ultra thick,-<, xshift=1.5cm] (2.3,2.8) -- (1.7,2.8);
\draw[red, ultra thick,-<, xshift=1.5cm] (1.2,2) -- (1.2,1.6);
\node[xshift=4.5cm] at (1.75,1.4) {$I_{N-1}$};
\draw[red, ultra thick, xshift=4.5cm] (1.2,0.2) -- (2.3,0.2) -- (2.3,2.8) -- (1.2,2.8) -- (1.2,0.2);
\draw[red, ultra thick,-<, xshift=4.5cm] (1.3,0.2) -- (1.9,0.2);
\draw[red, ultra thick,-<, xshift=4.5cm] (2.3,0.4) -- (2.3,1.6);
\draw[red, ultra thick,-<, xshift=4.5cm] (2.3,2.8) -- (1.7,2.8);
\draw[red, ultra thick,-<, xshift=4.5cm] (1.2,2) -- (1.2,1.6);
\end{tikzpicture}
\end{center}
\end{figure}

\noindent
As was discussed in detail in section~\ref{Sect:RelNonPlanarAmp4}, this integral prescription leads to the planar monodromy relation (\ref{PlanarNIntegrandRel}) (and its integrated form (\ref{AmplitudeNRelation})), in which also the contributions along the lower boundary $\tilde{a}^{(1)}(i_1,\ldots,i_{N-1}|1)$ are decorated with phase factors.  This is in contrast to the monodromy relations proposed in \cite{Tourkine:2016bak} in which no such phase factors appear. Since the authors of \cite{Tourkine:2016bak} have chosen to work in the open string channel (rather than the closed string channel we have considered thus far), we shall analyse the monodromy relation (\ref{AmplitudeNRelation}) also in this channel in section~\ref{Hongkong}. In doing so, we shall point out various important issues related to the above mentioned structure of branch cuts.

\subsubsection{Non--planar four--point amplitude}

To complete the discussion, we also consider the structure of branch cuts in the closed string non--planar four-point amplitude. 

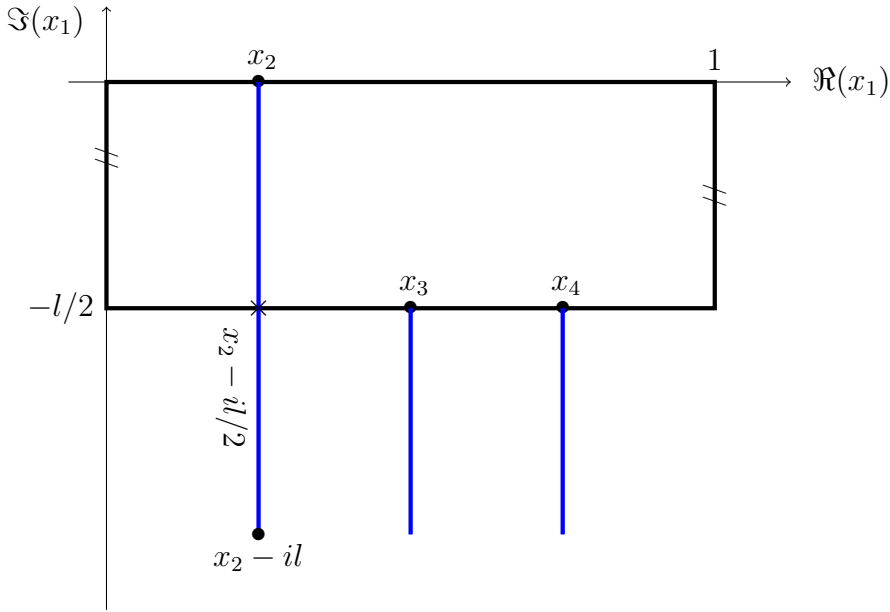
\begin{figure}[H]
\begin{center}
\begin{tikzpicture}
\draw[->] (-0.5,3) -- (9,3);
\draw[->] (0,-4) -- (0,4);
\draw[ultra thick] (0,0) rectangle (8,3); 
\node at (9.8,3) {$\Re (x_1)$};
\node at (-0.8,3.8) {$\Im (x_1)$};
\node at (-0.6,0) {$-l /2$};
\node at (8,3.3) {1};
\node at (2,3) {$\bullet$};
\node at (2.05,3.3) {$x_2$};
\node at (4,0) {$\bullet$};
\node at (4.05,0.3) {$x_3$};
\node at (6,0) {$\bullet$};
\node at (6.05,0.3) {$x_4$};
\node[rotate=90,scale=0.8] at (0,2) {{\small //}};
\node[rotate=90,scale=0.8] at (8,1.5) {{\small //}};
\draw[blue,ultra thick] (2,3) -- (2,-3);
\node[rotate=270] at (1.65,-1.05) {$x_2-il/2$};
\node at (2,0) {$\times$};
\draw[blue,ultra thick] (4,0) -- (4,-3);
\draw[blue,ultra thick] (6,0) -- (6,-3);
\node at (2,-3) {$\bullet$};
\node at (2,-3.3) {$x_2-il$};
\end{tikzpicture}
\end{center}
\end{figure}    

\noindent
As above, for a given kinematics we interpret the integrand of $a^{(1)}(1,2|3,4)$ (defined in (\ref{NonPlanar4Integrand})) 
\begin{align}
I^{(1)}(1,2|3,4)=\left(\frac{\theta_1(x_{21},il)}{\theta'_1(0,il)}\right)^{s_{12}}\,\left(\frac{\theta_1(x_{34},il)}{\theta'_1(0,il)}\right)^{s_{34}}\ \prod_{1\leq i\leq 2\atop 3\leq j\leq  4}\left(\frac{\theta_4(x_{ji},il)}{\theta'_1(0,il)}\right)^{s_{ij}} \ ,\label{Integ4Pt}
\end{align}
with $x_{2,3,4}\in {\bf R}$ fixed and (without loss of generality)
\begin{align}
0<x_2<x_3<x_4<1\,,
\end{align}
as a holomorphic function of $x_1$ in the (lower) complex (half--plane). This function has branch points at $x_2+inl$ and $x_{a}-\tfrac{il(2n+1)}{2}$ for $a=3,4$ and $n\in{\bf Z}$. 
In order to make the function (\ref{Integ4Pt}) single--valued, we need to introduce a set of branch cuts, which we pick in a similar fashion as in the planar case, \emph{i.e.} in the lower half-plane, parallel to the the imaginary axis, as schematically shown in the figure above. 
Upon integrating the point $x_1$ along the cylinder, these branch cuts need to be avoided. In fact, the contour integral introduced in section~\ref{Sect:StrategyNonPlanar} was designed to exactly this end:
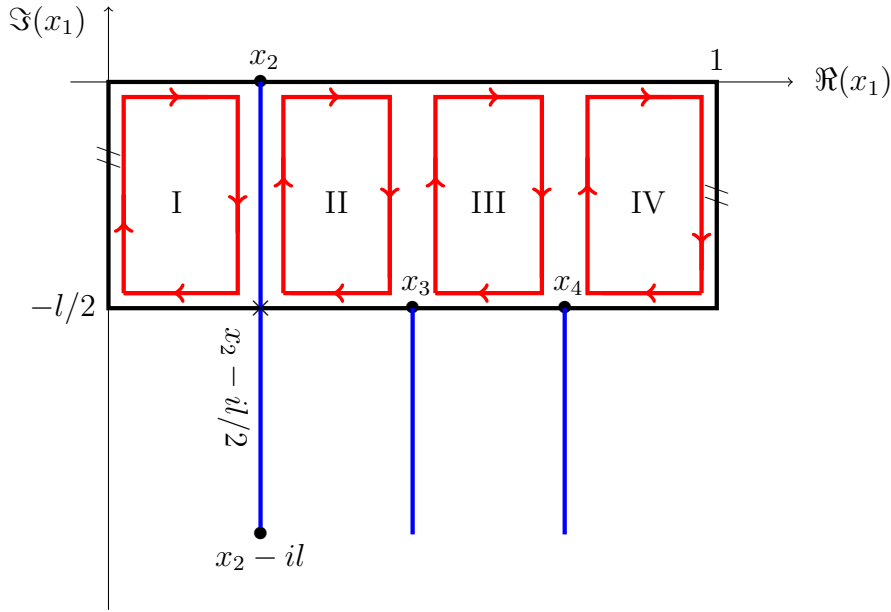
\begin{center}
\begin{tikzpicture}
\draw[->] (-0.5,3) -- (9,3);
\draw[->] (0,-4) -- (0,4);
\draw[ultra thick] (0,0) rectangle (8,3); 
\node at (9.8,3) {$\Re (x_1)$};
\node at (-0.8,3.8) {$\Im (x_1)$};
\node at (-0.6,0) {$-l /2$};
\node at (8,3.3) {1};
\node at (2,3) {$\bullet$};
\node at (2.05,3.3) {$x_2$};
\node at (4,0) {$\bullet$};
\node at (4.05,0.3) {$x_3$};
\node at (6,0) {$\bullet$};
\node at (6.05,0.3) {$x_4$};
\node[rotate=90,scale=0.8] at (0,2) {{\small //}};
\node[rotate=90,scale=0.8] at (8,1.5) {{\small //}};
\draw[blue,ultra thick] (2,3) -- (2,-3);
\node[rotate=270] at (1.65,-1.05) {$x_2-il/2$};
\node at (2,0) {$\times$};
\draw[blue,ultra thick] (4,0) -- (4,-3);
\draw[blue,ultra thick] (6,0) -- (6,-3);
\draw[red,ultra thick] (0.2,0.2) -- (1.7,0.2) -- (1.7,2.8) -- (0.2,2.8) -- (0.2,0.2);
\draw[red, ultra thick,-<] (1.7,1) -- (1.7,1.5);
\draw[red, ultra thick,-<] (1.3,2.8) -- (0.8,2.8); 
\draw[red, ultra thick,-<] (0.2,1.5) -- (0.2,1);
\draw[red, ultra thick,-<,yshift=0.2cm] (0.2,0) -- (1,0);
\node[yshift=0.2cm] at (0.9,1.2) {I};
\draw[red,ultra thick] (7.8,0.2) -- (6.3,0.2) -- (6.3,2.8) -- (7.8,2.8) -- (7.8,0.2);
\draw[red, ultra thick,-<] (6.5,0.2) -- (7.2,0.2);
\draw[red, ultra thick,-<] (6.3,2) -- (6.3,1.5);
\draw[red, ultra thick,-<] (7.4,2.8) -- (7,2.8);
\draw[red, ultra thick,-<] (7.8,0.5) -- (7.8,1);
\node[yshift=0.2cm] at (7.1,1.2) {IV};
\node[yshift=0.2cm] at (3,1.2) {II};
\draw[red, ultra thick] (2.3,0.2) -- (3.7,0.2) -- (3.7,2.8) -- (2.3,2.8) -- (2.3,0.2);
\draw[red, ultra thick,-<] (2.4,0.2) -- (3.1,0.2);
\draw[red, ultra thick,-<] (3.7,0.4) -- (3.7,1.6);
\draw[red, ultra thick,-<] (3.7,2.8) -- (2.9,2.8);
\draw[red, ultra thick,-<] (2.3,2) -- (2.3,1.6);
\node[yshift=0.2cm] at (5,1.2) {III};
\draw[red, ultra thick] (4.3,0.2) -- (5.7,0.2) -- (5.7,2.8) -- (4.3,2.8) -- (4.3,0.2);
\draw[red, ultra thick,-<] (4.3,0.2) -- (5,0.2);
\draw[red, ultra thick,-<] (5.7,0.4) -- (5.7,1.6);
\draw[red, ultra thick,-<] (5.7,2.8) -- (5,2.8);
\draw[red, ultra thick,-<] (4.3,2) -- (4.3,1.6);
\node at (2,-3) {$\bullet$};
\node at (2,-3.3) {$x_2-il$};
\end{tikzpicture}
\end{center}    

\noindent
As discussed in section~\ref{Sect:StrategyNonPlanar}, this choice of contour integrals leads to the monodromy relation~(\ref{PredictFormNonPlanar}), where also some of the vertical integral contributions remain. Notice, since there is no branch cut between the integrals II and III as well as III and IV, also the contributions $a^{(1)}_{\text{II}}(1,2|3,4)$, $a^{(1)}_{\text{III}}(1,2|3,4)$ and $a^{(1)}_{\text{IV}}(1,2|3,4)$ come with a common phase factor and pick up no additional phases. We will further study the structure of branch cuts for non-planar amplitudes in the following section.

\subsection{Open string channel amplitudes and loop momentum}\label{Hongkong}

In this subsection we want to discuss in the open string channel 
the objects $\tilde A^{(1)}(2,\ldots,N|1)$ and $
\hat A^{(1)}(2|1,3,\ldots,N)$  appearing in the monodromy relations \req{AmplitudeNRelation} and \req{MonoN}, respectively. We shall compare them with similar expressions in the recent literature \cite{Tourkine:2016bak} and point out significant differences and serious issues. 
However first, we would like to point out that  the integrands of \req{OpenStringChannelAmp} and \req{GENERIC1} do not represent holomorphic functions in the positions $\tilde z_i$ due to the presence of the second (manifest non--holomorphic) term in the Green's functions (\ref{Go}) and \req{GTo}. 
As a consequence, arguments using the virtue of analyticity become more delicate to apply and it is very important to carefully analyze the structure of branch cuts for each (partial) subamplitude. As discussed in the previous subsection this task can more conveniently be carried out in the closed string sector.

After transforming into the open string channel for $j\!=\!1$ the integrated version \req{Take} of 
\req{FNIntegrand} assumes the following form\footnote{Let us note that extending the integrand of  \req{OpenStringChannelAmp} to a holomorphic
  function in the vertex operator position $\tilde z_1$ and considering the shift
\be\label{SHIFT}
\tilde{z}_1\ \lra\  \tilde{z}_1+\h\ ,
\ee
which in the integrand of \req{OpenStringChannelAmp}  corresponds to 
taking $\Im \tilde{z}_1\ra \Im \tilde{z}_1-\tfrac{i}{2}$
gives also rise to \req{GENERIC2}.}
\begin{align}
\tilde A^{(1)}(2,\ldots,N|1)&=\delta(k_1+\ldots k_N) \int_0^\infty \fc{dt}{t^{1+\tfrac{D}{2}}}\ 
V_{\rm CKG}^{-1}\int_{\Jc_1}\prod_{i=1}^N d\tilde{z}_i\  P_N\ 
\lf(\prod_{j=2}^N\  e^{i\pi s_{1j} \fc{\Im(\tilde{z}_1-\tilde{z}_j)}{\Im \tilde{\tau}} }\ri)\nonumber\\ 
&\times \prod_{2\leq i<j\leq N} 
\lf(\fc{\theta_1(\tilde{z}_{ji},\tilde{\tau})}{\theta_1'(0,\tilde\tau)}\ri)^{s_{ij}} 
\prod_{2\leq j\leq N} \lf(\fc{\theta_2(\tilde{z}_{1j}-\tfrac{1}{2},\tilde{\tau})}{\theta_1'(0,\tilde\tau)}\ri)^{s_{1j}} \prod_{1\leq i<j\leq N} e^{-\pi s_{ij} \fc{\Im(\tilde{z}_j-\tilde{z}_i)^2}{\Im \tilde{\tau}} }\,,\label{GENERIC2}
\end{align}
with $\mathcal{J}_1$ defined in (\ref{Regions}). Like its closed string counter part, which has been discussed in section~\ref{Sect:PlanMonodromCuts}, the integrand of the amplitude (\ref{GENERIC2}) is not single--valued in the interior of the cylinder and  develops monodromies when $\Im(\tilde{z}_j-\tilde{z}_1)$ changes its sign. Therefore,  in order to obtain a well--defined (i.e. single--valued) result we need to introduce a set of branch cuts, which we choose to be located at $\Im(\tilde{z}_j-\tilde{z}_1)=0$. When performing the integration in (\ref{GENERIC2}) the latter need to be avoided and thus the 
$\Im \tilde z_1$--integral 
along the interval $0\leq \Im\tilde z_1\leq t$ has to be split into the $N-1$ regions: 
\begin{align}
0\leq \Im \tilde z_i\leq \Im \tilde z_1\leq \Im \tilde z_{i+1}\ \ \ ,\ \ \ i=1,\ldots,N-1\ .\label{IntegralRegionZ}
\end{align}
The integrand in each of these regions has to be supplemented by proper phase factors to guarantee that we integrate on the same branch. In \req{GENERIC2} we have restricted to the domain $0\leq \Im \tilde z_1\leq \Im \tilde z_2$, i.e. to the region $\Jc_1$.
The latter is the same  $\Jc_1$  as for the planar amplitude $A^{(1)}(1,\ldots,N)$ in (\ref{Regions}).
Note, that the $N-2$ other integration domains $\Im \tilde z_i\leq \Im \tilde z_1\leq \Im \tilde z_{i+1}$ with $i=2,\ldots,N-1$ give rise to the $N-2$ remaining contributions on the r.h.s. of the monodromy relation \req{AmplitudeNRelation}.
In total, together with the above mentioned phases the latter combine into a consistent 
$\Im\tilde z_1$--integral along $0\leq \Im\tilde z_1\leq t$.

There exists  a representation of \req{OpenStringChannelAmp} in terms of $D$--dimensional loop momentum $\ell$ \cite{DHoker:1988pdl}:
\begin{align}
A^{(1)}(1,\ldots,N)&=\delta(k_1+\ldots k_N)\ 2^{-D/2}\ \int_0^\infty \fc{dt}{t}\ V_{\rm CKG}^{-1}\int\limits_{\Jc_1}\prod_{i=1}^N d\tilde{z}_i\ P_N(\tilde{z}_1,\ldots,\tilde{z}_N,\tilde{\tau})\nonumber\\ 
&\times\int_{-\infty}^\infty d^D\ell\ \exp\lf\{-\h\pi \ap t \ell^2-2\pi i  \ap \ell
\sum_{i=1}^N k_i \tilde z_i\ri\}\ \prod_{1\leq i<j\leq N} \lf(\fc{\theta_1(\tilde z_{ji},\tilde\tau)}{\theta_1'(0,\tilde\tau)}\ri)^{s_{ij}}\ .\label{Loopm}
\end{align}
Indeed, the Gaussian integration of the loop momentum $\ell$ yields the correct factor
\begin{align}
2^{D/2}\ (\ap t)^{-D/2}\ \exp\lf\{\fc{\pi\ap}{t}\sum_{i<j}s_{ij}\ \tilde z_{ij}^2\ri\}\,,
\end{align}
such that (\ref{Loopm}) is identical to \req{OpenStringChannelAmp}. In a similar way one can express \req{GENERIC2} in terms of the loop momentum:
\begin{align}
\tilde A^{(1)}(2,\ldots,N|1)&=\delta(k_1+\ldots k_N)\ 2^{-D/2}\ \int_0^\infty \fc{dt}{t}\ V_{\rm CKG}^{-1}\int_{\Jc_1}\prod_{i=1}^N d\tilde{z}_i\ P_N(\tilde{z}_1,\ldots,\tilde{z}_N,\tilde{\tau})\nonumber\\ 
&\times\int_{-\infty}^\infty d^D\ell \ \exp\lf\{-\h\pi \ap t \ell^2-2\pi i  \ap \ell
\sum_{i=1}^N k_i \tilde z_i-\pi i \ap \ell k_1\ri\}\nonumber\\
&\times\prod_{2\leq i<j\leq N} \lf(\fc{\theta_1(\tilde z_{ji},\tilde\tau)}{\theta_1'(0,\tilde\tau)}\ri)^{s_{ij}}
\prod_{2\leq j\leq N} \lf(\fc{\theta_2(\tilde z_{j1}+\tfrac{1}{2},\tilde\tau)}{\theta_1'(0,\tilde\tau)}\ri)^{s_{1j}}\ .\label{LoopM}
\end{align}

Using the loop momentum representations \req{Loopm} and \req{LoopM} of the (partial) subamplitudes entitles to make a comparison with results in the recent work \cite{Tourkine:2016bak}. 
In contrast to \req{FourPointMonodromyRelation} for $N\!=\!4$ in  \cite{Tourkine:2016bak} according to eqs. (9) and (10) the following monodromy relation is stated: 
\begin{align}
A(1,2,3,4)+e^{\pi i \ap k_1k_2}\ A(2,1,3,4)+e^{\pi i \ap k_1(k_2+k_3)}\ A(2,3,1,4)=-A(2,3,4|1)[e^{-\pi i \ap \ell k_1}]\ .\label{TVmonodromy}
\end{align}
The object  $A(2,3,4|1)[e^{-\pi i \ap \ell k_1}]$ is explicitly defined in eq. (8) of \cite{Tourkine:2016bak} and corresponds to\footnote{Note, with $G(z_r,z_s)=-\ln\fc{\theta_1(z_{rs},\tilde\tau)}{\theta_1'(0,\tilde\tau)}$ we have $G(z_r+\tfrac{1}{2},z_s)=-\ln\fc{\theta_2(z_{rs},\tilde\tau)}{\theta_1'(0,\tilde\tau)}$.}: 
\begin{align}\label{SHTV}
A(2,3,4|1)[e^{-\pi i \ap \ell k_1}]=\tilde{A}^{(1)}(2,3,4|1)+\tilde{A}^{(1)}(3,4,2|1)+\tilde{A}^{(1)}(4,2,3|1)\ .
\end{align}
Thus, on the r.h.s. of (\ref{TVmonodromy}) the $\tilde z_1$--integration is performed along the whole boundary $0\leq \Im\tilde z_1\leq t$ without respecting additional phase factors when crossing into a different integration region as defined in (\ref{IntegralRegionZ}). In view of the branch cuts, which are required to render the integrands of (\ref{Loopm}) and (\ref{LoopM}) single--valued in the respective integral region (cf. the discussion above and in section~\ref{Sect:BranchCuts}) this integration is ill--defined. When imposing the monodromy relation \req{TVmonodromy}  $\tilde{z}_1$ is integrated along the boundary at $\Re(\tilde{z})=0$ and relative phase factors are introduced in \req{TVmonodromy} for the various integral regions (\ref{IntegralRegionZ}). While this is indeed compatible with crossing a branch cut as explained above, the absence of similar phase factors for the contributions from the boundary at $\Re(\tilde{z})=\tfrac{1}{2}$ seems to imply that the integrand of $A(2,3,4|1)[e^{-\pi i \ap l k_1}]$ does not live on the same branch as its counterparts $A(1,2,3,4),\ A(2,1,3,4)$ and $A(2,3,1,4)$ on the l.h.s. of (\ref{TVmonodromy}) in the respective integral region (\ref{IntegralRegionZ}).  
Indeed, by the methods exhibited in  section~\ref{Sect:AlphaExpansion} we have 
computed explicit $\ap$--expansions for all quantities in (\ref{TVmonodromy}) and 
could  not verify (\ref{TVmonodromy}) to higher orders in $\ap$.
Therefore, the r.h.s. of  (\ref{TVmonodromy}) implies improper integrations and hence represents an incorrect expression. Moreover, as we shall argue in subsection \ref{FTSubsection} this issue will also lead to wrong conclusions for the field theory limit of the monodromy relations (\ref{TVmonodromy}).

Next, we discuss the non--planar monodromy relation \req{MonoN}.
Treating the integrand of  \req{GENERIC1} (for $N_1=2$) as holomorphic function in the vertex 
operator position $\tilde{z}_1$ and 
considering the shift \req{SHIFT} gives rise to the following object in the open string channel
\begin{align}
\hat A^{(1)}(2|1,3,\ldots,N)&=\delta(k_1+\ldots k_N)\int_0^\infty \fc{dt}{t^{1+\tfrac{D}{2}}}\ 
V_{\rm CKG}^{-1}\int_{\Jc}\prod_{i=1}^N d\tilde{z}_i\ Q_N\ 
\lf(\prod_{j=2}^N\  e^{i\pi s_{1j} \fc{\Im(\tilde{z}_1-\tilde{z}_j)}{\Im \tilde{\tau}} }\ri)\  e^{-i\pi s_{12}}
\nonumber\\ 
&\times \prod_{1\leq i<j\leq N\atop i,j\neq 2} 
\lf(\fc{\theta_1(\tilde{z}_{ij},\tilde{\tau})}{\theta_1'(0,\tilde\tau)}\ri)^{s_{ij}}
\prod_{1\leq j\leq N\atop j\neq 2} \lf(\fc{\theta_2(\tilde{z}_{j2}-\tfrac{1}{2},\tilde{\tau})}{\theta_1'(0,\tilde\tau)}\ri)^{s_{1j}} \prod_{1\leq i<j\leq N} e^{-\pi s_{ij} \fc{\Im(\tilde{z}_j-\tilde{z}_i)^2}{\Im \tilde{\tau}} },\label{GENERIC3a}
\end{align}
with the integration region $\Jc=\{\cup_{i=1}^N x_i\in {\bf R}^N\ |\ 
 t\ \geq  x_3\geq\ldots\geq x_N\geq 0 \cap  0\leq x_1,x_2\leq t\}$.
The discussion on the structure of branch cuts from the previous subsection~\ref{Sect:BranchCuts} can be carried over to the open string channel. Hence, the only branch cut in the interior of the cylinder is located at $\Im (\tilde{z}_2-\tilde{z}_1)=0$. For the particular gauge choice 
$\tilde{z}_2=0$ we can integrate $\tilde{z}_1$ safely along the interior of the cylinder without crossing any further branch cuts and hence without the need to introduce any additional phase factors. Therefore, in the integrand of \req{GENERIC3a} the branch cut structure w.r.t. the coordinate $\tilde z_1$ is altered compared to \req{GENERIC2}.
Moreover, as already stressed above, the contributions from the integrals along $\Im(\tilde{z}_1)=0$ and $\Im (\tilde{z}_1)=t$ no longer cancel each other, thus yielding a non--trivial amount 
proportional to $B^{(1)}(2,1|3,\ldots,N)$ from the interior of the cylinder. We refer the reader to subsection~\ref{Sect:NonPlanarMonodromy4pt} for a more careful discussion of the monodromy relation  \req{Mono2a} together  with  explicit $\ap$--expansions of all (partial) amplitudes involved in this relation. 
For simplicity to confront our result \req{Mono2a} with the findings of \cite{Tourkine:2016bak} let us restrict to the case $N\!=\!4$. From the  relation (9) of \cite{Tourkine:2016bak} for the non--planar monodromy relation the following form is deduced: 
\begin{align}
A(1,2|3,4)&+A(2|1,3,4)[e^{-\pi i \ap \ell k_1}]+e^{\pi i \ap k_1k_3}\ A(2|3,1,4)[e^{-\pi i \ap \ell k_1}]\nonumber\\
&+e^{\pi i \ap k_1(k_3+k_4)}\ A(2|3,4,1)[e^{-\pi i \ap \ell k_1}]=0\ .\label{TVnonplanMonod}
\end{align}
Along our previous discussion for the planar case \req{TVmonodromy} we encounter a different structure of phase factors. In fact, according to our discussion from above for the gauge choice $\tilde z_2=0$  in \req{TVnonplanMonod} phases for the coordinate $\Im\tilde z_1$ passing the points $\tilde z_3$ and $\tilde z_4$ should not be there in front of the third and fourth term.
Moreover, the contributions proportional to $B^{(1)}(2,1|3,4)$ from the interior of the cylinder are missing. Both flaws are not compatible with the structure of the branch cuts of the non--planar amplitude. As discussed in section~\ref{Sect:StrategyNonPlanar} a cancellation of all such bulk terms (cf.  eq.~(\ref{BOUNDb}) for their closed string counterpart) is not possible. Indeed, we shall demonstrate in section~\ref{Sect:NonPlanarMonodromy4pt} that such terms are required to render the 
non--planar monodromy relation consistent as a power--series expansion in closed string Fourier coordinate 
$q=e^{2\pi i\tau}$. Furthermore, as we shall discuss in section~\ref{Sect:FieldTheoryLimit},  eq. (\ref{TVnonplanMonod})  does also not yield correct relations in the field theory limit.

To conclude, we have just discussed  the two examples \req{TVmonodromy} and \req{TVnonplanMonod}
following from  eq. (9) of \cite{Tourkine:2016bak} and evidenced that in the first example 
monodromy phases are missing while in the second case phases are not correctly placed and important bulk contributions are missing.

\section{One--loop superstring amplitude: $\bm{\ap}$--expansion and monodromy relations}\label{Sect:AlphaExpansion}

In this section we consider the $\alpha'$--expansions of the one--loop open superstring amplitudes introduced in section~\ref{Sect:OneLoopAmps} with the goal to analyze the underlying structure of the resulting elliptic periods
and provide explicit verifications and checks of the monodromy relations derived  in section~\ref{Sect:MonodromyOrientable}. 
Throughout this section we shall work in the closed string channel with modular parameter \req{modcl} and the vertex operator positions \req{vertcl}.

\subsection[Elliptic iterated integrals and $\ap$--expansion]{Elliptic iterated integrals and $\bm{\ap}$--expansion}
\label{ellipticsection}

Generically, the one--loop amplitudes \req{OpenStringChannelAmp} and \req{GENERIC1} have branch cuts  in the kinematic invariants $s_{ij}$ due to the fact that massless open strings are exchanged in the loop. These non--analytic contributions 
stem from the boundaries of the integration over the cylinder modulus. For fixed $\tilde\tau$ these effects can be isolated and subtracted. As a result we have analytical terms giving rise to higher derivative interactions in the effective action.
For finite values of the modulus $\tilde\tau$ the structure of the singularities in the kinematic invariants 
is given by the local operator product expansion of vertex operators governing the singularity structure
of two coinciding insertion points on the string world--sheet. Hence, for finite $\tilde\tau$ the poles to be expected are the same as for tree--level, \emph{i.e.} $N-3$--th poles in the kinematic invariants.
Therefore, for finite $\tau$ the integrands \req{ClosedStringChannelAmp} and \req{DefClosedChannelAmp} can be expanded w.r.t. $\ap$ yielding an absolutely convergent powers series in the $s_{ij}$ as introduced in (\ref{Mandel}). Specifically, the integrand of each amplitude can be written as a power series in the Green functions \req{G} and \req{GT} supplemented by the corresponding powers in $s_{ij}$ by using:
\begin{align}
\exp\lf\{\tfrac{1}{2}\sum_{1\leq i<j\leq N} s_{ij}\ G(z_j-z_i,\tau)\ri\}&=1+\tfrac{1}{2}\sum_{1\leq i<j\leq N} s_{ij}\ 
G(x_j-x_i,\tau)+\Oc(\ap^2)\ ,\label{expexp}\\
\exp\lf\{\tfrac{1}{2}\sum_{1\leq i\leq N_1\atop N_1+1\leq j\leq N_2} s_{ij}\ G_T(z_j-z_i+\tfrac{il}{2},\tau)\ri\}&=1+\tfrac{1}{2}\sum_{1\leq i\leq N_1\atop N_1+1\leq j\leq N_2} s_{ij}\ G_T(x_j-x_i,\tau)+\Oc(\ap^2)\ .\label{expexp1}
\end{align}
The (real) vertex positions $x_l$ are given in \req{positionsc} and take values in the interval 
$x_l\in[0,1]$. The integrals \req{ClosedStringChannelAmp} and \req{DefClosedChannelAmp} over vertex positions generically will lead to elliptic iterated integrals, i.e.  integrals along a certain path on an elliptic curve. 
These iterated integrals can be related to the multiple elliptic polylogarithms introduced by Brown and Levin after constraining to one of the canonical paths $A$ and $B$ to be explained below \cite{BL}.
The resulting periods are called  multiple elliptic zeta values (eMZVs).

Elliptic functions are the natural framework to describe higher loop string amplitudes
and their elliptic properties are related to genuine string effects. We should emphazise that in the region of finite $\tau$ we cannot probe the field theory properties, which will be discussed separately in section \ref{Sect:FieldTheoryLimit}. The latter emerge in the limit $\tilde\tau\ra i\infty$, i.e. $\tau\ra 0$ and in this region the underlying elliptic  functions cannot be related to any similar functions appearing at certain field theory computations. Thus, the appearance of elliptic functions in higher--loop string theory and in the description of certain field theory loop diagrams seems to be  unrelated.

Iterated integrals on the once--punctered complex elliptic curve $E_\tau^\times$ (corresponding to $E_\tau$ with  the origin removed) are related to multiple elliptic polylogarithms. The latter are multi--valued functions on $E_\tau$ with unipotent monodromy and are constructed by an averaging  procedure  over $q$ \cite{BL}. Together with the one--form
$\nu=2\pi i \ \fc{d\Im z}{\Im \tau}$ the classes $[dz],[\nu]$ form a basis
for $H^{(1)}(E_\tau^\times,{\bf C})$ with coordinate $z=r+s\tau,\ r,s\in{\bf R}$. A class of elliptic iterated integrals 
\be\label{eII}
\int_0^z \omega^{(n_1)}(z_1)\ \int_0^{z_1} \omega^{(n_2)}(z_2)\ \ldots 
\int_0^{z_{r-1}}\omega^{(n_r)}(z_r)\ 
\ee
 uses the family of one--forms $\omega^{(k)}$ on $E_\tau^\times$. The latter arise as the coefficients of 
 $\alpha^{k-1}$ in the expansion of the Eisenstein--Kronecker series
\be
\Omega(z,\alpha,\tau)=
\exp\lf\{2\pi i \alpha\ \fc{\Im(z)}{\Im(\tau)}\ri\} \ 
\fc{\theta_1'(0,\tau)\ \theta_1(z+\alpha,\tau)}{\theta_1(z,\tau)\theta_1(\alpha,\tau)}\ dz=\sum_{k=0}^\infty \omega^{(k)}(z,\tau)\ \alpha^{k-1}\ ,
\ee
which in turn can be expressed in terms  of the Green function $g(z,\tau):=\tilde G(z,\tau)$  given in \req{Go} and Eisenstein series $G_k=\sum\limits_{m,n\in{\bf Z}\atop (m,n)\neq (0,0)}(m\tau+n)^{-k},\ k\geq 2,\ \hat G_2=G_2-\frac{\pi}{\Im \tau}$ as:
\begin{align}
\omega^{(0)}(z,\tau)&=dz\ ,\nonumber\\
\omega^{(1)}(z,\tau)&=\partial g\ dz,\nonumber\\
\omega^{(2)}(z,\tau)&=\h\ \lf\{(\partial g)^2+\partial^2 g+\hat G_2\ri\}\ dz,\label{forms}\\
\omega^{(3)}(z,\tau)&=\fc{1}{6}\ \lf\{(\partial g)^3+3\ \partial g\ (\partial^2 g+\hat G_2)+\partial^3g-2\ G_3\ri\}\ dz,\nonumber\\
\vdots&\nonumber
\end{align}
Hence, the elliptic iterated integrals \req{eII} allow to implement
the Green function \req{G} with arguments \req{positionsc} as
\be\label{path1}
G(z_{ji},\tau)\simeq\int^{z_{ji}}  \omega^{(1)}(u,\tau)=\int^{z_{ji}} du\ \lf\{\ \partial\ln\theta_1(u,\tau)+2\pi i\ \fc{\Im(u)}{\Im(\tau)}\ \ri\}\ ,
\ee
if the integration on the elliptic curve $E^\times_\tau$ is restricted to a real path $\Im(u)=0$. 
The forms \req{forms} can be integrated along both homology cycles $A$ and $B$ of $E_\tau$. Therefore, 
the Green function \req{GT} can be obtained from
\be\label{path2}
G_T(z_{ji},\tau)\simeq\int^{z_{ji}+\tau/2} \omega^{(1)}(u,\tau)=\int^{z_{ji}} du\ \lf\{\ \partial\ln\theta_4(u,\tau)+2\pi i\ \fc{\Im(u)}{\Im(\tau)}\ \ri\}
\ee 
by deforming from the real path in \req{path1} to also include the interval $[0,\tau]$. 
The elliptic iterated integrals established in \cite{Broedel:2014vla} are of the 
form \req{eII} with an integration path along only real~$z_i$, i.e. $t_i\in [0,1]$
\be\label{eii}
\omega(n_1,\ldots,n_r)=
\int_0^1 \omega^{(n_r)}(t_r)\ \ldots  \int_0^{t_3}\omega^{(n_2)}(t_2)\  \int_0^{t_2}\omega^{(n_1)}(t_1)\ ,
\ee
which integrate to eMZVs (A--elliptic multiple zeta values). The latter have a Fourier expansion in $q$ as:
\be\label{eMZV}
\omega(n_1,\ldots,n_r)=\omega_0(n_1,\ldots,n_r)+\sum_{k=1}^\infty c_k\ q^k\ .
\ee
The first term is given by a MZV or integer powers of $2\pi i$,
 the sum $\sum_{i=1}^r n_i=w$ is called the weight of $\omega$ and $r$ its length. The Fourier coefficients $c_k$ are ${\bf Q}[(2\pi i)^{-1}]$--linear combinations of MZVs.
An other set of eMZVs can be defined\footnote{For this set there is also a homotopy invariant definition or completion of iterated elliptic integrals \req{eII}, which also uses the one--form $\nu$. The resulting eMZVs referring to the $B$--cycle are called B--elliptic multiple zeta values. We refer the reader to \cite{Matthes} for a recent detailed account on eMZVs.} by choosing the path of integration 
from $0$ to $\tau$, replacing the integration region $[0,1]$ by $[0,\tau]$ as in \req{path2} (Enriquez' B--elliptic multiple zeta values).
These integrals appear in the  modular transformation of eMZVs. 
The two sets of eMZVs, namely A--elliptic multiple and Enriquez' B--elliptic multiple zeta values, are obviously related to the two homology cycles $A$ and $B$ of the elliptic curve $E_\tau^\times$, respectively.

In \cite{Broedel:2014vla} only the planar case \req{path1} is discussed and all iterated integrals
\req{eII} are constrained to the real line.
However, our framework is more general since in \req{eII} we also allow for complex $z$
to also describe  the non--planar case \req{path2}. In other words, we also discuss 
the other integration path from $0$ to $\tau$ giving rise to Enriquez' B--elliptic multiple zeta values.
This other set of eMZVs is also probed by the boundary (bulk) terms necessary in the monodromy relations for non--planar amplitudes involving more than one state  at each  boundary, cf. subsection \ref{transcend}.

Using\footnote{The following manipulations hold for $x\in{\bf C}$ and generic modular parameter $\tau$.} eq. \req{Ma1} the propagator \req{G} can also be expressed as 
(with $q=e^{2\pi  i \tau}\equiv e^{-2\pi l}$)
\be
G(x,\tau)=\ln\lf(\fc{\sin(\pi x)}{\pi}\ri)+4\ \sum_{m\geq 1}\fc{q^m}{1-q^m}
\fc{\sin^2(m\pi x)}{m}+{\rm hc.}\ .
\ee
Furthermore, according to \req{Ma2} the scalar Green function \req{GT} in the closed string channel
can also be expressed as:  
\begin{align}
G_T(x,\tau)&=-\fc{1}{8}\ln |q|^2-2\ln(2\pi)+4\ \sum_{m\geq 1}\lf(\fc{q^{m/2}}{1-q^m}
\fc{\sin^2(m\pi x)}{m}+{\rm hc.}\ri)\nonumber\\
&+2\sum_{m\geq 1}\lf(\fc{q^{2m}}{1-q^{2m}}\fc{1}{m}+\fc{q^{m/2}}{1-q^{m}}\fc{(-1)^m}{m}+{\rm hc.}\ri)\ .
\label{GTa}
\end{align}
Finally, thanks to the identity
\be\label{NonTrivial}
\sum_{m=1}^\infty\fc{1}{m}\ \fc{q^m}{1-q^m}-\sum_{m=1}^\infty\fc{1}{m}\ \fc{q^{m/2}}{1-q^m}=
\sum_{m=1}^\infty\fc{1}{m}\ \fc{q^{2m}}{1-q^{2m}}+\sum_{m=1}^\infty\fc{1}{m}\ \fc{(-1)^m\ q^{m/2}}{1-q^{m}}=:\h\ Q_3\ ,
\ee
and $4\sin^2x-2=-2\cos(2x)$ the Green function \req{GTa} may also be cast into the form 
\begin{align}
G_T(x,\tau)&=-\fc{1}{8}\ln q-\ln(2\pi)-2\ \sum_{m\geq 1}\fc{q^{m/2}}{1-q^m}
\fc{\cos(2m\pi x)}{m}+2\sum_{n\geq 1}\Li_1(q^n)+{\rm hc.}\nonumber\\
&=-\fc{1}{24}\ln |q|^2-2\ \sum_{m\geq 1}\lf(\fc{q^{m/2}}{1-q^m}
\fc{\cos(2m\pi x)}{m}+{\rm hc.}\ri)-2\ln(2\pi)|\eta(q)|^2\ ,\label{G_T}
\end{align}
with the following definition of the polylogarithm: 
\be
\Li_a(x)=\sum_{m=1}^\infty\fc{x^m}{m^a}\ \ \ ,\ \ \ |x|<1\ .
\ee

In the following we shall study in detail open superstring four--point amplitudes, which are relevant for the monodromy relations (\ref{FourPointMonodromyRelation}) and (\ref{Mono2a}). For this in section \ref{Sect:AlphaPlanar} we start with the computation of the $\alpha'$--expansion of the planar and non--planar subamplitudes. These results allow us to explain in detail all technical details of the computation, which we shall use later on to perform explicit checks of the monodromy relations. For simplicity, we consider the particular gauge $z_1\!=\!0$ and integrate $z_2$ along a closed loop on the cylinder, for which  the non--planar monodromy relation takes a particularly simple form\footnote{Notice that the form of the non--planar monodromy relation seemingly depends on the choice of this gauge: in appendix~\ref{App:4PtOtherGauge} we have established a similar monodromy relation for the gauge $z_3=1-\frac{i\ell}{2}$, which shows a distinctively different structure of phase factors multiplying the various contributions. Similar to other problems already mentioned earlier, we leave a physical interpretation of the non--planar monodromy relation to a future publication \cite{Progress}.}.
Finally, we shall also comment on  the generalizations to arbitrary multiplicity $N$ and provide some transcendentality constraints on the terms of the $\ap$--expansion following from considering the monodromy relations.

\subsection[Four--point amplitudes: $\ap$--expansion of subamplitudes]{Four--point amplitude: $\bm{\ap}$--expansion of subamplitudes}\label{Sect:AlphaPlanar}

The partial planar four--point open superstring amplitude can be written as
\be\label{Planar}
A^{(1)}(1,2,3,4)=(s_{12}s_{14})\ A^{(0)}_{YM}(1,2,3,4)\ \int_0^\infty dl\ f(s_{12},s_{23})\ ,
\ee
with the partial SYM amplitude $A^{(0)}_{YM}(1,2,3,4)$ and the form factor
\be\label{FormPl}
f(s_{12},s_{23})= \int_0^1dx_4 \int_0^{x_4} dx_3\int_0^{x_3}dx_2\ \exp\lf\{\tfrac{1}{2}\sum_{1\leq i<j\leq 4} s_{ij}\ G(x_j-x_i,\tau)\ri\}
\ee
corresponding to the gauge choice $x_1=0$.
For finite $\tau$  the integrand of the amplitude \req{Planar} can be expanded w.r.t. $\ap$ by using
the series expansion \req{expexp}
\bea
\exp\lf\{\tfrac{1}{2}\sum_{1\leq i<j\leq 4} s_{ij}\ G(x_j-x_i,\tau)\ri\}&=&1+\tfrac{1}{2}\ s\ [\ G(x_{21})+G(x_{43})-G(x_{31})-G(x_{42})\ ]\\
&+&\tfrac{1}{2}\ u\ [\ G(x_{41})+G(x_{32})-G(x_{31})-G(x_{42})\ ]+\Oc(\ap^2)\ ,\nonumber
\eea
with the Mandelstam variables \req{Mandel}
\begin{align}\label{Manel4}
&s=s_{12}=s_{34}\,,&&t=s_{13}=s_{24}\,,&&u=s_{14}=s_{23}\ ,
\end{align}
which satisfy $s+t+u=0$.
To determine the linear order in $\ap$  of \req{Planar} amounts to consider the following
iterated integrals:
\begin{align}
&\int_0^1dx_4 \int_0^{x_4} dx_3\int_0^{x_3}dx_2=\fc{1}{6}\ ,\label{IteratedP}\\
&\h\int_0^1dx_4 \int_0^{x_4} dx_3\int_0^{x_3}dx_2\ [\ G(x_{2},\tau)+G(x_{43},\tau)-G(x_{3},\tau)-G(x_{42},\tau)\ ]\nonumber\\
 &=-\fc{3}{2\pi^2}\ \z_3-\fc{3}{\pi^2}\sum_{n=1}^\infty\Li_3(q^n)\ ,\nonumber\\
&\h \int_0^1dx_4 \int_0^{x_4} dx_3\int_0^{x_3}dx_2\ [\ G(x_{4},\tau)+G(x_{32},\tau)-G(x_{3},\tau)-G(x_{42},\tau)\ ]\nonumber\\
  &=-\fc{3}{2\pi^2}\ \z_3-\fc{3}{\pi^2}\sum_{n=1}^\infty\Li_3(q^n)\ .\nonumber
\end{align}
The individual integrals are computed in appendix~\ref{App:IteratedIntegs}.
Hence, for \req{Planar} we obtain the following expression:
\be
f(s,u)=\fc{1}{6}-(s+u)\ \lf\{\fc{3}{2\pi^2}\ \z_3+\fc{3}{\pi^2}\sum_{n=1}^\infty\Li_3(q^n)\ \ri\}+\Oc(\ap^2)\ .
\ee
Note, that this result has also been found in \cite{Broedel:2014vla} with the emphasis on elliptic MZVs.
Actually, we shall also perform the integration over the world--sheet modulus $\tau$. 
For this we use 
\be 
\int\limits_0^\infty  dl\ l^\eps\ \sum_{n=1}^\infty\Li_a(q^n)=\fc{1}{2\pi}\ \lf\{\ \fc{1}{\eps}\ \zeta(1+a)-\ln(2\pi)\ \zeta(1+a)+\zeta'(1+a)\ \ri\}+\Oc(\epsilon)\ ,
\ee
with some regularization parameter $\epsilon>-1$ and $\zeta'(s)=-\sum\limits_{n=1}^\infty\fc{\ln n}{n^s},\ s>1$.
With this information we find:
\begin{align}
\int_0^\infty dl\ l^\eps\ f(s,u)&=\int_0^\infty dl\ l^\eps\ \lf(\fc{1}{6}-(s+u)\ \fc{3}{2\pi^2}\ \z_3+\Oc(\ap^2)\ri)
\nonumber\\
&-(s+u)\ \lf\{\ \fc{\pi}{60}\ \fc{1}{\eps}-\fc{\pi}{60}\ln(2\pi)+\fc{3}{2\pi^3}\ \zeta'(4)\ \ri\}+\Oc(\ap^2)\ .\label{ResultP}
\end{align}
The first line of \req{ResultP} describes UV divergence related to the tadpole contribution 
 given by the $\ap$--derivative of the tree--level expression $(2\pi^2su)^{-1}\partial_{\alpha'}\fc{\Gamma(1+s)\Gamma(1+u)}{\Gamma(1+s+u)}$ \cite{Green:1981ya}.

Next, the partial non--planar four--point open superstring amplitude can be written as
\be\label{Nonplanar}
A^{(1)}(2,3,4|1)=(s_{12}s_{14})\ A^{(0)}_{YM}(1,2,3,4)\ \int_0^\infty dl\lf[\ g(s,u)+g(t,s)+g(u,t)\ \ri]\ ,
\ee
with the partial SYM amplitude $A^{(0)}_{YM}(1,2,3,4)$ and the form  factor
\begin{align}\label{FormNPl}
g(s,u)&=\int_0^1dx_4 \int_0^{x_4} dx_3\int_0^{x_3}dx_2\ 
\exp\lf\{\tfrac{1}{2}\sum_{2\leq i<j\leq 4} s_{ij}\ G(x_j-x_i,\tau)\ri\}\nonumber\\ 
&\times\exp\lf\{\tfrac{1}{2}\sum_{j=2}^4 s_{1j}\ G_T(x_j-x_1,\tau)\ri\}\ ,
\end{align}
corresponding to the gauge choice $x_1=0$.
For finite $\tau$  the integrand of the amplitude \req{Nonplanar} can be expanded w.r.t. $\ap$ by using
the series expansion \req{expexp1}
\bea
\exp\lf\{\tfrac{1}{2}\sum_{2\leq i<j\leq 4} s_{ij}\ G(x_j-x_i,\tau)\ri\}&&\hskip-0.75cm\exp\lf\{\tfrac{1}{2}\sum_{j=2}^{4} 
s_{1j}\ G_T(x_j-x_1,\tau)\ri\}\\
&=&1+\tfrac{1}{2}\ s\ [\ G_T(x_{21})+G(x_{43})-G_T(x_{31})-G(x_{42})\ ]\nonumber\\
&+&\tfrac{1}{2}\ u\ [\ G_T(x_{41})+G(x_{32})-G_T(x_{31})-G(x_{42})\ ]+\Oc(\ap^2)\ .\nonumber
\eea
Up to the linear order we need to compute the following iterated integrals
\begin{align}
&\h\int_0^1dx_4 \int_0^{x_4} dx_3\int_0^{x_3}dx_2\ [\ G_T(x_{2},\tau)+G_T(x_{43},\tau)-G_T(x_{3},\tau)-G(x_{42},\tau)\ ] \nonumber\\
&=-\fc{3}{4\pi^2}\ \z_3-\fc{3}{2\pi^2}\ \sum_{n=1}^\infty[\Li_3(q^n)+\Li_3(q^{n-1/2})]\ ,\nonumber\\
&\h\int_0^1dx_4 \int_0^{x_4} dx_3\int_0^{x_3}dx_2\ 
[\ G_T(x_{4},\tau)+G(x_{32},\tau)-G_T(x_{3},\tau)-G(x_{42},\tau)\ ]\nonumber\\ 
&=-\fc{3}{4\pi^2}\ \z_3-\fc{3}{2\pi^2}\ \sum_{n=1}^\infty[\Li_3(q^n)+\Li_3(q^{n-1/2})]\ ,\label{IteratedNP}
\end{align}
whose individual terms are again determined in the appendix.
Hence, we have
\be
g(s,u)=\fc{1}{6}-(s+u)\ \lf\{\fc{3}{4\pi^2}\ \z_3+\fc{3}{2\pi^2}\ \sum_{n=1}^\infty[\Li_3(q^n)+\Li_3(q^{n-1/2})]\ri\}+\Oc(\ap^2)\ ,
\ee
from which \req{Nonplanar} is determined up to order $\Oc(\ap^2)$.

Again, we shall also perform the integration over the world--sheet modulus $\tau$. 
With \be 
\int\limits_0^\infty  dl\ l^\eps\ \sum_{n=1}^\infty\Li_a(q^{n-1/2})=\fc{1}{2\pi}\ \lf\{\ \fc{1}{\eps}\ \zeta(1+a)+\ln\lf(\fc{2}{\pi}\ri)\ \zeta(1+a)+\zeta'(1+a)\ \ri\}+\Oc(\epsilon)\ ,
\ee
we find:
\begin{align}
\int_0^\infty dl\ l^\eps\ g(s,u)&=\int_0^\infty dl\ l^\eps\ \lf(\fc{1}{6}-(s+u)\ \fc{3}{4\pi^2}\ \z_3+\Oc(\ap^2)\ri)
\nonumber\\
&-(s+u)\ \lf\{\ \fc{\pi}{60}\ \fc{1}{\eps}-\fc{\pi}{60}\ln(\pi)+\fc{3}{2\pi^3}\ \zeta'(4)\ \ri\}+\Oc(\ap^2)\ .\label{ResultNP}
\end{align}
In \req{Nonplanar}, the relevant integral $\int_0^\infty dl\ l^\eps\ [g(s,u)+g(t,s)+g(u,t)]$ gives rise
to a divergent piece given by  the tree--level expression 
$\pi^{-2}\lf\{\fc{\Gamma(s)\Gamma(u)}{\Gamma(1+s+u)} +\fc{\Gamma(s)\Gamma(t)}{\Gamma(1+s+t)}+\fc{\Gamma(t)\Gamma(u)}{\Gamma(1+t+u)}\ri\}$ \cite{Green:1981ya}.

Finally, the partial double  non--planar four--point open superstring amplitude is given by the following integral
\be\label{DoubNP}
A^{(1)}(1,2|3,4)=2\ (s_{12}s_{14})\ A^{(0)}_{YM}(1,2,3,4)\ \int_0^\infty dl\  h(s,u)\ ,
\ee
with the partial SYM amplitude $A^{(0)}_{YM}(1,2,3,4)$ and the form  factor
\begin{align}
h(s,u)&=\int_0^{1}dx_2\int_0^1dx_3 \int_0^{x_3} dx_4\ 
\exp\lf\{\tfrac{1}{2}\ s_{12}\ G(x_2-x_1,\tau)+\tfrac{1}{2}\ s_{34}\ G(x_4-x_3,\tau)\ri\}\nonumber\\
&\times \exp\lf\{\tfrac{1}{2}\sum_{1\leq i<j\leq 4\atop (i,j)\notin \{(1,2),(3,4)\}} s_{ij}\ G_T(x_j-x_i,\tau)\ri\}\ ,
\label{FormNPl1}
\end{align}
corresponding to the gauge choice $x_1=0$.
Again, for finite $\tau$  the integrand of the amplitude \req{Nonplanar} can be expanded w.r.t. $\ap$ by using
the series expansion \req{expexp1}
\begin{align}
&\exp\lf\{\tfrac{1}{2}s_{12}\ G(x_{21},\tau)+\tfrac{1}{2}s_{34}\ G(x_{34})\ri\}\,\exp\lf\{\tfrac{1}{2}\sum_{1\leq i<j\leq 4\atop (i,j)\notin \{(1,2),(3,4)\}} s_{ij}\ G_T(x_{ji},\tau)\ri\}\\
&\hspace{1cm}=1+\tfrac{1}{2}\ s\ [\ G(x_{21})+G(x_{34})-G_T(x_{31})-G_T(x_{42})\ ]\nonumber\\
&\hspace{1.2cm}+\tfrac{1}{2}\ u\ [\ G_T(x_{41})+G_T(x_{32})-G_T(x_{31})-G_T(x_{42})\ ]+\Oc(\ap^2)\ .\nonumber
\end{align}
Up to the linear order we need to compute the following iterated integrals
\begin{align*}
&\int_0^1dx_2 \int_0^{1} dx_3\int_0^{x_3}dx_4=\h\ ,\\
&\h\int_0^1dx_2 \int_0^{1} dx_3\int_0^{x_3}dx_4\ [\ G(x_{21})+G(x_{34})-G_T(x_{31})-G_T(x_{42})\ ] \\
&=\fc{1}{8}\ \ln q+2\ \sum_{n=1}^\infty[\Li_1(q^n)-\Li_1(q^{n-1/2})]-Q_3=\fc{1}{8}\ \ln q\ ,\\
&\h\int_0^1dx_2 \int_0^{1} dx_3\int_0^{x_3}dx_4\ [\ G_T(x_{41})+G_T(x_{32})-G_T(x_{31})-G_T(x_{42})\ ]=0\ ,
\end{align*}
whose individual terms are determined in the appendix. Above we have used the identity
\req{NonTrivial}, which gives rise to:
\be\label{Q3}
Q_3=2\ \sum_{n=1}^\infty [\Li_1(q^{2n})-\Li_1(-q^{n-1/2})]=2\ \sum_{n=1}^\infty [\Li_1(q^{n})-\Li_1(q^{n-1/2})]\ .
\ee
Hence, we have:
\be
h(s,u)=\h+ \fc{s}{8}\ \ln q+\Oc(\ap^2)\ .\label{formh}
\ee
For $q\rightarrow 0$ we may compute the all order $\alpha'$ expression of \req{FormNPl1}
\begin{align}
&2^{2s}\ q^{s/4}\ \lf\{\int_0^{1}dx_2\ \sin(\pi x_2)^{s}\ri\}\  
\lf\{\int_0^1dx_3 \int_0^{x_3} dx_4\ \sin[\pi(x_3-x_4)]^s\ri\}\nonumber\\
&=2^{2s-1}\ q^{s/4}\ \gamma(s)^2=\h+\fc{s}{8}\ \ln q+s^2\ \lf(\ \fc{1}{4}\ \zeta_2+\fc{1}{64}\ (\ln q)^2\ \ri)+\Oc(\ap^3)\ ,\label{Gammanice1}
\end{align}
which reproduces the first two terms of \req{formh}.
Note, that up to a factor the two integrals in the brackets of \req{Gammanice1} actually give the same:
\be\label{actually}
\gamma(s):=\int_0^1 dz \sin(\pi z)^s=2\int_0^1 dx\int_0^1 dy\ x\ 
\sin[(\pi x(1-y)]^s=\fc{\Gamma(\tfrac{1}{2}+\tfrac{s}{2})}{
\sqrt\pi\ \Gamma(1+\tfrac{s}{2})}\ .
\ee

\subsection[Planar four--point monodromy relation]{Planar four--point monodromy relation}\label{Sect:Panar4PtMR}

Into the monodromy relation \req{FourPointMonodromyRelation} the non--planar partial
amplitude $\tilde{A}^{(1)}(2,3,4|1)$ enters. The latter is defined in
\req{take} with \req{F4Integrand} and can be written as (following the same form as (\ref{DoubNP})) 
\be\label{tildeAmp}
\tilde A^{(1)}(2,3,4|1):=\int_0^\infty  dl\ V_{\text{CKG}}^{-1}\ \tilde{a}^{(1)}(2,3,4|1)=(s_{12}s_{14})\ A^{(0)}_{YM}(1,2,3,4)\ \int_0^\infty dl\ \tilde g(s,u)\ ,
\ee
with the form factor
\begin{align}
\tilde g(s,u)&=\int_0^1dx_4 \int_0^{x_4} dx_3\int_0^{x_3}dx_2\ 
\exp\lf\{\tfrac{1}{2}\sum_{2\leq i<j\leq 4} s_{ij}\ G(x_j-x_i,\tau)\ri\}\nonumber\\
&\times \exp\lf\{\tfrac{1}{2}\sum_{j=2}^4 s_{1j}\ G_T(x_j-x_1,\tau)\ri\}\
e^{-i\pi s x_{21}}\ e^{-i\pi t x_{31}}\ e^{-i\pi u x_{41}}\ ,\label{formtg}
\end{align}
with the gauge choice $x_1=0$. The $\ap$--expansion of the form factor \req{formtg} can be computed in similar way as in the
previous subsection with the result:
\begin{align}
\tilde g(s,u)&=\fc{1}{6}+s\
\lf\{\fc{i\pi}{24}-\fc{3}{4\pi^2}\ \z_3-\fc{3}{2\pi^2}\ \sum_{n=1}^\infty[\Li_3(q^n)+\Li_3(q^{n-1/2})]\ri\}
\nonumber\\
&+u\ \lf\{-\fc{i\pi}{24}-\fc{3}{4\pi^2}\ \z_3-\fc{3}{2\pi^2}\ \sum_{n=1}^\infty[\Li_3(q^n)+\Li_3(q^{n-1/2})]\ri\}
+\Oc(\ap^2)\ .\label{Expgt}
\end{align}
Note, that the phases of the integrand \req{formtg} give rise to the pure imaginary term in \req{Expgt}. The latter are in fact required in order for the monodromy relation (\ref{FourPointMonodromyRelation}) to hold, which can be written in the form
\begin{align}
A^{(1)}(1,2,3,4)&+e^{\pi i s}\ A^{(1)}(2,1,3,4)+e^{\pi i (s+t)}\ 
A^{(1)}(2,3,1,4)\nonumber\\
&=\tilde A^{(1)}(2,3,4|1)+e^{\pi i s}\ \tilde A^{(1)}(3,4,2|1)+e^{\pi i (s+t)}\ \tilde A^{(1)}(4,2,3|1)\,,\label{Mono1}
\end{align}
where the planar amplitude can be found in \req{Planar}. In order to verify \req{Mono1} the form factors $f(s,u)$ and $\tilde{g}(s,u)$ have to obey the following relation:
\be
f(s,u)+e^{i\pi  s}\ f(s,t)+e^{i\pi  (s+t)}\ f(u,t)=
\tilde g(s,u)+e^{i\pi s}\ \tilde g(t,s)+e^{i\pi  (s+t)}\ \tilde g(u,t)\ .
\label{CMono1}
\ee

\subsection{Non--planar four--point monodromy relations}\label{Sect:NonPlanarMonodromy4pt}
\subsubsection{Monodromy relation}

We begin with a gauge that simplifies many of the following  been used throughout the previous sections, namely $z_1=0$ and we move point $z_2$ around the cylinder. At the same time we insert $z_3$ and $z_4$ on the upper boundary (\emph{i.e.} $\Im(z_3)=\Im(z_4)=\frac{l }{2}$) with $\Re(z_4)<\Re(z_3)$, as schematically shown in the following figure
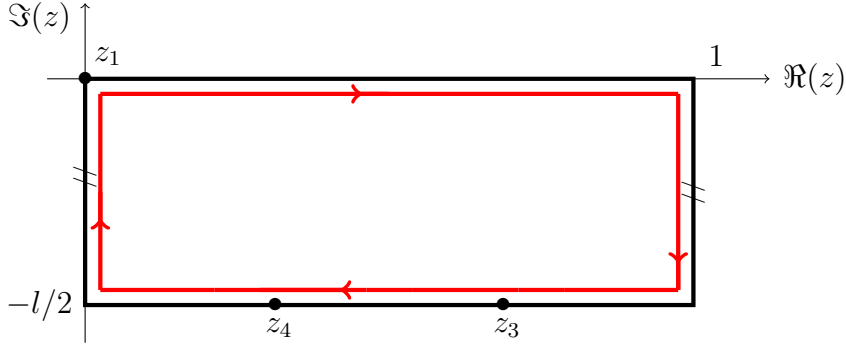
\begin{center}
\begin{tikzpicture}
\draw[->] (-0.5,3) -- (9,3);
\draw[->] (0,-0.5) -- (0,4);
\draw[ultra thick] (0,0) rectangle (8,3); 
\node at (9.6,3) {$\Re (z)$};
\node at (-0.6,3.8) {$\Im (z)$};
\node at (-0.6,0) {$-l /2$};
\node at (8.3,3.3) {1};
\node at (0,3) {$\bullet$};
\node at (0.3,3.3) {$z_1$};
\node at (2.5,0) {$\bullet$};
\node at (2.55,-0.3) {$z_4$};
\node at (5.5,0) {$\bullet$};
\node at (5.55,-0.3) {$z_3$};
\node[rotate=90,scale=0.8] at (0,1.7) {{\small //}};
\node[rotate=90,scale=0.8] at (8,1.5) {{\small //}};
\draw[red, ultra thick,yshift=0.2cm] (0.2,0) -- (1.7,0);
\draw[red, ultra thick,yshift=0.2cm] (2.3,0) -- (3.7,0);
\draw[red, ultra thick,yshift=0.2cm] (4.3,0) -- (5.7,0);
\draw[red, ultra thick,yshift=0.2cm] (6.3,0) -- (7.8,0);
%
\draw[red, ultra thick,yshift=0.2cm] (1.7,0) -- (2.3,0);
\draw[red, ultra thick,yshift=0.2cm] (3.7,0) -- (4.3,0);
\draw[red, ultra thick,yshift=0.2cm] (5.7,0) -- (6.3,0);
\draw[red, ultra thick,-<,yshift=0.2cm] (2.3,0) -- (3.5,0);
%
\draw[red, ultra thick,-<] (7.8,0.2) -- (7.8,0.7);
\draw[red, ultra thick,-<] (0.2,1.5) -- (0.2,1);
\draw[red, ultra thick,-<] (4,2.8) -- (3.5,2.8); 
\draw[red, ultra thick] (7.8,0.2) -- (7.8,2.8);
\draw[red, ultra thick] (0.2,0.2) -- (0.2,2.8);
\draw[red, ultra thick] (0.2,2.8) -- (7.8,2.8);
\end{tikzpicture}
\end{center}
Following the strategy explained in section~\ref{Sect:StrategyNonPlanar}, we divide the single contour for $z_2$ into three different contours, each of which staying away from the points $z_1$ and $z_{3,4}$:
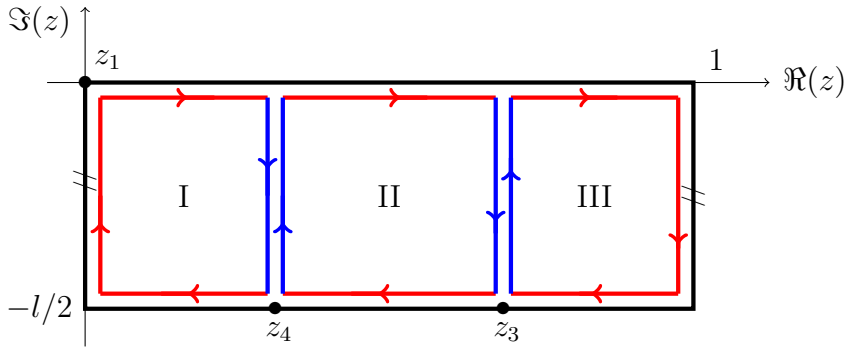
\begin{center}
\begin{tikzpicture}
\draw[->] (-0.5,3) -- (9,3);
\draw[->] (0,-0.5) -- (0,4);
\draw[ultra thick] (0,0) rectangle (8,3); 
\node at (9.6,3) {$\Re (z)$};
\node at (-0.6,3.8) {$\Im (z)$};
\node at (-0.6,0) {$-l /2$};
\node at (8.3,3.3) {1};
\node at (0,3) {$\bullet$};
\node at (0.3,3.3) {$z_1$};
\node at (2.5,0) {$\bullet$};
\node at (2.55,-0.3) {$z_4$};
\node at (5.5,0) {$\bullet$};
\node at (5.55,-0.3) {$z_3$};
\node[rotate=90,scale=0.8] at (0,1.7) {{\small //}};
\node[rotate=90,scale=0.8] at (8,1.5) {{\small //}};
%
\draw[red, ultra thick,yshift=0.2cm] (0.2,0) -- (2.4,0);
\draw[red, ultra thick,yshift=0.2cm] (2.6,0) -- (5.4,0);
\draw[red, ultra thick,yshift=0.2cm] (5.6,0) -- (7.8,0);
\draw[red, ultra thick,yshift=-0.2cm] (0.2,3) -- (2.4,3);
\draw[red, ultra thick,yshift=-0.2cm] (2.6,3) -- (5.4,3);
\draw[red, ultra thick,yshift=-0.2cm] (5.6,3) -- (7.8,3);
%

%
%
\draw[red, ultra thick] (7.8,0.2) -- (7.8,2.8);
\draw[blue, ultra thick] (5.6,0.2) -- (5.6,2.8);
\draw[blue, ultra thick] (5.4,0.2) -- (5.4,2.8);
\draw[blue, ultra thick] (2.6,0.2) -- (2.6,2.8);
\draw[blue, ultra thick] (2.4,0.2) -- (2.4,2.8);
\draw[red, ultra thick] (0.2,0.2) -- (0.2,2.8);
\node at (1.3,1.5) {I};
\node at (4,1.5) {II};
\node at (6.7,1.5) {III};
%
\draw[red, ultra thick,-<] (1,0.2) -- (1.5,0.2);
\draw[red, ultra thick,-<] (1.7,2.8) -- (1.2,2.8);
\draw[red, ultra thick,-<] (0.2,1.5) -- (0.2,1);
\draw[blue, ultra thick,-<] (2.4,1.5) -- (2.4,2);
\draw[red, ultra thick,-<] (3.5,0.2) -- (4,0.2);
\draw[blue, ultra thick,-<] (2.6,1.5) -- (2.6,1);
\draw[blue, ultra thick,-<] (5.6,2) -- (5.6,1.7);
\draw[red, ultra thick,-<] (4.5,2.8) -- (4,2.8);
\draw[blue, ultra thick,-<] (5.4,1) -- (5.4,1.3);
\draw[red, ultra thick,-<] (6.5,0.2) -- (6.8,0.2);
\draw[red, ultra thick,-<] (7,2.8) -- (6.5,2.8);
\draw[red, ultra thick,-<] (7.8,0.8) -- (7.8,1);
\end{tikzpicture}
\end{center}
where the integrands in each contour are chosen such that the blue lines cancel each other. More specifically, the contours are characterised by 
\begin{align}
&\text{contour I:} &&\Re(z_2)<\Re(z_4)<\Re(z_3)\,,\nonumber\\
&\text{contour II:} &&\Re(z_4)<\Re(z_2)<\Re(z_3)\,,\nonumber\\
&\text{contour III:} &&\Re(z_4)<\Re(z_3)<\Re(z_2)\,.\nonumber
\end{align}
Furthermore, since $z_2$ encounters no other point on the lower boundary, the integrand in each of the three regions is the same
\def\gf{{\frak g}}
\begin{align}
F(z_2,z_3,z_4)&=\exp\lf\{s_{12}\gf(z_2)+s_{34}\gf(z_{34})+s_{13}\gf_T(z_3)+s_{14}\gf_T(z_4)+s_{23}\gf_T(z_{32})+s_{24}\gf_T(z_{42})\ri\},
\end{align}
for $z_2\in{\bf C}$ (allowing for the possibility of $z_2$ being integrated on the lower as well as the upper boundary). Specifically, when $z_2$ is on the lower boundary we write
\begin{align}
F_T(z_2,z_3,z_4)&=\exp\lf\{s_{12}\ \gf(z_2-\tfrac{il}{2})+s_{34}\ \gf(z_{34})+
s_{13}\ \gf_T(z_3)+s_{14}\ \gf_T(z_4)\ri.\\
&+\lf.s_{23}\ \gf_T(z_{32}+\tfrac{il}{2})+s_{24}\ \gf_T(z_{42}+\tfrac{il}{2})\ri\}\nonumber\\
&=\exp\{\tfrac{1}{2}\pi i(-s+u+t)\}\ \exp\lf\{\pi i(s_{12}z_2-s_{23}z_{32}-s_{24}z_{42})\ri\}\nonumber \\
&\times \exp\lf\{s_{12} \gf_T(z_2)+s_{13}\gf_T(z_3)+s_{14}\gf_T(z_4)+
s_{34}\gf(z_{34})+s_{23}\gf(z_{32})+s_{24}\gf(z_{42})\ri\},\nonumber
\end{align}
for $z_2\in {\bf R}$ with $z_2\in[0,1]$. 
Above we have introduced the pure holomorphic parts of the Green's functions
\req{G} and \req{GT}, respectively:
\be
\gf(z,\tau)=\ln\fc{\theta_1(z,\tau)}{\theta_1'(0,\tau)}\ \ \ ,\ \ \ 
\gf_T(z,\tau)=\ln\fc{\theta_4(z,\tau)}{\theta_1'(0,\tau)}\ .
\ee
With this convention, the relevant integrals to compute for the various contours are as follows
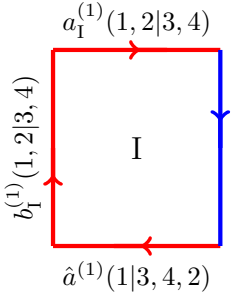
\begin{center}
\begin{tabular}{lcl}
\parbox{3cm}{
\begin{tikzpicture}
\draw[red, ultra thick,yshift=0.2cm] (0.2,0) -- (2.4,0);
\node[rotate=90] at (-0.2,1.5) {\footnotesize $b^{(1)}_{\text{I}}(1,2|3,4)$};
\draw[red, ultra thick] (0.2,0.2) -- (0.2,2.8);
\node at (1.3,3.2) {\footnotesize $a^{(1)}_{\text{I}}(1,2|3,4)$};
\draw[red, ultra thick,yshift=-0.2cm] (0.2,3) -- (2.4,3);
\node at (1.3,-0.2) {\footnotesize $\hat{a}^{(1)}(1|3,4,2)$};
\draw[blue, ultra thick] (2.4,0.2) -- (2.4,2.8);
\node at (1.3,1.5) {I};
\draw[red, ultra thick,-<] (1,0.2) -- (1.5,0.2);
\draw[red, ultra thick,-<] (1.7,2.8) -- (1.2,2.8);
\draw[red, ultra thick,-<] (0.2,1.5) -- (0.2,1);
\draw[blue, ultra thick,-<] (2.4,1.5) -- (2.4,2);
\end{tikzpicture}}
& ${}$ \hspace{1cm} ${}$&
\parbox{7cm}{\small\begin{align}
&a^{(1)}_{\text{I}}(1,2|3,4)=\int_0^1dz_3 \int_0^{z_3}dz_4\int_0^{z_4}dz_2\,F(z_2,z_3,z_4)\,,\nonumber\\
&\hat{a}^{(1)}(1|3,4,2)=\int_0^1dz_3 \int_0^{z_3}dz_4\int_0^{z_4}dz_2\,F_T(z_2,z_3,z_4)\,,\nonumber\\
&b^{(1)}_{\text{I}}(1,2|3,4)=-\frac{\tau}{2}\int_0^1dz_3 \int_0^{z_3}dz_4\int_0^{-1}dz_2\,F(\tfrac{\tau}{2}\,z_2,z_3,z_4)\,.\nonumber
\end{align}
}\\[70pt]
\parbox{3cm}
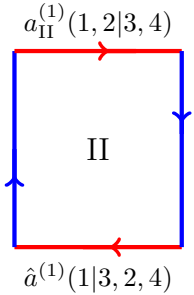
{\begin{tikzpicture}
\draw[red, ultra thick,yshift=0.2cm] (0.2,0) -- (2.4,0);
\node[rotate=90] at (-0.2,1.5) {\footnotesize $\phantom{B^{(1)}(1,2|3,4)}$};
\draw[blue, ultra thick] (0.2,0.2) -- (0.2,2.8);
\node at (1.3,3.2) {\footnotesize $a^{(1)}_{\text{II}}(1,2|3,4)$};
\draw[red, ultra thick,yshift=-0.2cm] (0.2,3) -- (2.4,3);
\node at (1.3,-0.2) {\footnotesize $\hat{a}^{(1)}(1|3,2,4)$};
\draw[blue, ultra thick] (2.4,0.2) -- (2.4,2.8);
\node at (1.3,1.5) {II};
\draw[red, ultra thick,-<,xshift=-2.4cm] (3.5,0.2) -- (4,0.2);
\draw[blue, ultra thick,-<,xshift=-2.4cm] (2.6,1.5) -- (2.6,1);
\draw[blue, ultra thick,-<,xshift=-3.2cm] (5.6,1.7) -- (5.6,2);
\draw[red, ultra thick,-<,xshift=-2.7cm] (4.5,2.8) -- (4,2.8);
\end{tikzpicture}}
& ${}$ \hspace{1cm} ${}$&
\parbox{7cm}{\small\begin{align}
&a^{(1)}_{\text{II}}(1,2|3,4)=\int_0^1dz_4 \int_0^{z_4}dz_2\int^{z_2}_{0}dz_3\,F(z_2,z_3,z_4)\,,\nonumber\\
&\hat{a}^{(1)}(1|3,2,4)=\int_0^1dz_3 \int_0^{z_4}dz_2\int_0^{z_2}dz_3\,F_T(z_2,z_3,z_4)\,,\nonumber
\end{align}
}\\[70pt]
\end{tabular}
\end{center}
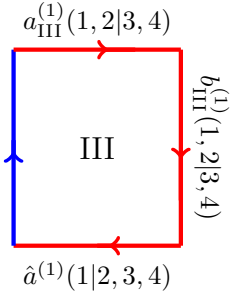
\begin{center}
\begin{tabular}{lcl}
\hspace{0.8cm}\parbox{3cm}{
\begin{tikzpicture}
\draw[red, ultra thick,yshift=0.2cm] (0.2,0) -- (2.4,0);
\node[rotate=270] at (2.8,1.5) {\footnotesize $b^{(1)}_{\text{III}}(1,2|3,4)$};
\draw[blue, ultra thick] (0.2,0.2) -- (0.2,2.8);
\node at (1.3,3.2) {\footnotesize $a^{(1)}_{\text{III}}(1,2|3,4)$};
\draw[red, ultra thick,yshift=-0.2cm] (0.2,3) -- (2.4,3);
\node at (1.3,-0.2) {\footnotesize $\hat a^{(1)}(1|2,3,4)$};
\draw[red, ultra thick] (2.4,0.2) -- (2.4,2.8);
\node at (1.3,1.5) {III};
\draw[blue, ultra thick,-<,xshift=-5.2cm] (5.4,1.6) -- (5.4,1.3);
\draw[red, ultra thick,-<,xshift=-5.2cm] (6.5,0.2) -- (6.8,0.2);
\draw[red, ultra thick,-<,xshift=-5.2cm] (7,2.8) -- (6.5,2.8);
\draw[red, ultra thick,-<,xshift=-5.4cm] (7.8,0.8) -- (7.8,1.5);
\end{tikzpicture}}
& ${}$ \hspace{1cm} ${}$&
\parbox{7cm}{\small\begin{align}
&a^{(1)}_{\text{III}}(1,2|3,4)=\int_0^1dz_2 \int_0^{z_2}dz_4\int_0^{z_4}dz_3\,F(z_2,z_3,z_4)\,,\nonumber\\
&\hat a^{(1)}(1|2,3,4)=\int_0^1dz_2 \int_0^{z_2}dz_4\int_0^{z_4}dz_3\,F_T(z_2,z_3,z_4)\,,\nonumber\\
&b^{(1)}_{\text{III}}(1,2|3,4)=-\frac{\tau}{2}\int_0^1dz_4 \int_0^{z_4}dz_3\int_0^{-1}dz_2\,F(\tfrac{\tau}{2}\,z_2+1,z_3,z_4)\,.\nonumber
\end{align}
}
\end{tabular}
\end{center}
Based on the discussion of section~\ref{Sect:RelNonPlanarAmp4}, we expect the following relation to hold among these contributions:
\begin{align}
a^{(1)}_{\text{I}}(1,2|3,4)&+a^{(1)}_{\text{II}}(1,2|3,4)+a^{(1)}_{\text{III}}(1,2|3,4)-a^{(1)}(1|2,3,4)-a^{(1)}(1|3,2,4)-a^{(1)}(1|3,4,2)\nonumber\\
&-b^{(1)}_{\text{I}}(1,2|3,4)+b^{(1)}_{\text{III}}(1,2|3,4)=0\,.\label{PredictGaugez1}
\end{align}
Given the explicit expressions for all contributions, we can check them explicitly to leading order in $\alpha'$. Indeed, we find analytically:
\begin{align}
a^{(1)}_{\text{I}}(1,2|3,4)+a^{(1)}_{\text{II}}(1,2|3,4)+a^{(1)}_{\text{III}}(1,2|3,4)&=\int_0^1dz_2 \int_0^{1}dz_3\int_0^{z_3}dz_4\,F(z_2,z_3,z_4)\nonumber\\
&=\frac{1}{2}+\frac{s}{8}\,\ln q+\mathcal{O}(\alpha^{\prime 2})\ ,
\end{align}
which agrees with \req{formh}, and: 
\begin{align}
\hat a^{(1)}(1|2,3,4)+\hat a^{(1)}(1|3,2,4)+\hat a^{(1)}(1|3,4,2)&=\int_0^1dz_2 \int_0^{1}dz_3\int_0^{z_3}dz_4\,F_T(z_2,z_3,z_4)\nonumber\\
&=\frac{1}{2}+\mathcal{O}(\alpha^{\prime 2})\ .\label{DoubleNonPlanContr}
\end{align}
Note, that the double--nonplanar contributions $a^{(1)}_{\text{I}}(1,2|3,4)+a^{(1)}_{\text{II}}(1,2|3,4)+a^{(1)}_{\text{III}}(1,2|3,4)$ depend on $\ln q=2\pi i\tau$ to linear order in $\alpha'$, while the single non--planar terms $\hat a^{(1)}(1|2,3,4)$, $\hat a^{(1)}(1|3,2,4)$ or $\hat a^{(1)}(1|3,4,2)$ have an integer power series expansion in $q$ (see also section~\ref{Sect:Panar4PtMR}) for further details. This indicates that there is no possible relation that involves only these boundary contributions, but {\it further terms} are needed, namely those integral contributions\footnote{These bulk terms are completely missing and not discussed at all in \cite{Tourkine:2016bak}. As a consequence, the eq.  (9) of \cite{Tourkine:2016bak} is wrong for $p\geq 2$.} where $z_2$ is integrated in the interior of the cylinder. Those required terms $B^{(1,2)}(1,2|3,4)$ are given by 
\begin{align}
&b^{(1)}_{\text{III}}(1,2|3,4)=e^{-i\pi s}\,b^{(1)}_{\text{I}}(1,2|3,4)\,,&&\text{with} &&b^{(1)}_{\text{I}}(1,2|3,4)=\frac{\tau}{4}+\mathcal{O}(\alpha^{\prime})\,,
\end{align}
and indeed are proportional to $\tau=\tfrac{\ln q}{2\pi i}$, thus being able to cancel the $\ln q$ term in (\ref{DoubleNonPlanContr}). Taking everything together, the left-hand side of (\ref{PredictGaugez1}) becomes to leading order in $\alpha'$
\begin{align}
\frac{1}{2}+\frac{s}{8}\,\ln q-\frac{1}{2}-\frac{\tau}{4}+(1-i\pi s)\,\frac{\tau}{4}+\mathcal{O}(\alpha^{\prime 2})=\mathcal{O}(\alpha^{\prime 2})\,,
\end{align}
thus validating (\ref{PredictGaugez1}) up to the order $\mathcal{O}(\alpha^{\prime 2})$. Furthermore, we have successfully checked (\ref{PredictGaugez1}) numerically up to order $\alpha^{\prime 3}$. Moreover, in the next subsection we shall also probe  higher 
orders in the $q$--series expansion of (\ref{PredictGaugez1}).

Notice that (\ref{PredictGaugez1}) contains no relative phase factors (of the type $e^{i\pi s_{12}}$) between the single and double non--planar boundary contributions. This is due to our gauge choice $z_1=0$, which means that the point $z_2$ is always to the right of $z_1$ (\emph{i.e.} $\Re(z_1)<\Re(z_2)$) and never 'steps over' point $z_1$ on the lower boundary. This changes drastically if we pick a different gauge, \emph{e.g.} $\Re(z_4)=1$, which indeed contains two different regions, namely $\Re(z_1)<\Re(z_2)$ and $\Re(z_2)<\Re(z_1)$. The distinction between these two regions leads  to a qualitatively different monodromy relation (see appendix~\ref{App:4PtOtherGauge} for a computation in the particular gauge $z_4=1-\frac{\tau}{2}$). This is an indication that the non-planar relations are not independent of the gauge choice thus making a physical interpretation difficult.

\subsubsection[$\alpha'$--expansion of non--planar subamplitudes and boundary term]{$\bm{\alpha'}$--expansion of non--planar subamplitudes and boundary terms}

By permuting the labels $1$ and $2$  the relation \req{Mono2a} can be translated into: 
\be
\h\ A^{(1)}(1,2|3,4)+\hat A^{(1)}(1|2,3,4)-(1-e^{-i\pi  s})\ B^{(1)}(1,2|3,4)=0\ .\label{Mono2}
\ee
This relation constitutes  the double non--planar amplitude \req{DoubNP} and the non--planar amplitude 
\begin{align}\label{hatAmp}
\hat A^{(1)}(1|2,3,4)&:=\int_0^\infty  dl\ V_{\text{CKG}}^{-1}\ \lf\{\ \hat a^{(1)}(1|2,3,4)
+\hat a^{(1)}(1|3,2,4)+\hat a^{(1)}(1|3,4,2)\ \ri\}\nonumber\\
&=(s_{12}s_{14})\ A^{(0)}_{YM}(1,2,3,4)\ \int_0^\infty dl\ \hat g(s,u)\ ,
\end{align}
with the form factor
\begin{align}
\hat g(s,u)&=-\int_0^1dx_3 \int_0^{x_3} dx_4\int_0^1dx_2\ 
\exp\lf\{\tfrac{1}{2}\sum_{2\leq i<j\leq 4} s_{ij}\ G(x_i-x_j,\tau)\ri\}\ \nonumber\\
&\times \exp\lf\{\tfrac{1}{2}\sum_{j=2}^4 s_{1j}\ G_T(x_j-x_1,\tau)\ri\}\ e^{i\pi s x_{21}}\ e^{i\pi t x_{24}}\ e^{i\pi u x_{23}}\ e^{-i\pi  s}\ ,
\label{formtg20}
\end{align}
for the gauge choice $x_1=0$. In addition, there is the boundary (bulk) contribution
\be\label{BOUNDA}
B^{(1)}(1,2|3,4)=(s_{12}s_{14})\ A^{(0)}_{YM}(1,2,3,4)\ \int_0^\infty dl\ b(s,u)\ ,
\ee
with the form factor:
\begin{align}
b(s,u)&=-\fc{\tau}{2}\ \int\limits_0^{-1}d\alpha\int_0^1dx_3 \int_0^{x_3} dx_4\ 
\exp\lf\{\fc{s}{2}\ G(x_{21},\tau)+\fc{s}{2}\ G(x_{34},\tau)\ri\}\nonumber\\
&\times \lf.\exp\lf\{\fc{u}{2}\ G_T(x_{41},\tau)+\fc{t}{2}\ G_T(x_{42},\tau)+\fc{t}{2}\ G_T(x_{31},\tau)+
\fc{u}{2}\ G_T(x_{32},\tau)\ri\}\ri|_{x_2=\alpha\fc{\tau}{2}}\ .
\label{FormNPl2}
\end{align}
Note, that in  \req{formtg20} the coordinate $x_2$ can safely be integrated along the whole boundary interval $[0,1]$  without passing any cuts. 
Sign flips originating from $\exp\{ \tfrac{t}{2} G(x_{24},\tau)\}$ and 
$\exp\{ \tfrac{u}{2} G(x_{23},\tau)\}$
are compensated by the phases $\exp\{i\pi t x_{24}\}$ and $\exp\{i\pi u x_{23}\}$, respectively. We also refer the reader to the discussion below \req{GENERIC3a}.
Furthermore, the gauge choice $x_1=0$ avoids additional cuts stemming from $x_2$ surpassing $x_1$.
Again, the $\ap$--expansion of the form factor \req{formtg20} can be computed in
similar way as in the previous subsection with the result
\begin{align}
\hat g(s,u)&=-\lf(\h +\fc{i\pi}{2}\ s+\Oc(\ap^2)\ri)\ e^{-i\pi s}=-\h+\Oc(\ap^2)\ ,\label{Expgt2}
\end{align}
in agreement with \req{DoubleNonPlanContr}.
The pure imaginary term in the bracket of \req{Expgt2} stems from the position dependent phases of the integrand \req{formtg20}.
Finally, the boundary term \req{FormNPl2} can be expanded w.r.t. $\ap$
\begin{align}
b(s,u)&=\fc{\tau}{4}+ s\ \lf(-\fc{i\pi}{24}-\fc{1}{16}\ln q+\fc{\tau}{32}\ln q+
\fc{1}{4\pi^2}\ \z_3+\fc{1}{2\pi^2}\ \sum_{n=1}^\infty[\Li_3(q^{n})-\Li_3(q^{n-1/2}) ] \ri)\nonumber\\
&+u\ \lf(\fc{1}{2\pi^2}\ \z_3+\fc{1}{\pi^2}\ 
\sum_{n=1}^\infty[\Li_3(q^{n})-\Li_3(q^{n-1/2}) ]\ri)+\Oc(\ap^2)\ ,\label{Expgt3}
\end{align}

\comment{ Similarly, as in \req{Gammanice1} for $q\rightarrow 0$ we may compute the all order $\alpha'$ expression of \req{FormNPl2} as
\begin{align}
&2^{2s}\ q^{s/4}\ \fc{\tau}{2}\ \int_0^{1}d\alpha\ \sin\lf(\pi \alpha\fc{\tau}{2}\ri)^{s}\ \int_0^1dx_4 \int_0^{x_4} dx_3\ \sin[\pi(x_4-x_3)]^s\\
&=2^{2s-1}\ q^{s/4}\ \gamma(s)\ \lf\{\ \h\ \gamma(s)-\fc{1}{\pi}\cos\lf(\fc{\pi\tau}{2}\ri)\ {}_2F_1\lf[{\h , \h-\tfrac{s}{2}\atop \fc{3}{2}},\cos\lf(\fc{\pi\tau}{2}\ri)^2\ri]\    \ri\}\nonumber\\
&=\fc{\tau}{4}\ +?\ldots?+\Oc(\ap^3)\ ,\label{Gamma2}
\end{align}
with $\gamma$ given in \req{actually}. Above the hypergeometric function 
we have treated as follows \cite{Kalmykov:2006pu}:
\begin{align}
{}_2F_1\lf[{\h , \h-\tfrac{s}{2}\atop \fc{3}{2}},\cos\lf(\fc{\pi\tau}{2}\ri)^2\ri]&=\sin\lf(\fc{\pi\tau}{2}\ri)^{1+s}\ 
{}_2F_1\lf[{1, 1+\tfrac{s}{2}\atop \fc{3}{2}},\cos\lf(\fc{\pi\tau}{2}\ri)^2
\ri]\nonumber\\
&= \h\  \sin\lf(\fc{\pi\tau}{2}\ri)^{s+1}\ \cos\lf(\fc{\pi\tau}{2}\ri)^{-2}\ 
\fc{1+q^{1/2}}{1-q^{1/2}}\\
&\hskip-4cm\times \lf\{\ln(-q^{1/2})+s\ \lf[\ -\Li_2(q^{1/2})-\ln (-q^{1/2})\ln(1-q^{1/2})
+\fc{1}{4}\ \ln^2(-q^{1/2})-\h \z_2\ \ri]+\Oc(\ap^2)\ri\}\nonumber
\end{align}
which reproduces the first two terms of \req{Expgt3}.}

\noindent
while, the double non--planar amplitude can be found in \req{DoubNP}.
In order to verify \req{Mono2} the form factors have to obey the following relation:
\be
h(s,u)-b(s,u)+e^{-\pi i s}\ b(s,u)+\hat g(s,u)=0\ .
\ee
With the form factors given in \req{formh}, \req{Expgt2} and \req{Expgt3}, respectively
this relation can be verified.

\subsection[$N$--point amplitude: $\ap$--expansion and monodromy relations]{$\bm{N}$--point amplitude: $\bm{\ap}$--expansion and monodromy relations}
\label{transcend}

After having exhibited in detail the computation of  $\ap$--expansions for the case $N\!=\!4$ we shall now comment on the generic case.
Again, to extract the $\ap$--expansion  one uses the expansions \req{expexp} and \req{expexp1} 
and computes the relevant elliptic iterated integrals over the Green functions \req{G} and \req{GT} and powers thereof, supplemented by the additional functions $P_N$ or $Q_N$. For $n=N-4\geq 1$ the functions $P_N$ and $Q_N$ are polynomials in at most $k$--th order derivatives of the Green functions \req{G} and \req{GT}, respectively and in the Eisenstein series $G_k$, with $k\leq n$. In lines of \req{path1} and \req{path2} the resulting integrals can be arranged as elliptic iterated integrals \req{eII} with generic paths on $E_\tau$ .

Some interesting constraints on the terms of the $\ap$--expansion follow from considering the monodromy relations.
Recall, the $\ap$--expansion of the tree--level superstring  amplitude  at order $\ap^k$  is given by polynomials of order $k$ in the kinematic invariants \req{Mandel} multiplied by MZVs of weight $k$
\cite{Stieberger:2009rr}. For a given basis  of MZVs of weight $k$ terms  at order $\ap^k$ referring to a basis element $\xi_k$, which does not contain powers of $\zeta_2$,  are constrained as: 
\begin{align}
\lf. A^{(0)}(1,2,\ldots,N)\ri|_{\xi_k}+\lf. A^{(0)}(2,1,\ldots,N)\ri|_{\xi_k}+\ldots+\lf. A^{(0)}(2,3,\ldots,N-1,1,N)\ri|_{\xi_k}=0\ ,\label{SitTree}\\
s_{12}\ \lf.A^{(0)}(2,1,\ldots,N)\ri|_{\xi_k}+\ldots+s_{1,2\ldots N-1}\ \lf. A^{(0)}(2,3,\ldots,N-1,1,N)\ri|_{\xi_k}=0\ .\label{SitTree1}
\end{align}
The above equations are valid subject to powers of $\zeta_2$, i.e. essentially taking $\pi\ra0$
in the $\ap$--expansion.
The identities \req{SitTree} and \req{SitTree1}, which follow from \req{TreeAmpRel}, assume the same form as the  field theory KK and BCJ relations, respectively.

Similar constraints can be imposed for integrands $a$ of higher--loop subamplitudes. 
For the planar \req{ClosedStringChannelAmp} and single ($N_1\!=\!1$) non--planar \req{DefClosedChannelAmp} one--loop $N$--point subamplitudes the $\ap$--expansion of their integrands is described by the power series \req{expexp} and \req{expexp1} whose coefficients at order $\ap^k$ are given by $k$ powers of the Greens functions 
\req{G} and \req{GT} multiplied by the polynomials $P_N$ and $Q_N$, respectively. As a consequence the integrations over the $N-1$ world--sheet positions provide multiple polylogarithms  of transcendentality degree $k+2$ divided by $\pi^2$, i.e. the associated elliptic iterated integral \req{eII}
integrates to multiple elliptic polylogarithms  \req{eMZV} of weight $k$. At a given order $\ap^k$ a minimal set of indecomposable eMZVs of weight $k$ and length $l$ can be defined such that any other eMZV of the same weight and length can be expressed as linear combination of elements from this set with coefficients being rational numbers, MZVs and integer powers of $2\pi i$ \cite{Broedel:2015hia}.
The phases of the monodromy relation \req{PlanarNIntegrandRel} as well as the additional position dependent phases of the integrands given in \req{GENERIC2} and \req{GENERIC3a} may contribute only  powers of $i\pi$. Hence, from \req{PlanarNIntegrandRel} for terms appearing at the order $\ap^k$ and referring to indecomposable eMZVs $\omega_k$ of weight $k$   we have the following  integrand relation
\be\label{RemiMono}
\lf.a^{(1)}(1,2,\ldots,N)\ri|_{\omega_k}+\lf.a^{(1)}(2,1,\ldots,N)\ri|_{\omega_k}+\ldots+
\lf.a^{(1)}(2,\ldots,N-1,1,N)\ri|_{\omega_k}=\lf.a^{(1)}(1|2,\ldots,N)\ri|_{\omega_k}
\ee
between the $\ap^k$ order of  planar amplitudes and the single non--planar amplitude.
E.g. for $N\!=\!4$ for indecomposable eMZVs of weight $k$ and a given length we have at one--loop $(k\geq 1$):
\be\label{RemiFT}
\lf.a^{(1)}(1,2,3,4)\ri|_{\omega_{k} }+\lf.a^{(1)}(2,1,3,4)\ri|_{\omega_{k}}+
\lf. a^{(1)}(2,3,1,4)\ri|_{\omega_{k}}=\lf. a^{(1)}(1|2,3,4)\ri|_{\omega_{k}}\ ,
\ee
which in the UV limit yields the following Gamma function identity:
\begin{align}
(2su)^{-1}&\lf.\partial_{\alpha'}\fc{\Gamma(1+s)\Gamma(1+u)}{\Gamma(1-t)}+
(2st)^{-1}\partial_{\alpha'}\fc{\Gamma(1+s)\Gamma(1+t)}{\Gamma(1-u)}+
(2ut)^{-1}\partial_{\alpha'}\fc{\Gamma(1+u)\Gamma(1+t)}{\Gamma(1-s)}\ri|_{\z_{2k+1}}\nonumber\\
&=\lf.\fc{\Gamma(s)\Gamma(u)}{\Gamma(1+s+u)} +\fc{\Gamma(s)\Gamma(t)}{\Gamma(1+s+t)}+\fc{\Gamma(t)\Gamma(u)}{\Gamma(1+t+u)}\ri|_{\z_{2k+1}}=0\ .
\end{align}
Again, the identity \req{RemiMono} is reminiscent from the analogous field theory relation \req{FTmonoPl} and for $N\!=\!4$ eq. \req{RemiFT} assumes the same form as \req{Dix1}.

Similarly, the monodromy relation \req{MonoN} for the double non--planar ($N_1\!=\!2$)  one--loop $N$--point subamplitude \req{DefClosedChannelAmp} yields constraints on the terms of the $\ap$--expansion of its integrand $a^{(1)}(1,2|3,\ldots,N)$ referring to  indecomposable eMZVs $\omega_k$ from a minimal set of indecomposable eMZVs of weight $k$ and length $l$ introduced at  order  $\ap^k$:
\be\label{RemiMono1}
\lf. a^{(1)}(1,2|3,\ldots,N)\ri|_{\omega_k}=-\lf.a^{(1)}(1|2,\ldots,N)\ri|_{\omega_k}\ .
\ee
 Again, the relation \req{RemiMono1} is understood to be modulo powers of 
$i\pi $. Then, the potential dilaton tadpole terms proportional to $\ln q=2\pi i\tau$ (cf. the discussion below \req{DoubleNonPlanContr}) and the bulk terms  as a whole  cancel  in \req{Mono2}.
Again, the identity \req{RemiMono1} is reminiscent from the analogous field theory relation 
\req{FTCOLOR} and for $N\!=\!4$ it assumes the same form as \req{Dix2}:
\begin{align}
\lf. a^{(1)}(1,2|3,4)\ri|_{\omega_k}&=-\lf. a^{(1)}(1|2,3,4)\ri|_{\omega_k}\nonumber \\
&=-\lf.a^{(1)}(1,2,3,4)\ri|_{\omega_{k} }-\lf.a^{(1)}(2,1,3,4)\ri|_{\omega_{k}}-
\lf. a^{(1)}(2,3,1,4)\ri|_{\omega_{k}}\ .
\end{align}

As anticipated in subsection \ref{ellipticsection} two sets of eMZVs  emerge from evaluating multiple elliptic polylogarithms along the canonical paths $A$ and $B$ on $E_\tau^\times$, respectively.
In \req{eII} the two paths of integration  describe planar \req{path1} and non--planar \req{path2} superstring amplitudes, respectively. Therefore both sets of eMZVs show up in the monodromy relation \req{PlanarNIntegrandRel}.
Together with the integrand of \req{MonoN} our one--loop open string monodromy equations 
 provide identities  between these two different  systems of eMZVs. 
 Eventually,  such relations should give further input in understanding the algebra of  eMZVs 
 \cite{Broedel:2015hia}.

Recall, the Drinfeld associator $\Phi_{KZ}$, which is a formal power series in the non--comm-utative 
variables $e_0$ and $e_1$, represents the generating series of MZVs and describes the regularized monodromy of the universal KZ equation, cf. e.g. \cite{Stieberger:2016xhs} and references therein.
On the other hand, eMZVs appear also as the coefficients of the KZB associator \cite{Enriquez}. The latter is given by a triple $\{\Phi_{KZ},A(\tau),B(\tau)\}$ of formal powers series in the non--commutative 
variables $e_0$ and $e_1$ and describes the regularized monodromy of the universal elliptic KZB equation 
w.r.t. to the paths $A$ and $B$ on $E_\tau^\times$.
The series  $A(\tau)$ and $B(\tau)$ can be constructed by iterated integration (Picard's method) of the  elliptic KZB equation along the two paths $A$ and $B$, respectively. In \cite{BL} the underlying elliptic KZB form is defined on $E_\tau^\times$ and specified by both the one--forms $\omega^{(k)}$, introduced   in \req{forms}, and $\nu$. The resulting coefficients of their power series give rise  to homotopy invariant iterated integrals encoding $A$--elliptic and $B$--elliptic 
MZVs, respectively. The latter are related to the two sets of eMZVs stemming from evaluating multiple elliptic polylogarithms \req{eII} along the canonical paths $A$ and $B$.
As a consequence our monodromy equations \req{PlanarNIntegrandRel} and (the integrand of) \req{MonoN} furnish relations between $A$--elliptic and $B$--elliptic 
MZVs of the KZB associator.

\section[One--loop  field theory limit and monodromy relations]{One--loop  field theory limit and monodromy relations}
\label{Sect:FieldTheoryLimit}

\subsection{Four--point amplitude: field theory limit and monodromy relations}

In this section we shall investigate the field theory limit and first $\ap$--correction
to our monodromy relations.
Throughout this section we will work in the open string channel with the parameters \req{modop}
and \req{positionso}.
To be most concrete we shall discuss the four--point $N\!=\!4$ case. We start with the one--loop expressions
\req{OpenStringChannelAmp} or \req{GENERIC1} whose field theory limit $\ap\ra0$ (controlled by 
$\Im\tilde{\tau}\ra\infty$) for $N\!=\!4$ can be written as \cite{Green:1982sw}
\be\label{BRINK}
A^{(1)}(1,2,3,4)=s_{12}s_{23}\ A^{(0)}_{YM}(1,2,3,4)\ f(s_{12},s_{23})\ ,
\ee
with
\begin{align}
f(s_{12},s_{23})&=C_D\ \int_0^1\fc{dw}{w}\ (-\ln w)^{-D/2}\nonumber\\[2mm]
&\times\ \int\limits_{w\leq \rho_1\leq\rho_2\leq\rho_3\leq 1} \fc{d\rho_i}{\rho_i}\ K\ \exp\lf\{\fc{1}{\ln w}\lf(s\ln \rho_1\ln\fc{\rho_3}{\rho_2}+u\ln\fc{\rho_2}{\rho_1}\ln\fc{w}{\rho_3}\ri)\ri\}\ ,\label{Brink}
\end{align}
with a constant $C_D=\tfrac{g_{10}^2}{\ap}\lf(\tfrac{\ap}{\pi R^2}\ri)^{(10-D)/2}$ encoding the compactification radius $R$,  the space--time dimension $D$, and the open string coupling $g_{10}$. Furthermore, the factor $K$ accounts for string effects, more precisely $K=1+\Oc(\ap)$ with contributions only polynomial in $w$. Finally, we have 
\begin{align}
\rho_i&=w^{\hat\nu_i}=e^{-2\pi \Im \tilde{z}_i}\ \ ,\ \ \ \ i=1,2,3
\label{rho1}\\
\rho_4&=w=e^{-2\pi \Im\tilde{\tau}}\ \ \ \mbox{i.e.:}\ \ \ \ln w=2\pi i\tilde{\tau}\ \ \ ,\ \ \ 0\leq w\leq 1\ ,\label{rho2}
\end{align}
with  $0\leq \Im \tilde{z}_i\leq \Im\tilde{\tau}$, i.e. $ w\leq\rho_i\leq 1$ and 
$\hat\nu_i=\tfrac{\Im \tilde z_i}{t}=\fc{x_i}{t},\ \hat \nu_4=1$ ($i=1,2,3$).
As demonstrated in \cite{Green:1982sw} for $K\ra 1$ with
\begin{align}
\lambda&=-\ln w\ ,\nonumber\\
\eta_1&=\fc{\ln\rho_1}{\ln w}\ \ \ ,\ \ \ \eta_i=\fc{\ln\rho_i/\rho_{i-1}}{\ln w}\ ,\ \ \ i=2,3,4\ ,\label{etas}
\end{align}
from the expression \req{Brink} one can extract the one--loop box contribution, which in Schwinger parameterization reads \cite{DHoker:1994gnm}
\begin{align}
f_{YM}(s,u)&=
\lf(\fc{\ap}{2\pi}\ri)^\gamma\int  \fc{d^Dk}{k^2\ (k+k_1)^2\ (k+k_1+k_2)^2\ (k-k_4)^2}\nonumber\\
&=\int_0^\infty \fc{d\lambda}{\lambda^{D/2-3}}\int_0^1\lf(\prod_{i=1}^4 d\eta_i\ri)\delta\lf(1-\sum_{i=1}^4\eta_i\ri)\ \exp\lf\{-\lambda\ (s\ \eta_1\eta_3+u\ \eta_2\eta_4)\ri\}\nonumber\\
&= \fc{\Gamma(-\gamma)}{1+\gamma}\ \int_0^1d\eta_4\int_0^{1-\eta_4}d\eta_3\ (1-\eta_4-\eta_3)^{\gamma+1}\ \fc{(\eta_4 u)^{\gamma+1}-(\eta_3 s)^{\gamma+1}}{\eta_4 u-\eta_3 s}\ ,\label{box}
\end{align}
with $\gamma=\tfrac{1}{2}D-4$. Evaluating the last integral w.r.t. $\gamma=\eps-2$, i.e. the dimensional regularization parameter $\eps=\h (4-D)$, gives \cite{Green:1982sw,Bern:1994zx}
\be\label{fFT}
f_{YM}(s,u)=-c_\Gamma\ \lf\{-\fc{2}{\eps^2}\lf[\lf(\fc{\mu^2}{-s}\ri)^\eps+\lf(\fc{\mu^2}{-u}\ri)^\eps\ri]+
\ln^2\lf(\fc{-s}{-u}\ri)+\pi^2\ri\}+\Oc(\eps)\ ,
\ee
with
\be\label{cGamma}
c_\Gamma=\fc{(4\pi)^\epsilon}{16\pi^2}\ \fc{\Gamma(1+\eps)\ \Gamma(1-\eps)^2}{\Gamma(1-2\eps)}\ ,
\ee
and the regularization scale $\mu$. 

For $N\!=\!4$ the field theory limit of the non--planar amplitudes \req{GENERIC1} boils down to the same integrand \req{Brink} except that the integration region  $w\leq \rho_1\leq\rho_2\leq\rho_3\leq 1$ is accordingly changed to the whole cube \cite{Green:1982sw}.
To this end, the lowest  orders in $\ap$ of the planar and non--planer  one--loop  four--point amplitude read
\begin{align}
A^{(1)}(1,2,3,4)&=A^{(1)}_{YM}(1,2,3,4)+\Oc(\ap)\ ,\nonumber\\
A^{(1)}(2,3,4|1)&=A^{(1)}_{YM}(1|2,3,4)+\Oc(\ap)\ ,\label{FTExpa1}\\
A^{(1)}(1,2|3,4)&=A^{(1)}_{YM}(1,2|3,4)+\Oc(\ap)\ ,\nonumber
\end{align}
with $A^{(1)}_{YM}$ the field theory one--loop subamplitudes\footnote{\label{Comment2} The factor of $2$ in the last equation follows by permuting the first and second open string vertex operators (i.e. $u\leftrightarrow t$), which takes into account the different alignments of $1$ and $2$  along one cylinder boundary
without affecting the group structure due to $\Tr(T^{a_1}T^{a_2})=\Tr(T^{a_2}T^{a_1})$, cf. also footnote \ref{Comment1}.} \cite{Gross:1970eg}
\begin{align}
A^{(1)}_{YM}(1,2,3,4)&=s_{12}s_{34}\ A^{(0)}_{YM}(1,2,3,4)\ f_{YM}(s_{12},s_{23})\ ,
\label{Feld1}\\
A^{(1)}_{YM}(1|2,3,4)&=-s_{12}s_{34}\ A^{(0)}_{YM}(1,2,3,4)\ [\ g_{YM}(s,u)+g_{YM}(t,s)+g_{YM}(u,t)\ ]\ ,\label{Feld2}\\
A^{(1)}_{YM}(1,2|3,4)&=2\ s_{12}s_{34}\ A^{(0)}_{YM}(1,2,3,4)\ h_{YM}(s,u)\ , \label{Feld3}
\end{align}
with the form factor \req{fFT} and the relations:
\begin{align}
g_{YM}(s,u)&=f_{YM}(s,u)\ ,\nonumber\\
h_{YM}(s,u)&=\tfrac{1}{2}\ \lf[\ f_{YM}(s,u)+f_{YM}(t,s)+f_{YM}(u,t)\ri.\nonumber\\
&+\lf.f_{YM}(s,t)+f_{YM}(t,u)+f_{YM}(u,s)\ \ri]\nonumber\\
&=\ f_{YM}(s,u)+f_{YM}(t,s)+f_{YM}(u,t)\ \ .\label{thanks}
\end{align}

In the following for $N\!=\!4$ we shall also consider the first order $\ap$--expansion of the
object \req{GENERIC2}. More precisely, we are only interested in the imaginary part of the latter.
The shift \req{SHIFT}, which translates into $\ln\rho_1\ra\ln\rho_1+i\pi$ and produces  in \req{GENERIC2} the additional phase factor 
\be\label{exposs}
\exp\lf\{-i\pi\lf[\ s \fc{\Im(\tilde{z}_2-\tilde{z}_3)}{\tilde{\tau}_2}+u \fc{\Im(\tilde{z}_4-\tilde{z}_3)}{\tilde{\tau}_2}\ri]\ri\}\ ,
\ee
in turn in the integrand of \req{Brink} gives rise to the additional phase factor: 
\be\label{expos}
\exp\lf\{i\pi\lf[s\ln\lf(\frac{\rho_3}{\rho_2}\ri)-u\ln\lf(\frac{w}{\rho_3}\ri)\ri]
\fc{1}{\ln w}  \ri\}=\exp\lf\{i\pi \lf(\eta_3 s-\eta_4 u\ri)\ri\}\ .
\ee
Hence, at linear order in $\ap$ the object \req{GENERIC2} develops an imaginary part, i.e. for $N\!=\!4$
we have 
\be\label{FTExpa2}
C_D^{-1}\ \tilde A^{(1)}(2,3,4|1)=\tilde A^{(1)}_{YM}(2,3,4|1)+i\pi\ \tilde A^{(1)}_{YM}(1,2,3,4)[k_1]+\Oc(\ap)\ ,
\ee
with the lowest order real part given by:
\be
\tilde A^{(1)}_{YM}(2,3,4|1)=s_{12}s_{23}\ A^{(0)}_{YM}(1,2,3,4)\  g_{YM}(s_{12},s_{23})\ .\label{Tri0}
\ee
In \req{FTExpa2} possible real terms of the order $\ap$ are not explicitly written. The lowest order imaginary part of \req{FTExpa2} is given by
\be
\tilde  A^{(1)}_{YM}(1,2,3,4)[k_1]=s_{12}s_{23}\ A^{(0)}_{YM}(1,2,3,4)\ \tilde  g_{YM}(s_{12},s_{23})\ ,\label{Tri1}
\ee
with:
\begin{align}
\tilde g_{YM}(s,u)&=
\int_0^\infty \fc{d\lambda}{\lambda^{D/2-3}}\int_0^1\lf(\prod_{i=1}^4 d\eta_i\ri)
\delta\lf(1-\sum_{i=1}^4\eta_i\ri)\lf(\eta_3s-\eta_4u\ri)\ 
e^{-\lambda\ (s\eta_1\eta_3+u\eta_2\eta_4)}\nonumber\\[2mm]
&=- \fc{\Gamma(-\gamma)}{1+\gamma}\  \int_0^1d\eta_4\int_0^{1-\eta_4}d\eta_3\ (1-\eta_3-\eta_4)^{1+\gamma}\ \lf[\ -(\eta_4 u)^{1+\gamma}+(\eta_3 s)^{1+\gamma}\ \ri]\ .\label{TRIA}
\end{align}
Here, the bracket $[k_1]$ refers to the shift \req{SHIFT} giving rise to the factor \req{expos}. Furthermore,  the first order term \req{Tri1} stems from the linear order of the exponentials \req{expos} and is not affected by taking the limit $K\,\ra\, 1$.
Note, that according to the definition of the integration region  \req{F4Integrand} in \req{FTExpa2} we have the following iteration
$z_1<z_2<z_3<z_4$ leading to the canonical color orderings in \req{Tri0} and \req{Tri1}.
Similar to \req{FTExpa2} we also seek the following expressions:
\begin{align}
C_D^{-1}\ \tilde A^{(1)}(3,4,2|1)&=\tilde A^{(1)}_{YM}(3,4,2|1)+i\pi\ \tilde A^{(1)}_{YM}(1,3,4,2)[k_1]+\Oc(\ap)\ ,\label{FTExpa2a}\\
C_D^{-1}\ \tilde A^{(1)}(4,2,3|1)&=\tilde A^{(1)}_{YM}(4,2,3|1)+i\pi\ \tilde A^{(1)}_{YM}(1,4,2,3)[k_1]+\Oc(\ap)\ .\label{FTExpa2b}
\end{align}
Now, according to the definition of the integration region  \req{F4Integrand} we have the following iteration
$z_1<z_3<z_4<z_2$ in the first case and 
$z_1<z_4<z_2<z_3$ in the second case, respectively. From \req{Tri1} by permuting labels we deduce:
\begin{align}
\tilde A^{(1)}_{YM}(1,3,4,2)[k_1]&=s_{13}s_{34}\ A_{YM}^{(0)}(1,3,4,2)\ \tilde 
g_{YM}(s_{13},s_{34})\nonumber\\
&=s_{12}s_{23}\ A_{YM}^{(0)}(1,2,3,4)\ \tilde 
g_{YM}(s_{13},s_{34})\ ,\label{Tri2}\\
\tilde A^{(1)}_{YM}(1,4,2,3)[k_1]&=s_{14}s_{24}\ A_{YM}^{(0)}(1,4,2,3)\ \tilde 
g_{YM}(s_{14},s_{24})\nonumber\\
&=s_{12}s_{23}\ A_{YM}^{(0)}(1,2,3,4)\ \tilde 
g_{YM}(s_{14},s_{24})\ .\label{Tri3}
\end{align}
With 
\begin{align}
\tilde A^{(1)}_{YM}(3,4,2|1)&=s_{12}s_{23}\ A^{(0)}_{YM}(1,2,3,4)\  g_{YM}(s_{13},s_{34})\ ,\nonumber\\
\tilde A^{(1)}_{YM}(4,2,3|1)&=s_{12}s_{23}\ A^{(0)}_{YM}(1,2,3,4)\  g_{YM}(s_{14},s_{24})\ ,
\end{align}
we obviously have:
\be\label{Note}
A^{(1)}_{YM}(2,3,4|1)=\tilde A^{(1)}_{YM}(2,3,4|1)+\tilde A^{(1)}_{YM}(3,4,2|1)+\tilde A^{(1)}_{YM}(4,2,3|1)\ .
\ee
On the other hand, we can show that:
\be\label{Thanks}
\tilde  A^{(1)}_{YM}(1,2,3,4)[k_1]+\tilde A^{(1)}_{YM}(1,3,4,2)[k_1]+\tilde A^{(1)}_{YM}(1,4,2,3)[k_1]=0\ .
\ee
The relation \req{Thanks} can be proven by using the explicit integral representation \req{TRIA} for the expressions \req{Tri1}, \req{Tri2} and \req{Tri3} entitling the following identity: 
\be\label{IDENT}
\tilde  g_{YM}(s,u)+\tilde  g_{YM}(t,s)+\tilde  g_{YM}(u,t)=0\ .
\ee

Our next goal is  verifying the monodromy relations \req{Mono1} and \req{Mono2}.
We shall consider their real and imaginary parts at lowest order in $\ap$, separately.
By using the  expressions \req{FTExpa1}, \req{FTExpa2}, \req{FTExpa2a} and \req{FTExpa2b} for the monodromy relation \req{Mono1} we obtain two equations. 
With \req{Note} the  real part of \req{Mono1} yields the one--loop field theory subamplitude relation \req{Dix1}, which translates into 
\be
f_{YM}(s,u)+f_{YM}(s,t)+f_{YM}(u,t)= g_{YM}(s,u)+g_{YM}(t,s)+ g_{YM}(u,t)\ ,
\ee
and holds true thanks to \req{thanks}.
The imaginary part of \req{Mono1} simply gives the amplitude relation \req{Thanks}.
In the next subsection we shall demonstrate, that \req{Thanks} becomes a trivial consequence
after rewriting the subamplitudes \req{Tri1}, \req{Tri2} and \req{Tri3} in terms of triangle diagrams.
To conclude, up to the linear order in $\ap$ we do not get
any new subamplitude relations than what is already known from field theory.

Let us now turn to the monodromy relation \req{Mono2}. The lowest order of the double non--planar amplitude
is given in \req{FTExpa1}.
The amplitude \req{hatAmp} in the open string channel assumes a similar form as \req{GENERIC3a}.
Indeed, for $N\!=\!4$ the latter may be obtained by considering the shift of the holomorphic coordinate $\tilde z_2$ as
$\tilde{z}_2\ \lra\  \tilde{z}_2+\h$, which
in the integrand of \req{GENERIC1}  corresponds to 
taking $\Im \tilde{z}_2\ra \Im \tilde{z}_2-\tfrac{i}{2}$ and gives rise to the object
\begin{align}
  \hat A^{(1)}(1|2,3,4)&=\delta(k_1+k_2+k_3+k_4)\ e^{-i\pi s_{12}}\nonumber\\
  &\times\int_0^\infty \fc{dt}{t^6}\ 
V_{\rm CKG}^{-1}\int_{\Jc}\prod_{i=1}^4 d\tilde{z}_i
\exp\lf\{\fc{i\pi}{\Im \tilde{\tau}} \Im\lf(s_{12} \tilde z_{21}+s_{23}\tilde z_{23}+s_{24}\tilde z_{24} \ri)\ri\}\nonumber\\ 
&\times \exp\lf\{\tfrac{1}{2}\sum_{2\leq i<j\leq 4} s_{ij}\ \tilde{G}(\tilde{z}_i-\tilde{z}_j,\tilde{\tau})\ri\}\ 
\exp\lf\{\tfrac{1}{2}\sum_{j=2}^4 s_{1j}\ \tilde{G}_T(\tilde{z}_j-\tilde{z}_1-\tfrac{1}{2},\tilde{\tau})\ri\},\label{GENERIC3}
\end{align}
with the integration region $\Jc$ specified below eq. \req{GENERIC3a}. In fact,  for $N\!=\!4$ eq. \req{GENERIC3} simply follows  from \req{GENERIC3a}  by permuting the labels $1$ and $2$.
The expression \req{GENERIC3} carries the additional phase factor 
\be\label{exposshat}
\exp\lf\{-i\pi\lf[\ s \fc{\Im(\tilde{z}_1-\tilde{z}_4)}{\tilde{\tau}_2}+u \fc{\Im(\tilde{z}_3-\tilde{z}_4)}{\tilde{\tau}_2}\ \ri]\ri\}\ e^{-i\pi s}\ ,
\ee
which in the integrand of \req{Brink} gives rise to the additional phase factor: 
\be\label{exposhat}
\exp\lf\{-i\pi \lf[s\ln\lf(\rho_1\ri)-u\ln\lf(\frac{w}{\rho_3}\ri)\ri] \fc{1}{\ln w} \ri\}=
\exp\lf\{i\pi \lf(\eta_4 u-\eta_1 s\ri)\ri\}\ .
\ee
Note, that in \req{GENERIC3} the vertex position $\tilde z_1$ is integrated along the whole boundary without crossing branch cuts, cf. the discussion in subsection \ref{Hongkong}. 
Again, for \req{GENERIC3} we consider the first order $\ap$--expansion of its imaginary part.
Similarly to \req{FTExpa2} at linear order in $\ap$ the object  \req{GENERIC3} develops an imaginary part and we have
\be\label{FTExpa2hat}
C_D^{-1}\ \hat A^{(1)}(1|2,3,4)=\hat A^{(1)}_{YM}(1|2,3,4)+i\pi\ \hat A^{(1)}_{YM}(1,2,3,4)[k_2]+\Oc(\ap)\ ,
\ee
with the lowest order real part comprising the single non--planar amplitude
\be
\hat A^{(1)}_{YM}(1|2,3,4)=A^{(1)}_{YM}(1|2,3,4)\ ,
\label{Tri0hat}
\ee
given in \req{Feld2}.
Again, in \req{FTExpa2hat} possible real terms of the order $\ap$ are not explicitly written. On the other hand, the imaginary part term of \req{FTExpa2hat} can be written as
\be\label{Zhuhai}
\hat A^{(1)}_{YM}(1,2,3,4)[k_2]=s_{12}s_{23}\ A_{YM}^{(0)}(1,2,3,4)\ 
\lf[\ \hat g_{YM}(s,u)+\hat g_{YM}(t,s)+\hat g_{YM}(u,t)\ \ri]\ ,
\ee
with the form factor:
\begin{align}
\hat g_{YM}(s,u)&=
\int_0^\infty \fc{d\lambda}{\lambda^{D/2-3}}\int_0^1\lf(\prod_{i=1}^4 d\eta_i\ri)
\delta\lf(1-\sum_{i=1}^4\eta_i\ri)\lf(\eta_4u-\eta_1s\ri)\ 
e^{-\lambda\ (s\eta_1\eta_3+u\eta_2\eta_4)}\ .\label{TRIA2a}
\end{align}
By comparing \req{TRIA2a} with \req{TRIA} we evidence
$\hat g_{YM}(s,u)=-\tilde g_{YM}(s,u)$, and  from the explicit integral representation \req{TRIA2a}
we derive:
\be\label{Thankshat}
\hat A^{(1)}_{YM}(1,2,3,4)[k_2]=0\ .
\ee

With these preparations the lowest $\ap$--order (field theory part) of the real part monodromy relation \req{Mono2}
gives rise to the identity
\be
\h\ A_{YM}^{(1)}(1,2|3,4)+A^{(1)}_{YM}(1|2,3,4)=0\ ,\label{FOUND}
\ee
translating into 
\be
h_{YM}(s,u)=g_{YM}(s,u)+g_{YM}(t,s)+ g_{YM}(u,t)\ ,
\ee
which is valid thanks to \req{thanks}.
Together with \req{Dix1} the identity \req{FOUND} yields the known field theory relation \req{Dix2}.
On the other hand, since 
\be\label{stilltoprove}
\Re\ B_{YM}^{(1)}(1,2|3,4)=0\ ,
\ee
the lowest $\ap$ order of the imaginary part monodromy relation \req{Mono2}
confirms the identity \req{Thankshat}.
The proof of the validity of \req{stilltoprove} goes as follows.
The integrand \req{FormNPl2} of the boundary term \req{BOUNDA} can be transformed into the open string channel
\begin{align}
b(s,u)&=-\fc{i}{2}\ 
\int_0^{-1}d\alpha\int_0^td x_3\int_0^{x_3}d x_4\ 
\exp\lf\{\tfrac{1}{2}s_{12}\ \tilde G(i x_{21},\tilde\tau)+\tfrac{1}{2}s_{34}\ \tilde G(i x_{34},\tilde\tau)\ri\}\nonumber\\
&\times \lf.\exp\lf\{\tfrac{1}{2}\sum_{1\leq i<j\leq 4\atop (i,j)\notin \{(1,2),(3,4)\}} s_{ij}\
\tilde G_T(i x_{ji},\tilde\tau)\ri\}\ri|_{x_2=i\fc{\alpha}{2}}\ \exp\lf\{\fc{i\pi\alpha}{t}\lf(s x_{14}+u x_{34}\ri)\ri\}\ ,
\label{BoundOpen}
\end{align}
which in terms of the coordinates  \req{rho1} and \req{rho2} and in the limit $\Im\tilde\tau\ra\infty$
(with $K\ra 1$) reduces to:
\begin{align}
B^{(1)}(1,2|3,4)&=-\fc{i}{2}\ (s_{12}s_{14})\ A_{YM}^{(0)}(1,2,3,4)\ \label{HONGKONG} \\[1mm]
&\times 
\int_0^1\fc{dw}{w}\ (-\ln w)^{-D/2}\hskip-0.5cm\int\limits_{w\leq \rho_1\leq\rho_3\leq 1}
\fc{d\rho_i}{\rho_i}\ \exp\lf\{\fc{1}{\ln w}\lf(s\ln \rho_1\ln\fc{\rho_3}{1}+u\ln\fc{1}{\rho_1}\ln\fc{w}{\rho_3}\ri)\ri\}\nonumber\\[2mm]
&\times\int_0^{-1}d\alpha\ e^{i\pi \alpha s}\ \exp\lf\{i\pi\alpha \lf[s\ln\lf(\rho_1\ri)-u\ln\lf(\frac{w}{\rho_3}\ri)\ri] \fc{1}{\ln w} \ri\}\ .\nonumber
\end{align}
The $\alpha$--integral has a real part of $\Oc(\ap^0)$, while the remaining integrals give also something real. Hence, we are able to confirm \req{stilltoprove}.
Therefore, in the field theory limit the bulk term \req{BoundOpen} does not contribute to the monodromy relation \req{Mono2}.

\subsection{Field theory amplitude relations  from string amplitudes}
\label{FTSubsection}

In $\mathcal{N}=4$ SYM the $N$--point one--loop gluon amplitude assumes the generic form \cite{Bern:1994zx}
\be\label{Dunb}
A^{(1)}_{YM}(1,\ldots,N)=c_\Gamma\ \lf\{\ V_N^g\ A^{(0)}_{YM}(1,\ldots,N)+i\ F_N^g\ \ri\}\ ,
\ee
with the tree--level amplitude $A^{(0)}_{YM}$, the factor $c_\Gamma$ defined in \req{cGamma}, and the functions
$V_N^g$ and $F_N^g$. For the cases $N=4,5$ the functions $V_N^g$ are universal
for any helicity configuration, while $F_N^g=0$ for these two cases. Generically, for $N\geq6$
the function $F_N^g$ only vanishes in the MHV case and for $N\!=\!4$ the expression \req{Dunb} reproduces \req{Feld1}.

With  $\int\limits_{-\infty}^\infty d^D k\ e^{-\lambda k^2}=\lf(\tfrac{\pi}{\lambda}\ri)^{D/2}$ the box--integral \req{box}
can be written as:
\be\label{HKG}
f_{YM}(s,u)=\pi^{-\tfrac{D}{2}}\hskip-0.15cm\int_0^\infty \fc{d\lambda}{\lambda^{-3}}\int d^D k'\int_0^1\lf(\prod_{i=1}^4 d\eta_i\ri)\delta\lf(1-\sum_{i=1}^4\eta_i\ri) \exp\lf\{-\lambda ({k'}^2+s \eta_1\eta_3+u \eta_2\eta_4)\ri\}.
\ee
To derive \req{HKG} by means of Schwinger parametrization one performs the  
shift $k'=k+\eta_2k_1+\eta_3(k_1+k_2)-\eta_4k_4$ on the momentum $k$.
The one--loop integral \req{HKG} is reminiscent of the loop momentum expression \req{Loopm}. 
Indeed, for $N\!=\!4$ after dropping the 
factor $\prod\limits_{1\leq i<j\leq N} \lf(\tfrac{\theta_1(\tilde z_{ji},\tilde\tau)}{\theta_1(0,\tilde\tau)}\ri)^{s_{ij}}$, which accounts for string effects the amplitude \req{Loopm} may be written as
\be
f_{YM}(s,u)\simeq 2^{-D/2}\ \int_0^\infty \fc{dt}{t}\ V_{\rm CKG}^{-1}\int_{\Jc_1}\prod_{i=1}^N d\tilde{z}_i \int d^D\ell'\ \exp\lf\{-\h\pi \ap t\lf( {\ell'}^2-4\ \sum_{i<j}s_{ij}\fc{\tilde z_i\tilde z_j}{t^2}\ri)\ri\}\ ,\label{loopm}
\ee
with $\ell'=\ell+2i\sum\limits_{i=1}^4 k_i \fc{\tilde z_i}{t}$.
By comparing \req{HKG} and \req{loopm} one sees, that $k$ plays the role of the loop momentum $\ell$, with
the identification of the Schwinger proper time $\lambda=\tfrac{1}{2}\pi\ap t$.
Furthermore, after gauge fixing the parameters $\eta_i$ can be related to
the coordinates $\tfrac{\tilde z_i}{t}$ in the loop momentum picture subject to the definitions \req{rho1}--\req{etas}.

The integral \req{TRIA} can explicitly be computed with the result:
\be\label{RESD}
\tilde g_{YM}(s,u)=\h\ \fc{(1+\gamma)\ \Gamma(1+\gamma)^2\ \Gamma(-\gamma)}{(2+\gamma)\ \Gamma(2\gamma+4)}\ \lf(s^{\gamma+1}-u^{\gamma+1}\ri)\ .
\ee
The explicit expression \req{RESD} confirms the identity \req{IDENT}. 
Eventually, evaluating \req{RESD} w.r.t. $\eps=\gamma+2$ gives:
\begin{align}
\tilde g_{YM}(s,u)&=\h\ \fc{(1+\gamma)\ \Gamma(1+\gamma)^2\ \Gamma(-\gamma)}{(2+\gamma)\ \Gamma(2\gamma+4)}\ \lf(s^{\gamma+1}-u^{\gamma+1}\ri)\nonumber\\
&=\fc{1}{\eps^2}\ \lf(\fc{1}{u}-\fc{1}{s}\ri)+\fc{1}{\eps}\ \lf(\fc{\gamma_E}{u}-\fc{\gamma_E}{s}+\fc{\ln u}{u}-\fc{\ln s}{s}\ri)\nonumber\\
&+\fc{\pi^2}{12}\ \lf(\fc{1}{s}-\fc{1}{u}\ri)-\fc{(\gamma_E+\ln s)^2}{2s}+\fc{(\gamma_E+\ln u)^2}{2u}
+\Oc(\eps)\ .
\end{align}

Recall, in the previous subsection we have considered in \req{GENERIC2} the shift \req{SHIFT} and extracted the form factor \req{TRIA}.
This has been achieved by introducing this shift into \req{Brink}
and taking its lowest order imaginary part $\tilde g_{1,YM}(s,u):=
\tilde g_{YM}(s,u)$.   
Similarly, we may discuss  the shifts 
$z_2\ra z_2+\frac{1}{2}$ and $z_3\ra z_3+\frac{1}{2}$, which in turn in \req{Brink} translate into  $\ln \rho_2\ra\ln \rho_2+i\pi $  and $\ln \rho_3\ra\ln \rho_3+i\pi$ and yield the factors $\exp\{i\pi  \lf(-\eta_1 s+\eta_4 u\ri)\}\exp(i\pi s)$ and $\exp\{i\pi \lf(\eta_1 s-\eta_2 u\ri)\}\exp(-i\pi s)$, respectively and give rise to the objects
\begin{align}
\tilde g_{2,YM}(s,u)&=
\int_0^\infty \fc{d\lambda}{\lambda^{D/2-3}}\int_0^1\lf(\prod_{i=1}^4 d\eta_i\ri)
\delta\lf(1-\sum_{i=1}^4\eta_i\ri)\lf(-\eta_1s+\eta_4u\ri)\ 
e^{-\lambda\ (s\eta_1\eta_3+u\eta_2\eta_4)}\ ,\label{TRIA2}\\
\tilde g_{3,YM}(s,u)&=\int_0^\infty \fc{d\lambda}{\lambda^{D/2-3}}\int_0^1
\lf(\prod_{i=1}^4 d\eta_i\ri)\delta\lf(1-\sum_{i=1}^4\eta_i\ri)\lf(\eta_1s-\eta_2u\ri)\ 
e^{-\lambda\ (s\eta_1\eta_3+u\eta_2\eta_4)}\ ,\label{TRIA3}
\end{align}
respectively. All the integrals \req{TRIA}, \req{TRIA2} and \req{TRIA3} are integrated w.r.t. to the canonical ordering $\Jc_1$. In lines of \req{Tri1} we have
\be
\tilde  A^{(1)}_{YM}(1,2,3,4)[k_i]=s_{12}s_{23}\ A^{(0)}_{YM}(1,2,3,4)\ \tilde g_{i,YM}(s,u)\ \ \ ,\ \ \ i=1,2,3\ ,\label{TriAll}
\ee
and we easily verify
\be\label{verify}
\tilde g_{2,YM}(s,u)=\tilde g_{YM}(u,s)=-\tilde g_{YM}(s,u)\ \ \ ,\ \ \ \tilde g_{3,YM}(s,u)=\tilde g_{YM}(s,u)\ ,
\ee
from which we deduce:
\begin{align}
\tilde A^{(1)}_{YM}(2,1,3,4)[k_1]&=s_{12}s_{13}\ A^{(0)}_{YM}(2,1,3,4)\ \tilde g_{YM}(t,s)\ ,\label{TTri2}\\[2mm]
\tilde A^{(1)}_{YM}(2,3,1,4)[k_1]&=s_{23}s_{13}\ A^{(0)}_{YM}(2,3,1,4)\ \tilde g_{YM}(u,t)\ .\label{TTri3}
\end{align}
Hence, the relations \req{Thanks} and \req{IDENT} entitle to:
\be\label{Mainz}
\tilde A^{(1)}_{YM}(1,2,3,4)[k_1]+\tilde A^{(1)}_{YM}(2,1,3,4)[k_1]+\tilde A^{(1)}_{YM}(2,3,1,4)[k_1]=0\ .
\ee
The amplitudes \req{TriAll} are related to combinations of triangle diagrams in $D$ dimensions. As a consequence the same is true for the amplitudes entering \req{Mainz}. To see this, let us first discuss the 
box diagram $\square_i$ with momentum insertion $lk_i$ in the numerator in $D$ dimensions. 
After introducing  Feynman parameters $\eta_i$ this diagram is described by
\begin{align}
\label{boxdiagrammomentuminsertion}
\square_i(s,u)&:=\lf(\fc{\ap}{2\pi}\ri)^\gamma\int d^{D}l\ \fc{lk_i}{l^2 (l+k_1)^2\ (l+k_1+k_2)^2\ (l-k_4)^2}=\h\int_0^\infty
\fc{d\lambda}{\lambda^{D/2-3}}\int_0^1\lf(\prod_{j=1}^4 d\eta_j\ri)  \\
&\times  \delta\lf(1-\sum_{i=1}^4\eta_i\ri)\ 
\lf[\ \eta_4 s_{4i}-\eta_3 (s_{1i}+s_{2i})-\eta_2 s_{1i}\ \ri]\ 
e^{-\lambda (s\eta_1\eta_3+u\eta_2\eta_4)}\ ,\ i=1,\ldots,4\ .\nonumber
\end{align}
By comparing the integrand of \req{boxdiagrammomentuminsertion} with \req{TRIA}, \req{TRIA2} and \req{TRIA3} we immediately
verify
\begin{align}
\tilde g_{1,YM}(s,u)&=-2\ \square_1(s,u)\ ,\nonumber\\
\tilde g_{2,YM}(s,u)&=-2\ \square_2(s,u)-s\ \square(s,u)\ ,\label{newform}\\
\tilde g_{3,YM}(s,u)&=-2\ \square_3(s,u)+s\ \square(s,u)\ ,\nonumber
\end{align}
with $\square$ the box integral in $D$ dimensions. The latter is given in \req{box}.
The relation \req{IDENT}, which can also be written as $\tilde g_{1,YM}(s,u)+
\tilde g_{2,YM}(s,t)+\tilde g_{3,YM}(u,t)=0$, in terms of \req{newform} then simply provides the following relation
between loop integrals in $D$ dimensions:
\be\label{IBP}
2\ \square_1(s,u)+2\ \square_2(s,t)+s\ \square(s,t)+2\ \square_3(u,t)-u\ \square(u,t)=0\ .
\ee
On the other hand, we can rewrite \req{boxdiagrammomentuminsertion} by introducing the denominators
\be
d_i=(l+q_i)^2\ \ \ , \ \ \ i=1,\ldots,4\ ,
\ee
with:
\be\begin{array}{rl}
q_1&=k_1\ ,\\
q_2&=k_1+k_2\ ,\\
q_3&=k_1+k_2+k_3=-k_4\ ,\\
q_4&=0\ .
\end{array}\ee
By using
\be
lq_i=\h\ (d_i-d_4-q_i^2)\ \ \ , \ \ \ i=1,\ldots,4\ ,
\ee
i.e.
\begin{align*}
lk_1&=\tfrac{1}{2}\ (d_1-d_4)\ ,\\
lk_2&=\tfrac{1}{2}\ (d_2-d_1-s)\ ,\\
lk_3&=\tfrac{1}{2}\ (d_3-d_2+s)\ ,
\end{align*}
the integral $\square_i$ for the box diagram with momentum insertion \req{boxdiagrammomentuminsertion}  can actually be decomposed
w.r.t. triangle diagrams $\Delta_i$ 
\be
\Delta_i=\int d^{D}l\ \fc{d_i}{d_1d_2d_3d_4}\ \ \ ,\ \ \ i=1,\ldots,4\ ,
\ee
and the box diagram $\square$ without momentum insertions as
\begin{align}
\square_1(s,u)&=\h\int d^{D}l\     \lf(\fc{1}{d_2d_3d_4}-\fc{1}{d_1d_2d_3}\ri)=
\h\ (\Delta_1-\Delta_4)\ ,\nonumber\\
\square_2(s,u)&=\h\int d^{D}l\     \lf(\fc{1}{d_1d_3d_4}-\fc{1}{d_2d_3d_4}-\fc{s}{d_1d_2d_3d_4}\ri)=
\h\ (\Delta_2-\Delta_1-s\square)\ ,\label{boxdeco}\\
\square_3(s,u)&=\h\int d^{D}l\     \lf(\fc{1}{d_1d_2d_4}-\fc{1}{d_1d_3d_4}+\fc{s}{d_1d_2d_3d_4}\ri)=
\h\ (\Delta_3-\Delta_2+s\square)\ ,\nonumber
\end{align}
which in turn yields:
\begin{align}
\tilde g_{1,YM}(s,u)&=- (\Delta_1-\Delta_4)\ ,\nonumber\\
\tilde g_{2,YM}(s,u)&=- (\Delta_2-\Delta_1)\ ,\label{Dreiecke}\\
\tilde g_{3,YM}(s,u)&=- (\Delta_3-\Delta_2)\ .\nonumber
\end{align}
Hence, subject to integral reduction the amplitudes \req{TriAll} and subsequently \req{Tri1} and \req{TTri2} and \req{TTri3} are simple combinations of two triangle diagrams in $D$ dimensions. As a consequence
the relations \req{Thanks} and \req{Thankshat} become  simple identities \req{IDENT} between triangle diagrams
\req{Dreiecke}
\be\label{IBPP}
\Delta_1(s,u)-\Delta_4(s,u)+\Delta_2(s,t)-\Delta_1(s,t)+\Delta_3(u,t)-\Delta_2(u,t)=0\ ,
\ee
which can easily be verified.

Loop level integrands share similar properties as tree--level amplitudes. In particular,
BCJ type relations can be
stated for one--loop planar integrands \cite{Boels:2011mn}.
While at tree--level such relations are established for the amplitudes at one--loop
these relations are stated for their integrands. Therefore, those relations 
depend on the loop momentum $l$ and are valid up to terms which vanish after loop integration. 
E.g. for the integrand \req{box} of the color ordered planar $\Nc=4$ SYM four--point amplitude \req{Feld1}
\be\label{boxintegrand}
I(1,2,3,4)=\fc{1}{l^2\ (l+k_1)^2\ (l+k_1+k_2)^2\ (l-k_4)^2}=\fc{1}{d_1d_2d_3d_4}
\ee
the following fundamental BCJ relation can be stated \cite{Boels:2011mn,Du:2012mt}
\be\label{Isermann}
s_{1l}\ I(1,2,3,4)+(s_{12}+s_{1l})\ I(2,1,3,4)+(s_{12}+s_{13}+s_{1l})\ I(2,3,1,4)=0\ ,
\ee
with $s_{1l}=(k_1+l)^2-k_1^2-l^2=2k_1l$.
Upon loop momentum integration the identity \req{Isermann} simply becomes
the relation \req{IBP} between loop integrals in $D$ dimensions.
Moreover, after decomposing the box integrals according to \req{boxdeco} the (integrated) identity \req{Isermann} turns into the trivial statement \req{IBPP} between triangle diagrams.

To summarize, the equations \req{Thanks} and \req{Thankshat} follow from the imaginary parts of the string monodromy relations 
 \req{Mono1} and \req{Mono2}, respectively in the limit $\ap\ra0$. 
 In field theory, these relations are just a consequence of
decomposing rational functions, i.e. trivial manipulations of rational functions by means of 
integral reductions \req{boxdeco} of certain box--diagrams into triangle diagrams.
Note, that full--fledged string amplitudes automatically incorporate such (integrated) relations as they
do not distinguish between different four--point amplitudes for generic
vertex operator positions at generic value of modular parameter $\tau$.
As a consequence, in the field theory limit $\ap\ra0$ for   one--loop amplitudes in $\Nc=4$ SYM we simply recover the well--known amplitude relation \req{FTmonoPl}.
On the other hand, considering higher orders in $\ap$ gives rise to relations between partial
subamplitudes describing higher--dimensional operators. Such constraints should be related to  the results in \cite{BjerrumBohr:2011xe}.
As we have shown in section \ref{transcend} a tower of constraining equations for higher order  
$\ap$--terms appears.

Let us now extract the field theory limit of the relations \req{TVmonodromy} and \req{TVnonplanMonod}.
Due to \req{Thanks} the imaginary part of the field theory limit of \req{SHTV}  yields the equation: 
$A(2,3,4|1)[\ell k_1]=0$. This statement implies, that  
the imaginary part of the field theory limit of \req{TVmonodromy} reads 
$ (k_1k_2) A_{YM}(2,1,3,4)+k_1(k_2+k_3) A_{YM}(2,3,1,4)\!=\!0$, which is clearly wrong.
On the other hand, the real part of the field theory limit of \req{TVnonplanMonod} 
gives $A_{YM}(1,2|3,4)+A_{YM}(2|1,3,4)+A_{YM}(2|3,1,4)+ A_{YM}(2|3,4,1)\!=\!0$, which clearly disagrees with eqs. \req{Dix1} and \req{Dix2}. Furthermore, the imaginary part of the field theory limit of \req{TVnonplanMonod}
provides $-A(2|1,3,4)[\ell k_1]-A(2|3,1,4)[\ell k_1]-A(2|3,4,1)[\ell k_1]+(k_1k_3) A_{YM}(2|3,1,4)+k_1(k_3+k_4) A_{YM}(2|3,4,1)\!=\!0$. Similar to \req{Thanks}   or \req{Thankshat} the first three terms sum up to zero leaving:
$(k_1k_3) A_{YM}(2|3,1,4)+k_1(k_3+k_4) A_{YM}(2|3,4,1)\!=\!0$, which again is incorrect.
Again,  the whole discrepancies and conflicts encountered above in the field theory limit 
of the relations \req{TVmonodromy} and \req{TVnonplanMonod} can be traced back
to missing phases in the monodromy relations.
Moreover, as a consequence also the field theory expression (14) of  \cite{Tourkine:2016bak} misses additional terms stemming from expanding monodromy phases.




\section{Monodromy relations for non--orientable amplitudes}\label{Sect:NonOrientable}

In this section we briefly discuss how to generalise the monodromy relations of $N$-point scattering amplitudes with a world-sheet cylinder (discussed in section~\ref{Sect:MonodromyOrientable}), to amplitudes with a M\"obius strip world-sheet. We explain how (in principle) such relations can be established in a similar fashion as for the cylinder, however, also point out new problems and issues that arise in the process.
\subsection{Amplitudes with M\"obius strip world--sheet}\label{Sect:AmpsMob}

We can generalize the discussion of $N$-point orientable scattering amplitudes of section~\ref{Sect:OneLoopAmps} to the case of non-orientable strings. In this paper we will limit ourselves to $N$-point amplitudes on the M\"obius strip. Using a similar notation as in (\ref{OneLoopAmpGen}), we can write the latter in the following way
\begin{align}
{\frak M}^{(1)}_N=\sum_{\sigma\in S_{N-1}}V^{-1}_{\text{CKG}}\,\left(\int_{\mathcal{J}_{\sigma}}\prod_{i=1}^{N}dz_i\right)\left\langle\prod_{i=1}^{N}:V_o(z_i):\right\rangle\,.\label{OneLoopAmpGenMoeb}
\end{align}
Indeed, representing the M\"obius strip as a sphere with a single boundary and a crosscap\\[10pt]
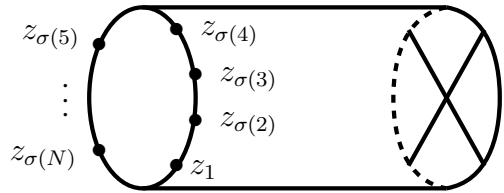
\begin{center}
\begin{tikzpicture}
\draw[ultra thick] (-1,0) ellipse (0.7cm and 1.2cm);
\draw[ultra thick,dashed] (3,1.2) to [out=190,in=90] (2.3,0) to [out=270,in=170] (3,-1.2);
\draw[ultra thick] (3,1.2) to [out=350,in=90] (3.7,0) to [out=270,in=10] (3,-1.2);
\draw[ultra thick] (-1,1.2) -- (3,1.2);
\draw[ultra thick] (-1,-1.2) -- (3,-1.2);
\node at (-0.55,-0.9) {$\bullet$};
\node at (-0.2,-0.95) {$z_1$};
\node at (-0.3,-0.3) {$\bullet$};
\node at (0.4,-0.35) {$z_{\sigma(2)}$};
\node at (-0.3,0.3) {$\bullet$};
\node at (0.4,0.25) {$z_{\sigma(3)}$};
\node at (-0.55,0.9) {$\bullet$};
\node at (0.15,0.85) {$z_{\sigma(4)}$};
\node at (-1.57,-0.7) {$\bullet$};
\node at (-2.3,-0.8) {$z_{\sigma(N)}$};
\node at (-1.57,0.7) {$\bullet$};
\node at (-2.2,0.8) {$z_{\sigma(5)}$};
\node[rotate=90] at (-2,0) {$\cdots$};
\draw[ultra thick] (2.5,-0.9) -- (3.5,0.9);
\draw[ultra thick] (2.5,0.9) -- (3.5,-0.9);
%
\end{tikzpicture}
\end{center}
${}$\\
the summation in (\ref{OneLoopAmpGenMoeb}) is over all $(N-1)!$ inequivalent orderings of the vertex insertions $z_i$ on the single boundary component, where the precise integral region $\mathcal{J}_\sigma$ shall be defined later on in (\ref{DefMobiusIntegral}) (and (\ref{DefMobiusIntegralc})). In the same way as for the cylinder, we consider a fixed ordering of the vertex points and strip-off a color factor:
\begin{align}
{\frak M}^{(1)}_N=\sum_{\sigma\in S_{N-1}}\,\text{Tr}(T^1\,T^{\sigma(2)}\ldots T^{\sigma(N)})\ 
M^{(1)}(1,\sigma(2),\ldots,\sigma(N))\ .
\end{align}
In the following we shall only focus on the contribution $M^{(1)}(1,\sigma(2),\ldots,\sigma(N))$, which (like the cylinder amplitude) can be described either in the open or in the closed string channel. For the former, we recall that the M\"obius strip can be parameterized in a similar way as the annulus
with the modular parameter $\tilde{\tau}_M=\tfrac{1}{2}+\tfrac{it}{2}$, cf. the next figure.
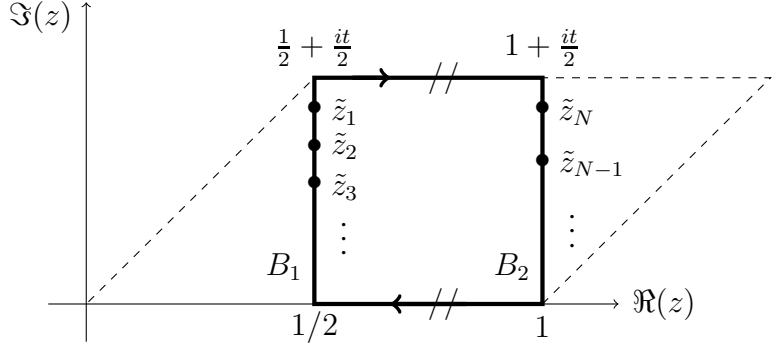
\begin{center}
\begin{tikzpicture}
\draw[dashed] (6,0) -- (9,3) -- (3,3) -- (0,0);
\draw[->] (-0.5,0) -- (7,0);
\draw[->] (0,-0.5) -- (0,4);
\draw[ultra thick] (3,0) rectangle (6,3);
\node at (7.6,0) {$\Re (z)$};
\node at (-0.6,3.8) {$\Im (z)$};
\node at (3,-0.3) {$1/2$};
\node at (6,-0.3) {$1$};
\node at (3,3.4) {$\tfrac{1}{2}+\tfrac{it}{2}$};
\node at (6,3.4) {$1+\tfrac{it}{2}$};
\node[xshift=2.2cm] at (2.5,0) {//};
\draw[ultra thick,->,xshift=2.5cm] (2.5,0) -- (1.5,0);
\node[xshift=2.2cm] at (2.5,3) {//};
\draw[ultra thick,<-,xshift=2.5cm] (1.5,3) -- (1,3);
\node[xshift=3cm] at (0,1.6) {$\bullet$};
\node[xshift=3cm] at (0.4,1.55) {$\tilde{z}_3$};
\node[xshift=3cm] at (0,2.1) {$\bullet$};
\node[xshift=3cm] at (0.4,2.1) {$\tilde{z}_2$};
\node[xshift=3cm] at (0,2.6) {$\bullet$};
\node[xshift=3cm] at (0.4,2.55) {$\tilde{z}_1$};
\node[rotate=90] at (3.4,0.9) {$\cdots$};
\node[rotate=90] at (6.4,1) {$\cdots$};
\node at (6,2.6) {$\bullet$};
\node at (6.45,2.55) {$\tilde{z}_{N}$};
\node at (6,1.9) {$\bullet$};
\node at (6.65,1.85) {$\tilde{z}_{N-1}$};
\node at (2.6,0.5) {$B_1$};
\node at (5.6,0.5) {$B_2$};
\end{tikzpicture}
\end{center}


\noindent
The main difference, however, is that the two horizontal parts are oppositely oriented in such a way that the two vertical lines $B_{1,2}$ form two parts of the single boundary component. The vertex positions are parametrised as 
\begin{align}
\tilde{z}_i=\left\{\begin{array}{lcl} \tfrac{1}{2}+\tfrac{ix_i}{2} & \text{for} & x_i\in B_1 \\[4pt] 1+\tfrac{ix_i}{2} & \text{for} & x_i\in B_2\end{array}\right.&&\text{with} &&x_i\in[0,t] \,,&&i=1,\ldots,N
\end{align}
while the integration regions are given by
\begin{align}
&\mathcal{J}_\sigma=\bigcup_{a=1}^N {\cal J}^a_{\sigma}\cup {\cal J}^{\prime a}_{\sigma}\,,&& \text{with} &&\sigma\in S_{N-1}\,.\label{DefMobiusIntegral}
\end{align}
Here the points $\{\tilde{z}_1,\ldots,\tilde{z}_a\}$ are inserted on $B_1$ and the points $\{\tilde{z}_{a+1},\ldots,\tilde{z}_N\}$ on $B_2$ and:
\begin{align}\label{RegionsMob}
{\cal J}^a_{\sigma}&=\lf\{\ \bigcup_{i=1}^{a}x_i\in E_{\tilde{\tau}_M}^{a}\ |\ 0\leq x_1\leq x_{\sigma(2)} \leq\ldots\leq x_{\sigma(a)} \leq t\ \ri\}\ ,\ \sigma\in S_{N_1-1}\ ,
\nonumber\\
{\cal J}^{\prime a}_{\sigma}&=\,\lf\{\ \bigcup_{i=a+1}^{N}x_i\in E_{\tilde{\tau}_M}^{N-a}\ |\ 0\leq x_{a+1}\leq x_{\sigma(a+2)} \leq\ldots\leq x_{\sigma'(N)} \leq t\ \ri\}\ ,\ \sigma'\in S_{N_2-1}\ .
\end{align}
The corresponding amplitude can be written in the form \cite{Green:1987mn}
\begin{align}
M^{(1)}(1,\ldots, N)&=\delta(k_1,\ldots,k_N)\int_0^\infty \frac{dt}{t^6}\ V_{\text{CKG}}^{-1}\ \int_{\mathcal{J}_1}\,\prod_{i=1}^Nd\tilde{z}_i\ R_N(\tilde{z}_1,\ldots,\tilde{z}_N,\tilde{\tau}_M)\nonumber\\
&\times \text{exp}\left\{\tfrac{1}{2}\sum_{1\leq i<j\leq N}s_{ij}\,\tilde{G}(\tilde{z}_{ji},\tilde{\tau}_M)\right\}\,,\label{OpenStringChannelMobius}
\end{align}
with $\tilde{G}(\tilde{z},\tilde{\tau}_M)$ given as in (\ref{Go}). 

Similar to the case of scattering amplitudes with a cylinder world-sheet, we can also formulate (\ref{OpenStringChannelMobius}) in the closed string channel. To this end we introduce the new modular parameter $l$ through $\tfrac{i}{2t}=2il$ and map all points
\begin{align}
z\longmapsto \frac{z}{1-2\tau_M}\,,
\end{align}
such that we obtain the following figure
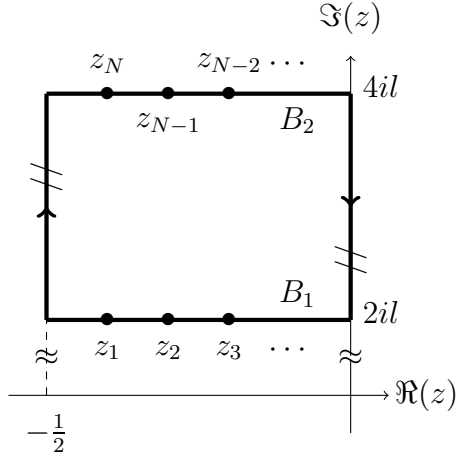
\begin{center}
\begin{tikzpicture}[decoration={
    markings,
    mark=at position 0.5 with {\arrow{>}}}
    ] 
\draw (0,-0.5) -- (0,0.475);
\node at (0,0.5) {$\approx$};
\draw[->] (0,0.525) -- (0,4.5);
\draw[->] (-4.5,0) -- (0.5,0);
\node at (1,0) {$\Re (z)$};
\node at (0,5) {$\Im (z)$};
\draw[ultra thick] (-4,1) -- (0,1);
\draw[ultra thick] (0,4) -- (-4,4);
\draw[ultra thick,postaction={decorate}]  (0,4) -- (0,1);
\draw[ultra thick,postaction={decorate}] (-4,1) -- (-4,4);
\node at (0.4,1.1) {$2il$};
\node at (0.4,4.1) {$4il$};
\node at (-4,-0.5) {$-\tfrac{1}{2}$};
\draw[dashed] (-4,-0) -- (-4,0.475);
\node at (-4,0.5) {$\approx$};
\draw[dashed] (-4,0.525) -- (-4,1);
\node[rotate=90] at (0,1.8) {$\slash\slash$};
\node[rotate=90] at (-4,2.9) {$\slash\slash$};
\node at (-3.2,1) {$\bullet$};
\node at (-3.2,0.6) {$z_1$};
\node at (-2.4,1) {$\bullet$};
\node at (-2.4,0.6) {$z_2$};
\node at (-1.6,1) {$\bullet$};
\node at (-1.6,0.6) {$z_3$};
\node at (-0.8,0.6) {$\cdots$};
\node at (-3.2,4) {$\bullet$};
\node at (-3.2,4.4) {$z_N$};
\node at (-2.4,4) {$\bullet$};
\node at (-2.4,3.6) {$z_{N-1}$};
\node at (-1.6,4) {$\bullet$};
\node at (-1.6,4.4) {$z_{N-2}$};
\node at (-0.8,4.4) {$\cdots$};
\node at (-0.7,1.35) {$B_1$};
\node at (-0.7,3.65) {$B_2$};
\end{tikzpicture}
\end{center}
The position of the vertex insertion points in the closed string channel is
\begin{align}
&z_i=\left\{\begin{array}{lcl}-\tfrac{x_i}{2}+2il & \text{for} & x_i\in B_1 \\ -\tfrac{x_i}{2}+4il & \text{for} & x_i\in B_1 \end{array}\right.&&\text{with} &&x_i\in [0,1]\,, &&i=1,\ldots,N\,.
\end{align}
Furthermore, the $N$-point amplitude takes the form
\begin{align}
M^{(1)}(1,\ldots,N)&=\,\delta(k_1+\ldots+k_N)\int_0^\infty dl\ V^{-1}_{\text{CKG}}\ \int_{\mathcal{I}_1}\prod_{i=1}^Ndz_i\ R_N(z_1,\ldots,z_N,\tau)\nonumber\\
&\hspace{1.5cm}\times \exp\left\{\tfrac{1}{2}\sum_{1\leq i<j\leq N}s_{ij}G(z_{ji},\tau_M)\right\}\nonumber\\
&=\delta(k_1+\ldots+k_N)\int dl\ V^{-1}_{\text{CKG}}\ \int_{\mathcal{I}_1}\prod_{i=1}^N dz_i\ \prod_{i<j}\left|\frac{\theta_1(z_{ji},2il-\tfrac{1}{2})}{\theta'_1(0,2il-\tfrac{1}{2})}\right|^{s_{ij}} \,.\label{ClosedStringChannelAmpMob2}
\end{align}
Here the integral measure is given by
\begin{align}
&\mathcal{I}_\sigma=\bigcup_{a=1}^N {\cal I}^a_{\sigma}\cup {\cal I}^{\prime a}_{\sigma}\,,&& \text{with} &&\sigma\in S_{N-1}\,,\label{DefMobiusIntegralc}
\end{align}
where the points $\{z_1,\ldots,z_a\}$ are inserted on $B_1$ and the points $\{z_{a+1},\ldots,z_N\}$ on $B_2$ and (with $\tau_M=2il-\tfrac{1}{2}$):
\begin{align}\label{RegionsMobc}
{\cal I}^a_{\sigma}&=\lf\{\ \bigcup_{i=1}^{a}x_i\in E_{\tau_M}^{a}\ |\ 0\leq x_1\leq x_{\sigma(2)} \leq\ldots\leq x_{\sigma(a)} \leq 1\ \ri\}\ ,\ \sigma\in S_{N_1-1}\ ,
\nonumber\\
{\cal I}^{\prime a}_{\sigma}&=\,\lf\{\ \bigcup_{i=a+1}^{N}x_i\in E_{\tau_M}^{N-a}\ |\ 0\leq x_{a+1}\leq x_{\sigma(a+2)} \leq\ldots\leq x_{\sigma'(N)} \leq 1\ \ri\}\ ,\ \sigma'\in S_{N_2-1}\ .
\end{align}
Finally, we would like to remark that for the amplitudes $M^{(1)}(1,\ldots,N)$ (in the open as well as in the closed string channel) the distinction between 'planar' and 'non-planar' -- as in the case of cylinder amplitudes -- does not apply: indeed, the M\"obius strip only has a single boundary (which is split into two components $B_{1,2}$ in the above representation), and the amplitude $M^{(1)}(1,\ldots,N)$ is a summation over all distributions (subject to a cyclic ordering) of the vertex insertion points over these two components.



\subsection{Contour integrals}

For simplicity we consider the case of a four-point amplitude on the M\"obius strip (with a fixed cyclic ordering $z_1<z_2<z_3<z_4$), which shall highlight all novel features and problems related to the topology of the world-sheet. 
\begin{align}
M^{(1)}(i_1,i_2,i_3,i_4)&=\delta(k_1+k_2+k_3+k_4)\int dl\ V^{-1}_{\text{CKG}} \int_{\mathcal{I}_1}\prod_{a=1}^4 dz_{i_a}\prod_{a<b}\left(\frac{\theta_1(z_{i_bi_a},2il-\tfrac{1}{2})}{\theta'_1(0,2il-\tfrac{1}{2})}\right)^{s_{ij}} +\text{hc.}.\label{4PtMobiusPt1}
\end{align}
Since the first (holomorphic) part of this $4$-point amplitude looks very similar to its counter-part (\ref{ClosedStringChannelAmp}) in the case of a cylinder world-sheet, we can employ the same strategy to find monodromy relations: we can integrate one of the vertex insertion points over a closed curve on the M\"obius strip in such a way as to avoid the branch cuts when two insertions collide. However, due to the particular topology of the (non-orientable) world-sheet, there are several issues that warrant particular attention. 

The first such problem concerns the distribution of the insertions points over the two components $B_{1,2}$. While in the cylinder case it was possible to confine all insertion points to a single boundary (a configuration that we called the planar case), it is no longer possible to constrain all insertion points to e.g. $B_1$ on the M\"obius strip. Instead, we have to consider all possible distributions of the insertion points over $B_1$ and $B_2$ that are compatible with the cyclic ordering under consideration. In order to keep the discussion tractable, we shall work in the gauge $z_4=4il-\tfrac{1}{2}$, in which case there are three different such contributions that need to be considered, which are schematically shown in the following series of figures 
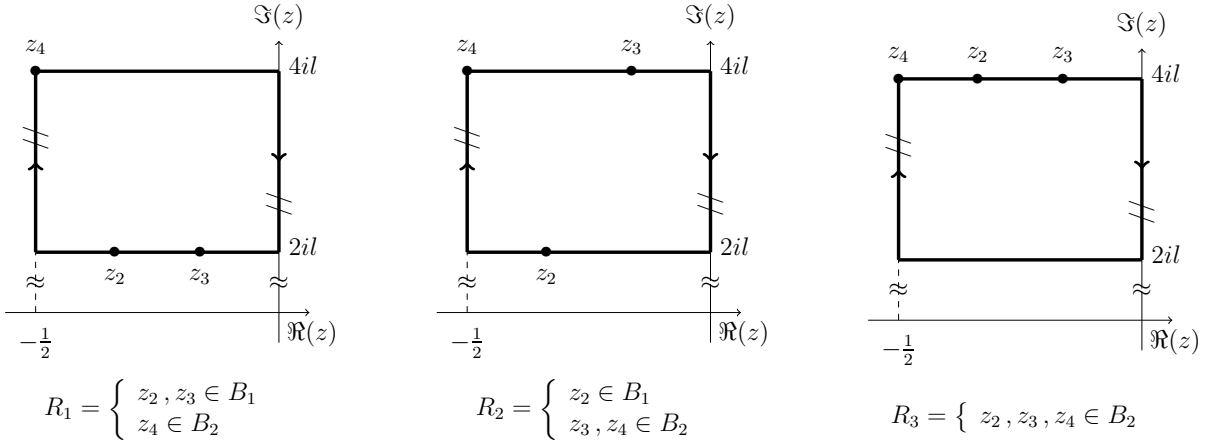
\begin{center}
\scalebox{0.8}{\parbox{5.5cm}{\begin{tikzpicture}[decoration={
    markings,
    mark=at position 0.5 with {\arrow{>}}}
    ] 
\draw (0,-0.5) -- (0,0.475);
\node at (0,0.5) {$\approx$};
\draw[->] (0,0.525) -- (0,4.5);
\draw[->] (-4.5,0) -- (0.5,0);
\node at (0.55,-0.35) {$\Re (z)$};
\node at (0,4.9) {$\Im (z)$};
\draw[ultra thick] (-4,1) -- (0,1);
\draw[ultra thick] (0,4) -- (-4,4);
\draw[ultra thick,postaction={decorate}]  (0,4) -- (0,1);
\draw[ultra thick,postaction={decorate}] (-4,1) -- (-4,4);
\node at (0.4,1.1) {$2il$};
\node at (0.4,4.1) {$4il$};
\node at (-4,-0.5) {$-\tfrac{1}{2}$};
\draw[dashed] (-4,-0) -- (-4,0.475);
\node at (-4,0.5) {$\approx$};
\draw[dashed] (-4,0.525) -- (-4,1);
\node[rotate=90] at (0,1.8) {$\slash\slash$};
\node[rotate=90] at (-4,2.9) {$\slash\slash$};
\node at (-2.7,1) {$\bullet$};
\node at (-2.7,0.6) {$z_2$};
\node at (-1.3,1) {$\bullet$};
\node at (-1.3,0.6) {$z_3$};
\node at (-4,4) {$\bullet$};
\node at (-4,4.4) {$z_4$};
\node at (-2,-1.6) {$R_1=\left\{\begin{array}{l}z_2\,,z_3\in B_1 \\ z_4\in B_2\end{array}\right.$};
\end{tikzpicture}}}
\hspace{1cm}
\scalebox{0.8}{\parbox{5.5cm}{\begin{tikzpicture}[decoration={
    markings,
    mark=at position 0.5 with {\arrow{>}}}
    ] 
\draw (0,-0.5) -- (0,0.475);
\node at (0,0.5) {$\approx$};
\draw[->] (0,0.525) -- (0,4.5);
\draw[->] (-4.5,0) -- (0.5,0);
\node at (0.55,-0.35) {$\Re (z)$};
\node at (0,4.9) {$\Im (z)$};
\draw[ultra thick] (-4,1) -- (0,1);
\draw[ultra thick] (0,4) -- (-4,4);
\draw[ultra thick,postaction={decorate}]  (0,4) -- (0,1);
\draw[ultra thick,postaction={decorate}] (-4,1) -- (-4,4);
\node at (0.4,1.1) {$2il$};
\node at (0.4,4.1) {$4il$};
\node at (-4,-0.5) {$-\tfrac{1}{2}$};
\draw[dashed] (-4,-0) -- (-4,0.475);
\node at (-4,0.5) {$\approx$};
\draw[dashed] (-4,0.525) -- (-4,1);
\node[rotate=90] at (0,1.8) {$\slash\slash$};
\node[rotate=90] at (-4,2.9) {$\slash\slash$};
\node at (-2.7,1) {$\bullet$};
\node at (-2.7,0.6) {$z_2$};
\node at (-1.3,4) {$\bullet$};
\node at (-1.3,4.4) {$z_3$};
\node at (-4,4) {$\bullet$};
\node at (-4,4.4) {$z_4$};
\node at (-2,-1.6) {$R_2=\left\{\begin{array}{l}z_2\in B_1 \\ z_3\,,z_4\in B_2\end{array}\right.$};
\end{tikzpicture}}}
\hspace{1cm}
\scalebox{0.8}{\parbox{5.5cm}{\begin{tikzpicture}[decoration={
    markings,
    mark=at position 0.5 with {\arrow{>}}}
    ] 
\draw (0,-0.5) -- (0,0.475);
\node at (0,0.5) {$\approx$};
\draw[->] (0,0.525) -- (0,4.5);
\draw[->] (-4.5,0) -- (0.5,0);
\node at (0.55,-0.35) {$\Re (z)$};
\node at (0,4.9) {$\Im (z)$};
\draw[ultra thick] (-4,1) -- (0,1);
\draw[ultra thick] (0,4) -- (-4,4);
\draw[ultra thick,postaction={decorate}]  (0,4) -- (0,1);
\draw[ultra thick,postaction={decorate}] (-4,1) -- (-4,4);
\node at (0.4,1.1) {$2il$};
\node at (0.4,4.1) {$4il$};
\node at (-4,-0.5) {$-\tfrac{1}{2}$};
\draw[dashed] (-4,-0) -- (-4,0.475);
\node at (-4,0.5) {$\approx$};
\draw[dashed] (-4,0.525) -- (-4,1);
\node[rotate=90] at (0,1.8) {$\slash\slash$};
\node[rotate=90] at (-4,2.9) {$\slash\slash$};
\node at (-2.7,4) {$\bullet$};
\node at (-2.7,4.4) {$z_2$};
\node at (-1.3,4) {$\bullet$};
\node at (-1.3,4.4) {$z_3$};
\node at (-4,4) {$\bullet$};
\node at (-4,4.4) {$z_4$};
\node at (-2,-1.6) {$R_3=\left\{\begin{array}{l} z_2\,,z_3\,,z_4\in B_2\end{array}\right.$};
\end{tikzpicture}}}
\end{center}
For $z_1<z_2$, the 4-point amplitude on the M\"obius strip is the sum over all three contributions since in this case
\begin{align}
\mathcal{I}_1=\underbrace{\left(\mathcal{I}^3_1\cup \mathcal{I}^{\prime 3}_1\right)}_{=R_1}\cup \underbrace{\left(\mathcal{I}^2_1\cup \mathcal{I}^{\prime 2}_1\right)}_{=R_2}\cup \underbrace{\left(\mathcal{I}^1_1\cup \mathcal{I}^{\prime 1}_1\right)}_{=R_3}\,,
\end{align}
where we used the notation (\ref{DefMobiusIntegralc}) and (\ref{RegionsMobc}).
The next step is to consider an integral of the point $z_1$ over a closed curve on the M\"obius strip that avoids the branch points when $z_1$ collides with $z_{2,3,4}$. To this end (as in the cylinder case) we consider a series of local contour integrals over regions in which the integrand of $M^{(1)}(1,2,3,4)$ is well defined. A set of local integrals in the three regions introduced above is shown in the following figures
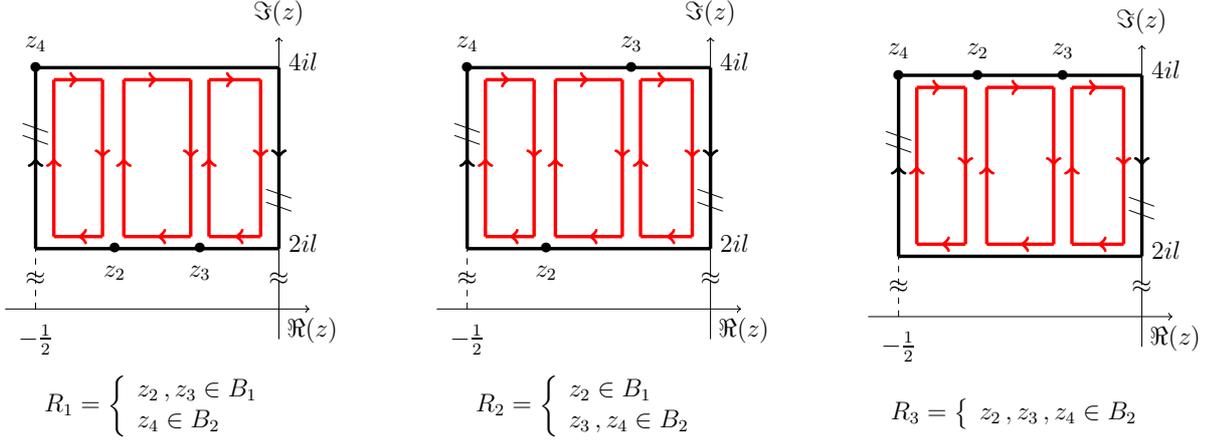
\begin{center}
\scalebox{0.8}{\parbox{5.5cm}{\begin{tikzpicture}[decoration={
    markings,
    mark=at position 0.5 with {\arrow{>}}}
    ] 
\draw (0,-0.5) -- (0,0.475);
\node at (0,0.5) {$\approx$};
\draw[->] (0,0.525) -- (0,4.5);
\draw[->] (-4.5,0) -- (0.5,0);
\node at (0.55,-0.35) {$\Re (z)$};
\node at (0,4.9) {$\Im (z)$};
\draw[ultra thick] (-4,1) -- (0,1);
\draw[ultra thick] (0,4) -- (-4,4);
\draw[ultra thick,postaction={decorate}]  (0,4) -- (0,1);
\draw[ultra thick,postaction={decorate}] (-4,1) -- (-4,4);
\node at (0.4,1.1) {$2il$};
\node at (0.4,4.1) {$4il$};
\node at (-4,-0.5) {$-\tfrac{1}{2}$};
\draw[dashed] (-4,-0) -- (-4,0.475);
\node at (-4,0.5) {$\approx$};
\draw[dashed] (-4,0.525) -- (-4,1);
\node[rotate=90] at (0,1.8) {$\slash\slash$};
\node[rotate=90] at (-4,2.9) {$\slash\slash$};
\node at (-2.7,1) {$\bullet$};
\node at (-2.7,0.6) {$z_2$};
\node at (-1.3,1) {$\bullet$};
\node at (-1.3,0.6) {$z_3$};
\node at (-4,4) {$\bullet$};
\node at (-4,4.4) {$z_4$};
\node at (-2,-1.6) {$R_1=\left\{\begin{array}{l}z_2\,,z_3\in B_1 \\ z_4\in B_2\end{array}\right.$};
\draw[red, ultra thick,postaction={decorate}] (-2.9,1.2) -- (-3.7,1.2); 
\draw[red, ultra thick,postaction={decorate}] (-3.7,1.2) -- (-3.7,3.8);
\draw[red, ultra thick,postaction={decorate}] (-3.7,3.8) -- (-2.9,3.8); 
\draw[red, ultra thick,postaction={decorate}] (-2.9,3.8) -- (-2.9,1.2); 
\draw[red, ultra thick,postaction={decorate}] (-2.55,1.2) -- (-2.55,3.8); 
\draw[red, ultra thick,postaction={decorate}] (-2.55,3.8) -- (-1.45,3.8); 
\draw[red, ultra thick,postaction={decorate}] (-1.45,3.8) -- (-1.45,1.2); 
\draw[red, ultra thick,postaction={decorate}] (-1.45,1.2) -- (-2.55,1.2); 
\draw[red, ultra thick,postaction={decorate}] (-1.15,1.2) -- (-1.15,3.8); 
\draw[red, ultra thick,postaction={decorate}] (-1.15,3.8) -- (-0.3,3.8); 
\draw[red, ultra thick,postaction={decorate}] (-0.3,3.8) -- (-0.3,1.2); 
\draw[red, ultra thick,postaction={decorate}] (-0.3,1.2) -- (-1.15,1.2); 
\end{tikzpicture}}}
\hspace{1cm}
\scalebox{0.8}{\parbox{5.5cm}{\begin{tikzpicture}[decoration={
    markings,
    mark=at position 0.5 with {\arrow{>}}}
    ] 
\draw (0,-0.5) -- (0,0.475);
\node at (0,0.5) {$\approx$};
\draw[->] (0,0.525) -- (0,4.5);
\draw[->] (-4.5,0) -- (0.5,0);
\node at (0.55,-0.35) {$\Re (z)$};
\node at (0,4.9) {$\Im (z)$};
\draw[ultra thick] (-4,1) -- (0,1);
\draw[ultra thick] (0,4) -- (-4,4);
\draw[ultra thick,postaction={decorate}]  (0,4) -- (0,1);
\draw[ultra thick,postaction={decorate}] (-4,1) -- (-4,4);
\node at (0.4,1.1) {$2il$};
\node at (0.4,4.1) {$4il$};
\node at (-4,-0.5) {$-\tfrac{1}{2}$};
\draw[dashed] (-4,-0) -- (-4,0.475);
\node at (-4,0.5) {$\approx$};
\draw[dashed] (-4,0.525) -- (-4,1);
\node[rotate=90] at (0,1.8) {$\slash\slash$};
\node[rotate=90] at (-4,2.9) {$\slash\slash$};
\node at (-2.7,1) {$\bullet$};
\node at (-2.7,0.6) {$z_2$};
\node at (-1.3,4) {$\bullet$};
\node at (-1.3,4.4) {$z_3$};
\node at (-4,4) {$\bullet$};
\node at (-4,4.4) {$z_4$};
\node at (-2,-1.6) {$R_2=\left\{\begin{array}{l}z_2\in B_1 \\ z_3\,,z_4\in B_2\end{array}\right.$};
\draw[red, ultra thick,postaction={decorate}] (-2.9,1.2) -- (-3.7,1.2); 
\draw[red, ultra thick,postaction={decorate}] (-3.7,1.2) -- (-3.7,3.8);
\draw[red, ultra thick,postaction={decorate}] (-3.7,3.8) -- (-2.9,3.8); 
\draw[red, ultra thick,postaction={decorate}] (-2.9,3.8) -- (-2.9,1.2); 
\draw[red, ultra thick,postaction={decorate}] (-2.55,1.2) -- (-2.55,3.8); 
\draw[red, ultra thick,postaction={decorate}] (-2.55,3.8) -- (-1.45,3.8); 
\draw[red, ultra thick,postaction={decorate}] (-1.45,3.8) -- (-1.45,1.2); 
\draw[red, ultra thick,postaction={decorate}] (-1.45,1.2) -- (-2.55,1.2); 
\draw[red, ultra thick,postaction={decorate}] (-1.15,1.2) -- (-1.15,3.8); 
\draw[red, ultra thick,postaction={decorate}] (-1.15,3.8) -- (-0.3,3.8); 
\draw[red, ultra thick,postaction={decorate}] (-0.3,3.8) -- (-0.3,1.2); 
\draw[red, ultra thick,postaction={decorate}] (-0.3,1.2) -- (-1.15,1.2); 
\end{tikzpicture}}}
\hspace{1cm}
\scalebox{0.8}{\parbox{5.5cm}{\begin{tikzpicture}[decoration={
    markings,
    mark=at position 0.5 with {\arrow{>}}}
    ] 
\draw (0,-0.5) -- (0,0.475);
\node at (0,0.5) {$\approx$};
\draw[->] (0,0.525) -- (0,4.5);
\draw[->] (-4.5,0) -- (0.5,0);
\node at (0.55,-0.35) {$\Re (z)$};
\node at (0,4.9) {$\Im (z)$};
\draw[ultra thick] (-4,1) -- (0,1);
\draw[ultra thick] (0,4) -- (-4,4);
\draw[ultra thick,postaction={decorate}]  (0,4) -- (0,1);
\draw[ultra thick,postaction={decorate}] (-4,1) -- (-4,4);
\node at (0.4,1.1) {$2il$};
\node at (0.4,4.1) {$4il$};
\node at (-4,-0.5) {$-\tfrac{1}{2}$};
\draw[dashed] (-4,-0) -- (-4,0.475);
\node at (-4,0.5) {$\approx$};
\draw[dashed] (-4,0.525) -- (-4,1);
\node[rotate=90] at (0,1.8) {$\slash\slash$};
\node[rotate=90] at (-4,2.9) {$\slash\slash$};
\node at (-2.7,4) {$\bullet$};
\node at (-2.7,4.4) {$z_2$};
\node at (-1.3,4) {$\bullet$};
\node at (-1.3,4.4) {$z_3$};
\node at (-4,4) {$\bullet$};
\node at (-4,4.4) {$z_4$};
\node at (-2,-1.6) {$R_3=\left\{\begin{array}{l} z_2\,,z_3\,,z_4\in B_2\end{array}\right.$};
\draw[red, ultra thick,postaction={decorate}] (-2.9,1.2) -- (-3.7,1.2); 
\draw[red, ultra thick,postaction={decorate}] (-3.7,1.2) -- (-3.7,3.8);
\draw[red, ultra thick,postaction={decorate}] (-3.7,3.8) -- (-2.9,3.8); 
\draw[red, ultra thick,postaction={decorate}] (-2.9,3.8) -- (-2.9,1.2); 
\draw[red, ultra thick,postaction={decorate}] (-2.55,1.2) -- (-2.55,3.8); 
\draw[red, ultra thick,postaction={decorate}] (-2.55,3.8) -- (-1.45,3.8); 
\draw[red, ultra thick,postaction={decorate}] (-1.45,3.8) -- (-1.45,1.2); 
\draw[red, ultra thick,postaction={decorate}] (-1.45,1.2) -- (-2.55,1.2); 
\draw[red, ultra thick,postaction={decorate}] (-1.15,1.2) -- (-1.15,3.8); 
\draw[red, ultra thick,postaction={decorate}] (-1.15,3.8) -- (-0.3,3.8); 
\draw[red, ultra thick,postaction={decorate}] (-0.3,3.8) -- (-0.3,1.2); 
\draw[red, ultra thick,postaction={decorate}] (-0.3,1.2) -- (-1.15,1.2); 
\end{tikzpicture}}}
\end{center}

\noindent
Each of the local integrals $I_{R_a}$ (for $a=1,\ldots,4$) is well defined in the sense that we can locally define a normal vector to the (oriented) surface that is encircled. Moreover, since (by construction) the contours do not incircle any poles, they vanish identically, giving rise to relations between the various contributions. 

Regarding the horizontal contributions, they correspond to (parts) of different color orderings of the four--point amplitude (\ref{4PtMobiusPt1}) for different arguments $(i_1,i_2,i_3,i_4)$. 

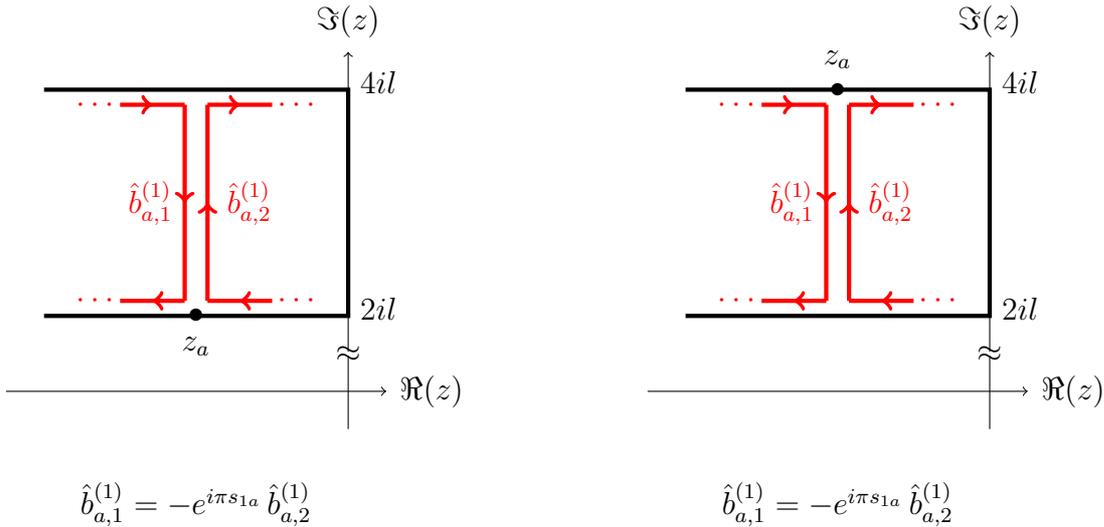
\begin{center}
\begin{tikzpicture}[decoration={
    markings,
    mark=at position 0.5 with {\arrow{>}}}
    ] 
\draw (0,-0.5) -- (0,0.475);
\node at (0,0.5) {$\approx$};
\draw[->] (0,0.525) -- (0,4.5);
\draw[->] (-4.5,0) -- (0.5,0);
\node at (1.1,0) {$\Re (z)$};
\node at (0,4.9) {$\Im (z)$};
\draw[ultra thick] (-4,1) -- (0,1) -- (0,4) -- (-4,4); 
\node at (0.4,1.1) {$2il$};
\node at (0.4,4.1) {$4il$};
\node at (-2,1) {$\bullet$};
\node at (-2,0.6) {$z_a$};
\draw[red, ultra thick,postaction={decorate}] (-2.15,1.2) -- (-3,1.2); 
\node[red] at (-3.3,1.2) {$\cdots$};
\draw[red, ultra thick,postaction={decorate}] (-2.15,3.8) -- (-2.15,1.2); 
\draw[red, ultra thick,postaction={decorate}] (-3,3.8) -- (-2.15,3.8); 
\node[red] at (-3.3,3.8) {$\cdots$};
\node[red] at (-2.6,2.5) {$\hat{b}^{(1)}_{a,1}$};
\draw[red, ultra thick,postaction={decorate}] (-1.85,3.8) -- (-1,3.8); 
\node[red] at (-0.65,3.8) {$\cdots$};
\draw[red, ultra thick,postaction={decorate}] (-1.85,1.2) -- (-1.85,3.8); 
\draw[red, ultra thick,postaction={decorate}] (-1,1.2) -- (-1.85,1.2); 
\node[red] at (-0.65,1.2) {$\cdots$};
\node[red] at (-1.3,2.5) {$\hat{b}^{(1)}_{a,2}$};
\node at (-2,-1.5) {$\hat{b}^{(1)}_{a,1}=-e^{i\pi s_{1a}}\,\hat{b}^{(1)}_{a,2}$};
\end{tikzpicture}
\hspace{2cm}
\begin{tikzpicture}[decoration={
    markings,
    mark=at position 0.5 with {\arrow{>}}}
    ] 
\draw (0,-0.5) -- (0,0.475);
\node at (0,0.5) {$\approx$};
\draw[->] (0,0.525) -- (0,4.5);
\draw[->] (-4.5,0) -- (0.5,0);
\node at (1.1,0) {$\Re (z)$};
\node at (0,4.9) {$\Im (z)$};
\draw[ultra thick] (-4,1) -- (0,1) -- (0,4) -- (-4,4); 
\node at (0.4,1.1) {$2il$};
\node at (0.4,4.1) {$4il$};
\node at (-2,4) {$\bullet$};
\node at (-2,4.4) {$z_a$};
\draw[red, ultra thick,postaction={decorate}] (-2.15,1.2) -- (-3,1.2); 
\node[red] at (-3.3,1.2) {$\cdots$};
\draw[red, ultra thick,postaction={decorate}] (-2.15,3.8) -- (-2.15,1.2); 
\draw[red, ultra thick,postaction={decorate}] (-3,3.8) -- (-2.15,3.8); 
\node[red] at (-3.3,3.8) {$\cdots$};
\node[red] at (-2.6,2.5) {$\hat{b}^{(1)}_{a,1}$};
\draw[red, ultra thick,postaction={decorate}] (-1.85,3.8) -- (-1,3.8); 
\node[red] at (-0.65,3.8) {$\cdots$};
\draw[red, ultra thick,postaction={decorate}] (-1.85,1.2) -- (-1.85,3.8); 
\draw[red, ultra thick,postaction={decorate}] (-1,1.2) -- (-1.85,1.2); 
\node[red] at (-0.65,1.2) {$\cdots$};
\node[red] at (-1.3,2.5) {$\hat{b}^{(1)}_{a,2}$};
\node at (-2,-1.5) {$\hat{b}^{(1)}_{a,1}=-e^{i\pi s_{1a}}\,\hat{b}^{(1)}_{a,2}$};
\end{tikzpicture}
\end{center}

\noindent
Concerning the vertical contributions, we need to combine the contour integrals $R_{1,2,3}$ in such a way that the latter mutually cancel. 
In doing so, we use the fact that two such bulk integrals are related to each other by a phase factor that depends on the insertion points, i.e. for a generic contribution (with $a=2,3$) which follows in the same fashion as in the case of the cylinder amplitudes (taking into account that $z_1$ and $z_a$ are always inserted on the same boundary). The main difference, however, compared to the latter is the fact that an additional minus sign is present for the case $a=4$ due to the topology of the M\"obius strip: $\hat{b}_{4,1}^{(1)}=e^{i\pi s_{14}}\,\hat{b}^{(1)}_{4,2}$, leading to further complications for the monodromy relations: in particular, due to this minus sign not all bulk contributions $\hat{b}^{(1)}_{a,1}$ and $\hat{b}^{(1)}_{a,2}$ cancel each other (much in the same way as for the non-planar cylinder amplitudes). The physical interpretation of these terms is not clear to as at the moment. Because of this (and owing to the complexity of the final monodromy relation) we refrain from giving further details of this relation in the present paper.


\section[Monodromy relations and amplitudes in non--commutative background]{Monodromy relations and amplitudes in non--com--mutative background}\label{Sect:NonCommutative}

In this section we argue that the new objects (\ref{F4Integrand}) (that we found in the planar monodromy relation (\ref{FourPointMonodromyRelation})) have an interpretation in terms of non-planar four-point functions with a non-trivial (but constant) gauge-field flux switched on.  
Indeed, such correlation functions correspond to amplitudes computing the open--string exchange between two parallel Dp--branes. The general form for an $M$--point amplitude with $N\leq M$ vertex insertions on one boundary (and $M-N$ on the other) in the open string channel was given \emph{e.g.} in \cite{Liu:2000ad}:
{\allowdisplaybreaks\begin{align}
&A^{(1),\text{brane}}(k_i|k_m)=-i\ \sqrt{\text{det}\,G}\,\frac{g_o^4}{4\alpha'^2}\,(2\alpha')^4\ (2\pi)^{p+1}\ \delta^{(p+1)}\ \left(\sum_{q=1}^Mk_q\right)\,\mathcal{K}\,\int_0^\infty\frac{dt}{2t}\,(8\pi^2\alpha't)^{-\frac{p+1}{2}}\nonumber\\
&\times \,\exp\left\{\frac{k_\mu(\Theta G\Theta)^{\mu\nu}k_\nu}{8\pi\alpha' t}\right\}\left(\prod_{q=1}^M\int_0^{2\pi t}dx_q\right)\ \prod_{1\leq i<j\leq N}\left|2\pi i\,e^{-\frac{x_{ij}^2}{4\pi t}}\,\frac{\theta_1\left(\frac{ix_{ij}}{2\pi},it\right)}{\theta'_1(0,it)}\right|^{2\alpha' k_{i}\cdot k_{j}}\nonumber\\
&\times\prod_{N+1\leq m<n\leq M}\left|2\pi i\,e^{-\frac{x_{mn}^2}{4\pi t}}\,\frac{\theta_1\left(\frac{ix_{mn}}{2\pi},it\right)}{\theta'_1(0,it)}\right|^{2\alpha' k_{m}\cdot k_{n}}\ \prod_{{i=1,\ldots,N}\atop{m=N+1,\ldots, M}}\left|2\pi\,e^{-\frac{x_{mi}^2}{4\pi t}}\,\frac{\theta_2\left(\frac{ix_{mi}}{2\pi},it\right)}{\theta'_1(0,it)}\right|^{2\alpha' k_{m}\cdot k_{i}}\nonumber\\
&\times \exp\left\{-\frac{i}{2}\sum_{1\leq i<j\leq N}(k_i\times k_j)\left[\frac{x_{ij}}{\pi t}-\text{sign}(x_{ij})\right]+\frac{i}{2}\sum_{N+1\leq m<n\leq M}(k_m\times k_n)\left[\frac{x_{mn}}{\pi t}-\text{sign}(x_{mn})\right]\right\}.\label{4PointNoncomm}
\end{align}}
Here the indices $i,j$ run over all open string insertions on the first boundary (\emph{i.e.} from $1,\ldots,N$), while the indices $m,n$ run over all insertions on the second boundary (\emph{i.e.} from $N+1,\ldots,M$). Furthermore, we have introduced the vector and star products
\begin{align}
&k_i\cdot k_j:=k_{i\mu} G^{\mu\nu} k_{j\nu}\,&&\text{and}&&k_i\times k_j:=k_{i\mu}\Theta^{\mu\nu}k_{j\nu}\ ,
\end{align}
respectively. All kinematical factors are comprised into the factor ${\cal K}$, which will not play any role in the following.
The open string coupling $g_o$ is related to the YM coupling $g$ as $g_o=(2\ap)^{1/2}g$, while the open string metric $G^{\mu\nu}$ and the anti--symmetric matrix $\Theta^{\mu\nu}$ are related to the closed string metric $g^{\mu\nu}$ and the anti--symmetric tensor $B^{\mu\nu}$ as:
\begin{align}
&G^{\mu\nu}=\left(\frac{1}{g+B}\,g\,\frac{1}{g-B}\right)^{\mu\nu}\,,&&\text{and:} &&\Theta^{\mu\nu}=(-2\pi \ap)\left(\frac{1}{g+B}\,B\,\frac{1}{g-B}\right)^{\mu\nu}\ .
\end{align}
Finally, we have used the notation for the non--planar momentum:
\begin{align}
k_\mu=\sum_{i=1}^Nk_{i\mu}=-\sum_{m=N+1}^Mk_{r\mu}\ .
\end{align}
In order to compare (\ref{4PointNoncomm}) to  (\ref{F4Integrand}), we first have to change the former to the dual closed string channel. To this end, we introduce $l =1/t$
{\allowdisplaybreaks
\begin{align}
&A^{(1),\text{brane}}(k_i|k_m)=-i\sqrt{\text{det}\,G}\,\frac{g_o^4}{4\alpha'^2}\ (2\alpha')^4\ (2\pi)^{p+1}\ \delta^{(p+1)}\left(\sum_{q=1}^Mk_q\right)\ \mathcal{K}\ \int_0^\infty dl\ \fc{dl}{2l}\ \lf(\fc{8\pi^2\ap}{l}\ri)^{-\fc{p+1}{2}}\nonumber\\
&\hskip1cm\times \,\exp\left\{\frac{k_\mu(\Theta G\Theta)^{\mu\nu}k_\nu\,l }{8\pi\alpha'}\right\}\left(\prod_{q=1}^M\int_0^{\frac{2\pi}{l } }dx_q\right)\ \prod_{1\leq i<j\leq N}\left|\frac{\theta_1\left(\frac{x_{ij}l }{2\pi},il \right)}{\theta'_1(0,il )}\right|^{2\alpha' k_{i}\cdot k_{j}}\nonumber\\
&\hskip1cm\times\prod_{N+1\leq m<n\leq M}\left|\frac{\theta_1\left(\frac{x_{mn}l }{2\pi},il \right)}{\theta'_1(0,il )}\right|^{2\alpha' k_{m}\cdot k_{n}}\ \prod_{{i=1,\ldots,N}\atop{m=N+1,\ldots,M}}\left|\frac{\theta_4\left(\frac{x_{mi}l }{2\pi},il \right)}{\theta'_1(0,il )}\right|^{2\alpha' k_{m}\cdot k_{i}}\nonumber\\
&\times \exp\left\{-\frac{i}{2}\sum_{1\leq i<j\leq N}(k_i\times k_j)\left[\frac{x_{ij}l }{\pi}-\text{sign}(x_{ij})\right]+\frac{i}{2}\sum_{N+1\leq m<n\leq M}(k_m\times k_n)\left[\frac{x_{mn}l }{\pi}-\text{sign}(x_{mn})\right]\right\},\label{4PointNoncommDual}
\end{align}}
where we have also used the on--shell condition: 
\begin{align}
\sum_{i<j}k_i\cdot k_j+\sum_{m<n}k_m\cdot k_n+\sum_{i,m}k_m\cdot k_i=0\,.
\end{align}
Finally, for all $x_i,x_m$ we perform the change of coordinates
\begin{align}
&x_i\to \frac{2\pi x_i}{l }\,,&&x_m\to\frac{2\pi x_m}{l }\ ,
\end{align}
to find:
\begin{align}
&A^{(1),\text{brane}}(k_i|k_m)=-i\sqrt{\text{det}\,G}\,\frac{g_o^4}{4\alpha'^2}\,(2\alpha')^4\ (2\pi)^{p-3}\ \delta^{(p+1)}\left(\sum_{q=1}^Mk_q\right)\ \mathcal{K}\ 
\int_0^\infty\frac{dl }{2}\,l^{-1-M}\,\left(\frac{8\pi^2\alpha'}{l }\right)^{-\frac{p+1}{2}}\nonumber\\
&\hskip1cm\times \exp\lf\{\frac{k_\mu(\Theta G\Theta)^{\mu\nu}k_\nu\,l }{8\pi\alpha'}\right\} \ \left(\prod_{q=1}^M\int_0^{1}dx_q\right)\ \prod_{i<j}\left|\frac{\theta_1\left(x_{ij},il \right)}{\theta'_1(0,il )}\right|^{2\alpha' k_{i}\cdot k_{j}} \prod_{m<n}\left|\frac{\theta_1\left(x_{mn},il \right)}{\theta'_1(0,il )}\right|^{2\alpha' k_{m}\cdot k_{n}} \nonumber\\
&\hskip1cm\times\prod_{i,m}\left|\frac{\theta_4\left(x_{mi},il \right)}{\theta'_1(0,il )}\right|^{2\alpha' k_{m}\cdot k_{i}}  \exp\left\{-i\sum_{i<j}(k_i\times k_j)\left[x_{ij}-\frac{\text{sign}(x_{ij})}{2}\right]\right\}\nonumber\\
&\hskip1cm\times \exp\left\{i\sum_{m<n}(k_m\times k_n)\left[x_{mn}-\frac{\text{sign}(x_{mn})}{2}\right]\right\}\ .\label{4PointNoncommDualCords}
\end{align}
We now specify (\ref{4PointNoncommDualCords}) to the case with three insertions on the lower boundary ($N=3$) and a single one on the upper boundary ($M=4$), such that $i=2,3,4$ and $m=1$. The insertions on the lower boundary are ordered, \emph{i.e.} $x_2<x_3<x_4$
\begin{align}
&A^{(1),\text{brane}}(2,3,4|1)=-i\sqrt{\text{det}\,G}\,\frac{g_o^4}{4\alpha'^2}\,(2\alpha')^4(2\pi)^{p-3}\,\delta^{(p+1)}\left(\sum_{q=1}^4k_q\right)\,\mathcal{K}\,\int_0^\infty\frac{dl }{2}\,l^{-5}\nonumber\\
&\times \left(\frac{8\pi^2\alpha'}{l }\right)^{-\frac{p+1}{2}}\ \exp\left\{\frac{k_\mu(\Theta G\Theta)^{\mu\nu}k_\nu\,l }{8\pi\alpha'}\right\}\  \left(\prod_{q=1}^4\int_0^{1}dx_q\right)\ \prod_{2\leq i<j\leq 4}\left|\frac{\theta_1\left(x_{ij},il \right)}{\theta'_1(0,il )}\right|^{2\alpha' k_{i}\cdot k_{j}}\nonumber\\
&\times\prod_{i=2}^4\left|\frac{\theta_4\left(x_{1i},il \right)}{\theta'_1(0,il )}\right|^{2\alpha' k_{1}\cdot k_{i}}\ \text{exp}\left\{-i\sum_{2\leq i<j\leq 4}(k_i\times k_j)\left[x_{ij}-\frac{1}{2}\right]\right\}\ .\label{4PointNoncommSpec}
\end{align} 
Thus, in order to compare to (\ref{F4Integrand}), with ${\cal K}\equiv P_4$ and $D=p+1=10$ we particularly have to match the phase factors
\begin{align}
-i\pi\sum_{i=2}^4s_{1i}\ x_{i}\stackrel{!}{=}-i\sum_{2\leq i<j\leq 4}(k_i\times k_j)\ x_{ij}\ ,
\end{align}
which becomes:
\begin{align}
2\pi\alpha'\sum_{i=1}^4(k_1\cdot k_i)\ x_i=\sum_{2\leq i<j\leq 4}(k_i\times k_j)\ x_{ij}\ .\label{RelPhase1}
\end{align}
Writing furthermore
\begin{align}
\sum_{i=2}^4\sum_{j=2}^4k_{i\mu}\Theta^{\mu\nu}k_{j\nu}(x_i-x_j)&=\left[\sum_{2\leq i<j\leq 4}+\sum_{2\leq j<i\leq 4}\right]k_{i\mu}\Theta^{\mu\nu}k_{j\nu}(x_i-x_j)\nonumber\\
&=2\sum_{2\leq i<j\leq 4}k_{i\mu}\Theta^{\mu\nu}k_{j\nu}(x_i-x_j)\,,\nonumber
\end{align}
shows that (\ref{RelPhase1}) becomes
\begin{align}
2\pi\alpha'\sum_{i=1}^4(k_1\cdot k_i)x_i&=\frac{1}{2}\sum_{i=2}^4\sum_{j=2}^4(k_i\times k_j)(x_i-x_j)=-\frac{1}{2}\sum_{i=2}^2k_{i\mu}\Theta^{\mu\nu}k_{1,\nu}x_i+\frac{1}{2}\sum_{j=2}^4k_{1\mu}\Theta^{\mu\nu}k_{j\nu}x_j\nonumber\\
&=-\sum_{i=2}^4k_{i\mu}\Theta^{\mu\nu}k_{1\nu}x_i\ .
\end{align}
where we used $\sum\limits_{i=2}^4k_i=-k_1$. Therefore we have: 
\begin{align}
0&=2\pi\alpha'\sum_{j=2}^4 k_{1\mu}G^{\mu\nu}k_{j\nu}x_j+\sum_{i=2}^4k_{i\mu}\Theta^{\mu\nu}k_{1\nu}x_i=\sum_{j=2}^4k_{1\mu} \left(2\pi\alpha' G^{\mu\nu}-\Theta^{\mu\nu}\right) k_{j\nu}x_j\ .
\end{align}
This relation needs to hold for any $x_j$ and therefore we can split this into 3 separate equations
\begin{align}
&k_{1\mu}\left(2\pi\alpha' G^{\mu\nu}-\Theta^{\mu\nu}\right)k_{j\nu}=0\ \ \ ,\ \ \ j=2,3,4\,.
\end{align}
We propose to solve this relation by imposing
\begin{align}
k_{1\mu}\ (2\pi\alpha' G^{\mu\nu}-\Theta^{\mu\nu})=0\ ,\label{SingleEquNonCom}
\end{align}
which is a single condition for the anti-symmetric matrix $\Theta^{\mu\nu}$. To show that (\ref{SingleEquNonCom}) indeed has a solution, we set
\begin{align}
&k_{1\mu}=(k,k,0,\ldots,0)\,,&&G^{\mu\nu}=\text{diag}(-1,1,\ldots,1)\,,\label{ChoiceMom}
\end{align}
where the choice of $k_{1\mu}$ can always be achieved by Lorentz transformations. For (\ref{ChoiceMom}) the following anti-symmetric matrix is a solution to (\ref{SingleEquNonCom})
\begin{align}
\Theta^{\mu\nu}=2\pi\alpha'\left(\begin{array}{ccc} 0 & -1 & \\ 1 & 0 & \\ & & \vartheta^{mn}\end{array}\right)\,,\label{SolutionTheta}
\end{align}
where $\vartheta^{mn}$ is an arbitrary anti-symmetric $(p-2)\times (p-2)$ matrix. Indeed, since we have
\begin{align}
&k_{1\mu}G^{\mu\nu}=\left(\begin{array}{c}-k \\ k \\ 0 \\\vdots \\ 0\end{array}\right)\,,&&k_{1\mu}\Theta^{\mu\nu}=2\pi\alpha'\left(\begin{array}{c}-k \\ k \\ 0 \\\vdots \\ 0\end{array}\right)\,,
\end{align}
condition (\ref{SingleEquNonCom}) is indeed satisfied.
Furthermore, with $g^{\mu\nu}=G^{\mu\nu}-(2\pi\ap)^{-2}\ (\Theta G\Theta)^{\mu\nu}$ and the choices
\req{ChoiceMom} and \req{SolutionTheta} we find  $k_\mu g^{\mu\nu}k_\nu=0$ and
$k_\mu(\Theta G\Theta)^{\mu\nu}k_\nu=0$ (note $k=-k_1$). As a consequence
 in \req{4PointNoncommSpec} the exponential becomes:
\be
\exp\left\{\frac{k_\mu(\Theta G\Theta)^{\mu\nu}k_\nu\,l }{8\pi\alpha'}\right\}=1\ .
\ee 

This shows that the quantities $\tilde{A}^{(1)}_4(2,3,4|1)$ (and similar versions with permutations of all elements) appearing in (\ref{FourPointMonodromyRelation}) can be interpreted in terms of four--point amplitudes in the presence of non--trivial gauge--field flux of the form (\ref{4PointNoncomm}). However, since (\ref{SingleEquNonCom}) can be read as an eigenvalue equation for $M^{\mu\nu}=2\pi\alpha' G^{\mu\nu}-\Theta^{\mu\nu}$ with eigenvalue $0$, the choice (\ref{SolutionTheta}) means that $M^{\mu\nu}$ is degenerate and not invertible. A physical interpretation of this flux shall be given elsewhere.

\section{On monodromy relations for higher loop amplitudes}\label{Sect:HigherLoop}

As a generalization of the monodromy relations on the annulus, in this section we provide an outlook for similar relations on surfaces with more than two boundaries. We mostly focus on graphical representations of the monodromy relations based on particular contour integrals on Riemann surfaces with more than two boundary components (but no handles or cross-caps). Explicit expressions for different amplitude components and a more systematic study of the ensuing monodromy relations will be presented elsewhere. Our discussion of Riemann surfaces with multiple boundary components mostly follows \cite{Blau:1987pn,Bianchi:1989du,Antoniadis:2005sd}. 
\subsection[Riemann surfaces with $n\geq 2$ boundaries]{Riemann surfaces with $\bm{n\geq 2}$ boundaries}
In order to describe Riemann surfaces with $n\geq 2$ boundary components (and no handles or cross-caps), we begin with their double covers, \emph{i.e.} closed Riemann surface with genus $g=n-1$. Such a surface $\tilde{\Sigma}_{n-1}$ can be endowed with a basis of \emph{homology cycles}. In the following we show an example with genus $g=n-1=3$:
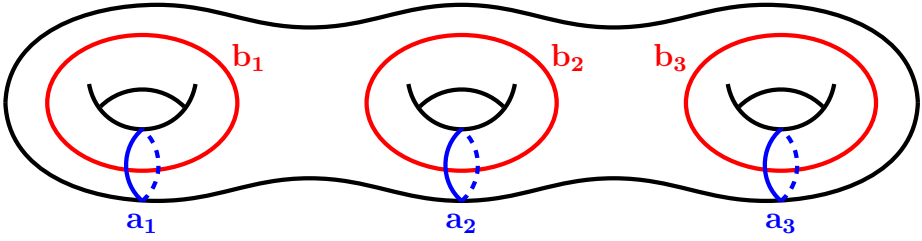
\begin{center}
\begin{tikzpicture}
\draw[ultra thick] (-6,0) to [out=88, in=180] (-4,1.3) to [out=0, in=180] (-2,1) to [out=0, in=180] (0,1.3) to [out=0, in=180] (2,1) to [out=0, in=180] (4,1.3) to [out=0, in=92] (6,0);
\draw[ultra thick] (-6,0) to [out=272, in=180] (-4,-1.3) to [out=0, in=180] (-2,-1) to [out=0, in=180] (0,-1.3) to [out=0, in=180] (2,-1) to [out=0, in=180] (4,-1.3) to [out=0, in=268] (6,0);
\draw[ultra thick] (-4.9,0.25) to [out=280, in=180] (-4.2,-0.35) to [out=0, in=260] (-3.5,0.25);
\draw[ultra thick] (-4.75,-0.05) to [out=45, in=135]  (-3.65,-0.05);
\draw[ultra thick] (3.5,0.25) to [out=280, in=180] (4.2,-0.35) to [out=0, in=260] (4.9,0.25) ;
\draw[ultra thick] (3.65,-0.05) to [out=45, in=135] (4.75,-0.05);
\draw[ultra thick] (-0.7,0.25) to [out=280, in=180] (0,-0.35) to [out=0, in=260] (0.7,0.25) ;
\draw[ultra thick] (-0.55,-0.05) to [out=45, in=135] (0.55,-0.05);
\draw[ultra thick,red] (0,0) ellipse (1.25cm and 0.9cm);
\draw[ultra thick,red] (-4.2,0) ellipse (1.25cm and 0.9cm);
\draw[ultra thick,red] (4.2,0) ellipse (1.25cm and 0.9cm);
\draw[ultra thick, blue] (0,-1.3) to [out=140, in=220] (0,-0.35);
\draw[ultra thick, dashed, blue] (0,-1.3) to [out=40, in=320] (0,-0.35);
\draw[ultra thick, blue] (-4.2,-1.3) to [out=140, in=220] (-4.2,-0.35);
\draw[ultra thick, dashed, blue] (-4.2,-1.3) to [out=40, in=320] (-4.2,-0.35);
\draw[ultra thick, blue] (4.2,-1.3) to [out=140, in=220] (4.2,-0.35);
\draw[ultra thick, dashed, blue] (4.2,-1.3) to [out=40, in=320] (4.2,-0.35);
\node[blue] at (-4.2,-1.6) {$\mathbf{a_1}$};
\node[blue] at (0,-1.6) {$\mathbf{a_2}$};
\node[blue] at (4.2,-1.6) {$\mathbf{a_3}$};
\node[red] at (-2.8,0.6) {$\mathbf{b_1}$};
\node[red] at (1.4,0.6) {$\mathbf{b_2}$};
\node[red] at (2.75,0.6) {$\mathbf{b_3}$};
\end{tikzpicture}
\end{center}
These circles are normalised such that
\begin{align}
&\int_{\mathbf{a}_i}\omega_{j}=\delta_{ij}\,,&&\int_{\mathbf{b}_i}\omega_{j}=\tau_{ij}\,,\label{PeriodMatrix}
\end{align} 
where $\omega_i$ are a basis of $n-1$ complex differentials on $\tilde{\Sigma}_{n-1}$ and $\tau_{ij}$ is the symmetric $(n-1)\times (n-1)$ complex period matrix. Concerning the $\mathbf{b}_i$ we can introduce the basis $\gamma_i$ with the property $\gamma_i\bar{\gamma}_i^{-1}=\sum_{k=1}^i\mathbf{b}_k$ which will turn out more useful later on.

In order to obtain a surface that has only boundary components, but no handles, we consider a ${\bf Z}_2$ involution which identifies the cycles in the following way
\begin{align}
&\left(\begin{array}{c}\mathbf{a}_i \\ \mathbf{b}_j\end{array}\right)=\mathcal{I}\cdot \left(\begin{array}{c}\mathbf{a}_i \\ \mathbf{b}_j\end{array}\right)\,,&&\text{with} &&\mathcal{I}=\left(\begin{array}{cc}1\!\!1 & 0 \\ 0 & -1\!\!1 \end{array}\right)\,.\label{WorldSheetInvolution}
\end{align}
In this way, we obtain a two-dimensional surface $\Sigma_n$ with $n$ distinct boundary components:
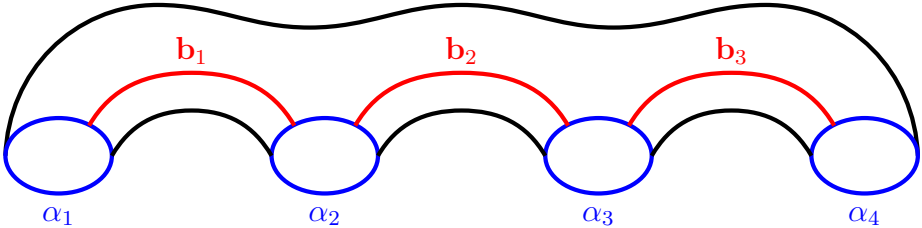
\begin{center}
\begin{tikzpicture}
\draw[ultra thick] (-6,0) to [out=88, in=180] (-4,2) to [out=0, in=180] (-2,1.7) to [out=0, in=180] (0,2) to [out=0, in=180] (2,1.7) to [out=0, in=180] (4,2) to [out=0, in=92] (6,0);
\draw[ultra thick, blue] (-5.3,0) ellipse (0.7cm and 0.5cm);
\draw[ultra thick, blue] (-1.8,0) ellipse (0.7cm and 0.5cm);
\draw[ultra thick, blue] (1.8,0) ellipse (0.7cm and 0.5cm);
\draw[ultra thick, blue] (5.3,0) ellipse (0.7cm and 0.5cm);
\draw[ultra thick] (-4.6,0) to [out=60, in=180] (-3.55,0.6) to [out=0, in=120] (-2.5,0);
\draw[ultra thick] (-1.1,0) to [out=60, in=180] (0,0.6) to [out=0, in=120] (1.1,0);
\draw[ultra thick] (2.5,0) to [out=60, in=180] (3.55,0.6) to [out=0, in=120] (4.6,0);
\node[blue] at (-5.3,-0.8) {$\alpha_1$};
\node[blue] at (-1.8,-0.8) {$\alpha_2$};
\node[blue] at (1.8,-0.8) {$\alpha_3$};
\node[blue] at (5.3,-0.8) {$\alpha_4$};
\draw[red, ultra thick] (-4.9,0.4) to [out=60, in=180] (-3.55,1.1) to [out=0, in=120] (-2.2,0.4);
\node[red] at (-3.55,1.4) {$\mathbf{b}_1$};
\draw[red, ultra thick] (-1.4,0.4) to [out=60, in=180] (0,1.1) to [out=0, in=120] (1.4,0.4);
\node[red] at (0,1.4) {$\mathbf{b}_2$};
\draw[red, ultra thick] (2.2,0.4) to [out=60, in=180] (3.55,1.1) to [out=0, in=120] (4.9,0.4);
\node[red] at (3.55,1.4) {$\mathbf{b}_3$};
\end{tikzpicture}
\end{center}
The boundary components are given by the fixed points of the involution (\ref{WorldSheetInvolution}), \emph{i.e.} combinations of the original $\mathbf{a}_i$ cycles of the double cover. Explicitly, for $I=1,\ldots,n $ we denote the distinct boundary components by $\alpha_I$, which are given as:
\begin{align}
\alpha_1&=\mathbf{a}_1\,,\nonumber\\
\alpha_i&=\mathbf{a}_i+\mathbf{a}_{i-1}^{-1}\ ,\hspace{1cm} i=2,\ldots, n-1\ ,\\
\alpha_n&=\mathbf{a}_{n-1}^{-1}\ .\nonumber
\end{align}
For explicit computations (along the line of the previous sections for the cylinder), it is necessary to cut the surface $\Sigma_n$ open to obtain a contractible region. To this end, we need to choose a base point $P_0$ and introduce a basis of cycles $\gamma_i$ (replacing the cycles $\mathbf{b}_i$) 
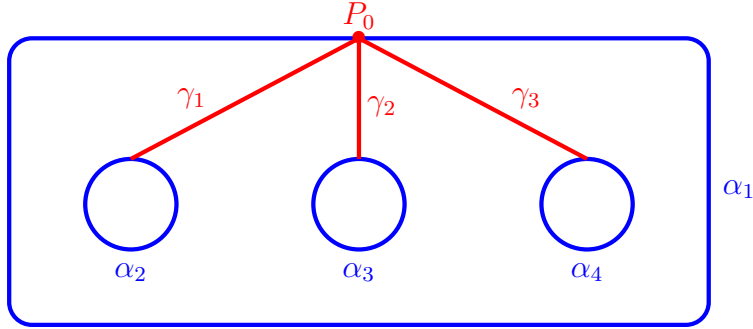
\begin{center}
\begin{tikzpicture}
\draw[rounded corners=0.3cm, ultra thick, blue] (-4.6,2) rectangle (4.6,-1.8);
\draw[ultra thick, blue] (0,-0.2) circle (0.6cm);
\draw[ultra thick, blue] (-3,-0.2) circle (0.6cm);
\draw[ultra thick, blue] (3,-0.2) circle (0.6cm);
\draw[ultra thick, red] (0,2) -- (-3,0.4);
\draw[ultra thick, red] (0,2) -- (0,0.4);
\draw[ultra thick, red] (0,2) -- (3,0.4);
\node[blue] at (5,0) {$\alpha_1$};
\node[blue] at (-3,-1.1) {$\alpha_2$};
\node[blue] at (0,-1.1) {$\alpha_3$};
\node[blue] at (3,-1.1) {$\alpha_4$};
\node[red] at (-2.2,1.2) {$\gamma_1$};
\node[red] at (0.3,1.1) {$\gamma_2$};
\node[red] at (2.2,1.2) {$\gamma_3$};
\node[red] at (0,2) {$\bullet$};
\node[red] at (0,2.3) {$P_0$};
\end{tikzpicture}
\end{center}
This allows us to draw the fundamental polygon of the Riemann surface (for simplicity, we only consider the case $n=4$, which, however, can be generalized in a straight--forward way)
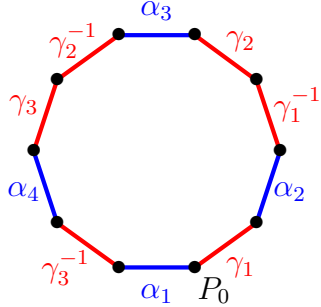
\begin{center}
\begin{tikzpicture}
\draw[blue, ultra thick] (-1,0) -- (0,0);
\draw[red, ultra thick] (0,0) -- (0.81,0.59);
\draw[blue, ultra thick] (0.81,0.59) -- (1.12,1.54);
\draw[red, ultra thick] (1.12,1.54) -- (0.81,2.48);
\draw[red, ultra thick] (0.81,2.48) -- (0,3.08);
\draw[blue, ultra thick] (0,3.08) -- (-1,3.08);
\draw[red, ultra thick] (-1,3.08) -- (-1.81,2.49);
\draw[red, ultra thick] (-1.81,2.49) -- (-2.12,1.54);
\draw[blue, ultra thick] (-2.12,1.54) -- (-1.81,0.59);
\draw[red, ultra thick] (-1.81,0.59) -- (-1,0);
\node at (-1,0) {$\bullet$};
\node at (0,0) {$\bullet$};
\node at (0.25,-0.3) {$P_0$};
\node at (0.81,0.59) {$\bullet$};
\node at (1.12,1.54) {$\bullet$};
\node at (0.81,2.48) {$\bullet$};
\node at (0,3.08) {$\bullet$};
\node at (-1,3.08) {$\bullet$};
\node at (-1.81,2.49) {$\bullet$};
\node at (-2.12,1.54) {$\bullet$};
\node at (-1.81,0.59) {$\bullet$};
\node[blue] at (-0.5,-0.35) {$\alpha_1$};
\node[blue] at (1.25,1.0) {$\alpha_2$};
\node[blue] at (-0.5,3.4) {$\alpha_3$};
\node[blue] at (-2.25,1.0) {$\alpha_4$};
\node[red] at (0.6,0.0) {$\gamma_1$};
\node[red] at (1.35,2.15) {$\gamma_1^{-1}$};
\node[red] at (0.6,3.0) {$\gamma_2$};
\node[red] at (-1.6,3.0) {$\gamma_2^{-1}$};
\node[red] at (-2.25,2.15) {$\gamma_3$};
\node[red] at (-1.7,0.0) {$\gamma_3^{-1}$};
\end{tikzpicture}
\end{center}

\subsection[Monodromy relations for planar $N$--point function]{Monodromy relations for planar 
$\bm{N}$--point function}

We now consider the case of an $N$--point function with vertex insertions at points $z_1,\ldots,z_N$ on the boundaries. We first limit ourselves to the case where $N-1$ insertions (at $z_1,\ldots,z_N$) are inserted on the same boundary component $\alpha_1$ and we integrate the remaining insertion along a closed contour.\footnote{This generalizes the planar monodromy relations on the cylinder discussed in section~\ref{Sect:RelNonPlanarAmp4}, which we expect not to depend on the parametrisation of the cylinder (\emph{i.e.} the choice of the base point $P_0$).} Schematically, this is depicted in the following figure, where the integral path for the point $z_1$ is depicted in green, which is a simply connected, contractible loop:
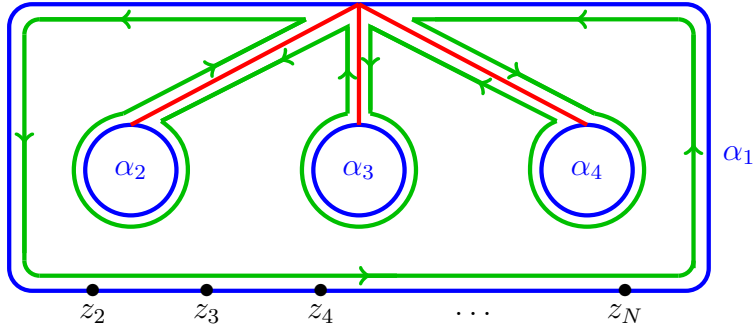
\begin{center}
\begin{tikzpicture}
\draw[rounded corners=0.3cm, ultra thick, blue] (-4.6,2) rectangle (4.6,-1.8);
\node at (-3.5,-1.8) {$\bullet$};
\node at (-3.5,-2.1) {$z_2$};
\node at (-2,-1.8) {$\bullet$};
\node at (-2,-2.1) {$z_3$};
\node at (-0.5,-1.8) {$\bullet$};
\node at (-0.5,-2.1) {$z_4$};
\node at (1.5,-2.1) {$\ldots$};
\node at (3.5,-1.8) {$\bullet$};
\node at (3.5,-2.1) {$z_N$};
\draw[ultra thick, blue] (0,-0.2) circle (0.6cm);
\draw[ultra thick, blue] (-3,-0.2) circle (0.6cm);
\draw[ultra thick, blue] (3,-0.2) circle (0.6cm);
\draw[ultra thick, red] (0,2) -- (-3,0.4);
\draw[ultra thick, red] (0,2) -- (0,0.4);
\draw[ultra thick, red] (0,2) -- (3,0.4);
\draw[green!75!black, ultra thick] (0.7,1.8) -- (4.2,1.8); 
\draw[green!75!black, ultra thick] (4.2,1.8) to [out=0, in=90] (4.4,1.6);
\draw[green!75!black, ultra thick] (4.4,1.6) -- (4.4,-1.5);
\draw[green!75!black, ultra thick] (4.4,-1.4) to [out=270, in=0] (4.2,-1.6);
\draw[green!75!black, ultra thick] (4.2,-1.6) -- (-4.2,-1.6);
\draw[green!75!black, ultra thick] (-4.2,-1.6) to [out=180, in=270] (-4.4,-1.4);
\draw[green!75!black, ultra thick] (-4.4,-1.4) -- (-4.4,1.6);
\draw[green!75!black, ultra thick] (-4.4,1.6) to [out=90, in=180] (-4.2,1.8);
\draw[green!75!black, ultra thick] (-4.2,1.8) -- (-0.7,1.8);
\draw[green!75!black, ultra thick] (-0.7,1.8) -- (-3.1,0.55);
\draw[green!75!black, ultra thick] (-3.1,0.55) to [out=190, in=90] (-3.75,-0.2) to [out=270, in=180] (-3,-0.95) to [out=0, in=270] (-2.25,-0.2) to [out=90, in=325] (-2.6,0.45);
\draw[green!75!black, ultra thick] (-2.6,0.45) -- (-0.15,1.7) -- (-0.15,0.55);
\draw[green!75!black, ultra thick] (-0.15,0.55) to [out=190, in=90] (-0.75,-0.2) to [out=270, in=180] (0,-0.95) to [out=0, in=270] (0.75,-0.2) to [out=90, in=350] (0.15,0.55);
\draw[green!75!black, ultra thick] (0.15,0.55) -- (0.15,1.7) -- (2.6,0.45);
\draw[green!75!black, ultra thick] (3.1,0.55) to [out=350, in=90] (3.75,-0.2) to [out=270, in=0] (3,-0.95) to [out=180, in=270] (2.25,-0.2) to [out=90, in=200] (2.6,0.45);
\draw[green!75!black, ultra thick] (3.1,0.55) -- (0.7,1.8);
\draw[green!75!black, ultra thick, -<] (4.4,1.6) -- (4.4,0);
\draw[green!75!black, ultra thick, -<] (-4.4,-1.3) -- (-4.4,0.3);
\draw[green!75!black, ultra thick, -<] (-4.2,1.8) -- (-3,1.8);
\draw[green!75!black, ultra thick, -<] (1,1.8) -- (3,1.8);
\draw[green!75!black, ultra thick, -<] (0.5,-1.6) -- (0,-1.6);
\draw[green!75!black, ultra thick, -<] (-0.7,1.8)-- (-2,1.13);
\draw[green!75!black, ultra thick, -<] (-2.6,0.45)-- (-0.9,1.315);
\draw[green!75!black, ultra thick, -<] (-0.15,1.7)-- (-0.15,1);
\draw[green!75!black, ultra thick, -<] (0.15,0.6)-- (0.15,1.3);
\draw[green!75!black, ultra thick, -<] (0.15,1.7)-- (1.7,0.9);
\draw[green!75!black, ultra thick, -<] (3.1,0.55) -- (2,1.13);
\node[blue] at (5,0) {$\alpha_1$};
\node[blue] at (-3,-0.2) {$\alpha_2$};
\node[blue] at (0,-0.2) {$\alpha_3$};
\node[blue] at (3,-0.2) {$\alpha_4$};
\end{tikzpicture}
\end{center}
As in the case of the annulus, we need to provide a prescription for how to deal with the points $z_2,\ldots,z_N$. To this end, we apply the same strategy as in section~\ref{Sect:4PtPlanar}, namely we divide the contour integral into $N$ disjoint integrals, each of which stays away from the points $z_2,\ldots,z_N$ and thus vanishes by itself. We then combine these individual integrals in such a way that only the parts along the boundary components remain and the contributions in the interior of the diagram vanish. To describe this procedure, we first select a boundary component $\alpha'$ (different than $\alpha_1$) that is connected to $\alpha_1$ by a curve $\gamma$ and choose a set of ordered points $\{z'_2, z'_3,\ldots,z'_N\}$ on this boundary. We can use the curves which  connect the points $z_i$ and $z'_i$ (for $i=2,\ldots,N$) to tessellate the above contour integral into $N$ smaller ones, as is depicted for the case $N\!=\!4$ (and $\alpha'=\alpha_3$) in the following picture
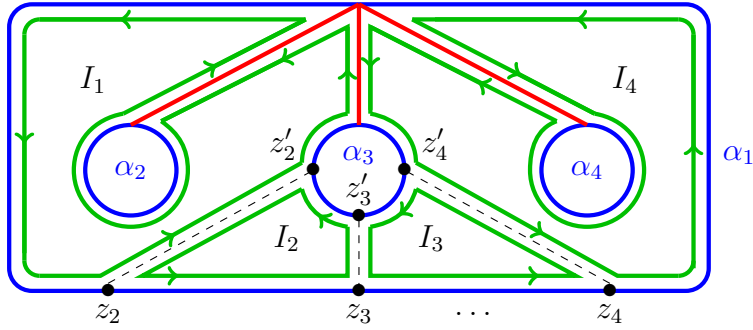
\begin{center}
\begin{tikzpicture}[decoration={
    markings,
    mark=at position 0.5 with {\arrow{>}}}
    ] 
\draw[rounded corners=0.3cm, ultra thick, blue] (-4.6,2) rectangle (4.6,-1.8);
\node at (-3.3,-1.8) {$\bullet$};
\node at (-3.3,-2.1) {$z_2$};
\node at (0,-1.8) {$\bullet$};
\node at (0,-2.1) {$z_3$};
\node at (1.5,-2.1) {$\ldots$};
\node at (3.3,-1.8) {$\bullet$};
\node at (3.3,-2.1) {$z_4$};
\draw[ultra thick, blue] (0,-0.2) circle (0.6cm);
\draw[ultra thick, blue] (-3,-0.2) circle (0.6cm);
\draw[ultra thick, blue] (3,-0.2) circle (0.6cm);
\draw[ultra thick, red] (0,2) -- (-3,0.4);
\draw[ultra thick, red] (0,2) -- (0,0.4);
\draw[ultra thick, red] (0,2) -- (3,0.4);
\draw[green!75!black, ultra thick] (0.7,1.8) -- (4.2,1.8); 
\draw[green!75!black, ultra thick] (4.2,1.8) to [out=0, in=90] (4.4,1.6);
\draw[green!75!black, ultra thick] (4.4,1.6) -- (4.4,-1.5);
\draw[green!75!black, ultra thick] (4.4,-1.4) to [out=270, in=0] (4.2,-1.6);
\draw[green!75!black, ultra thick] (-4.2,-1.6) to [out=180, in=270] (-4.4,-1.4);
\draw[green!75!black, ultra thick] (-4.4,-1.4) -- (-4.4,1.6);
\draw[green!75!black, ultra thick] (-4.4,1.6) to [out=90, in=180] (-4.2,1.8);
\draw[green!75!black, ultra thick] (-4.2,1.8) -- (-0.7,1.8);
\draw[green!75!black, ultra thick] (-0.7,1.8) -- (-3.1,0.55);
\draw[green!75!black, ultra thick] (-3.1,0.55) to [out=190, in=90] (-3.75,-0.2) to [out=270, in=180] (-3,-0.95) to [out=0, in=270] (-2.25,-0.2) to [out=90, in=325] (-2.6,0.45);
\draw[green!75!black, ultra thick] (-2.6,0.45) -- (-0.15,1.7) -- (-0.15,0.55);
\draw[green!75!black, ultra thick] (0.15,0.55) -- (0.15,1.7) -- (2.6,0.45);
\draw[green!75!black, ultra thick] (3.1,0.55) to [out=350, in=90] (3.75,-0.2) to [out=270, in=0] (3,-0.95) to [out=180, in=270] (2.25,-0.2) to [out=90, in=200] (2.6,0.45);
\draw[green!75!black, ultra thick] (3.1,0.55) -- (0.7,1.8);
\draw[green!75!black, ultra thick, -<] (4.4,1.6) -- (4.4,0);
\draw[green!75!black, ultra thick, -<] (-4.4,-1.3) -- (-4.4,0.3);
\draw[green!75!black, ultra thick, -<] (-4.2,1.8) -- (-3,1.8);
\draw[green!75!black, ultra thick, -<] (1,1.8) -- (3,1.8);
%
\draw[green!75!black, ultra thick, -<] (-0.7,1.8)-- (-2,1.13);
\draw[green!75!black, ultra thick, -<] (-2.6,0.45)-- (-0.9,1.315);
\draw[green!75!black, ultra thick, -<] (-0.15,1.7)-- (-0.15,1);
\draw[green!75!black, ultra thick, -<] (0.15,0.6)-- (0.15,1.3);
\draw[green!75!black, ultra thick, -<] (0.15,1.7)-- (1.7,0.9);
\draw[green!75!black, ultra thick, -<] (3.1,0.55) -- (2,1.13);
\node[blue] at (5,0) {$\alpha_1$};
\node[blue] at (-3,-0.2) {$\alpha_2$};
\node[blue] at (0,0) {$\alpha_3$};
\node[blue] at (3,-0.2) {$\alpha_4$};
\draw[green!75!black, ultra thick] (-0.15,0.55) to [out=190, in=90] (-0.75,-0.1);
\draw[green!75!black, ultra thick] (0.75,-0.1) to [out=90, in=350] (0.15,0.55);
\draw[dashed] (-3.3,-1.7) -- (-0.6,-0.2);
\node at (-0.6,-0.2) {$\bullet$};
\node at (-1,0.15) {$z'_2$};
\draw[dashed] (0,-1.8) -- (0,-0.8);
\node at (0,-0.8) {$\bullet$};
\node at (0,-0.45) {$z'_3$};
\draw[dashed] (3.3,-1.7) -- (0.6,-0.2);
\node at (0.6,-0.2) {$\bullet$};
\node at (1,0.15) {$z'_4$};
\draw[green!75!black, ultra thick,postaction={decorate}] (-4.2,-1.6) -- (-3.4,-1.6) -- (-0.75,-0.1);
\draw[green!75!black, ultra thick,postaction={decorate}]  (0.75,-0.1) -- (3.4,-1.6) -- (4.2,-1.6);
\draw[green!75!black, ultra thick,postaction={decorate}] (-0.75,-0.45) -- (-2.85,-1.6) -- (-0.15,-1.6) -- (-0.15,-0.95);
\draw[green!75!black, ultra thick,postaction={decorate}] (-0.15,-0.95) to [out=170,in=285] (-0.75,-0.45);
\draw[green!75!black, ultra thick,postaction={decorate}] (0.15,-0.95) -- (0.15,-1.6) -- (2.85,-1.6) -- (0.75,-0.45);
\draw[green!75!black, ultra thick,postaction={decorate}] (0.75,-0.45) to [out=255,in=10] (0.15,-0.95)  ;
\node[ultra thick] at (-3.5,1) {$I_1$};
\node[ultra thick] at (-0.95,-1.1) {$I_2$};
\node[ultra thick] at (0.95,-1.1) {$I_3$};
\node[ultra thick] at (3.5,1) {$I_4$};
\end{tikzpicture}
\end{center}
The individual contour integrals $I_{1},\ldots, I_N$ contain no pole in their interior and therefore vanish:
\begin{align}
&I_j=0\ \ \ ,\ \ \ j=1,\ldots, N\ .
\end{align}
Their integrands are determined by the monodromies around the points $z_2,\ldots,z_N$, which in turn are determined by the (local) behaviour of the Greens-functions $G(z)=-G(-z)$, in the same way as discussed in section~\ref{Sect:OneLoopAmps} at one loop. Specifically, we have the following relations:
\begin{align}
F_1=e^{-i\pi s_{12}} F_2=e^{-i\pi (s_{12}+s_{13})}F_3=\ldots=e^{-i\pi \sum\limits_{j=2}^{N-1}s_{1j}}F_{N-1}=F_N\ .
\end{align}
Therefore, in the following combination of the contour integrals
\begin{align}
I_1+e^{i\pi s_{12}} I_2+\ldots+e^{-i\pi \sum\limits_{j=2}^{N-1}s_{1j}}I_{N-1}+I_N=0\,,
\end{align}
all contributions along the lines $\overline{z_j z'_j}$ (for $j=2,\ldots, N$) cancel, leaving only the contributions along the boundary components. This gives rise to the following monodromy relation among the $N$--point amplitudes
\begin{align}
&A^{(g)}(1,2,\ldots,N)+e^{i\pi s_{12}}A^{(g)}(2,1,\ldots,N)+\ldots+e^{i\pi \sum\limits_{j=2}^{N-1}s_{1j}}A^{(g)}(2,\ldots, 1,N)\\
&=\tilde{A}^{(g)}(2,3,\ldots,N|1)+e^{i\pi s_{12}} \tilde{A}^{(g)}(3,4,\ldots,N,2|1)+\ldots+e^{i\pi \sum\limits_{j=2}^{N-1}s_{1j}} \tilde{A}^{(g)}(N,2,\ldots,N-1|1)\ ,\nonumber
\end{align}
which generalizes (\ref{AmplitudeNRelation}). Here $A^{(g)}(1,2,\ldots,N)$ are the particular color ordered $N$--point $n$--loop amplitudes with all insertions on the same boundary component $\alpha_1$. Furthermore, the remaining objects $\tilde{A}^{(g)}(2,3,\ldots,N|1)$ corresponds to $N$--point amplitudes, where the position of $z_1$ is shifted by $\gamma$ and integrated along (a portion of) the boundary $\alpha$ (while all remaining points remain on $\alpha_1$).


\subsection{Non--planar amplitudes and invariance of the base point}

In this section we repeat (part of) the above discussion for non-planar monodromy relations. By this we have in mind relations which involve (pieces of) $N$--point amplitudes with at most $N-2$ points inserted on a single boundary component. Our main example is again the four--point relation with one point ($z_2$) inserted on the boundary $\alpha_1$, two further points ($z_3$ and $z_4$) inserted on a different boundary component (which we choose to be $\alpha_3$) and the last point ($z_1$) being integrated over the interior of a Riemann surface with $n>2$ boundary components. 

To repeat the discussion of the previous section, we first need to find a tessellation of the Riemann surface. To this end, we first need to choose a set of 'mirror points' $\{z'_2,z'_3,z'_4\}$ on different boundaries. One (non--unique) choice for the latter is shown in the following figure, which also shows how the contour integral of $z_1$ can be divided into four small contours staying away from the insertion points $z_2,z_3$ and $z_4$.
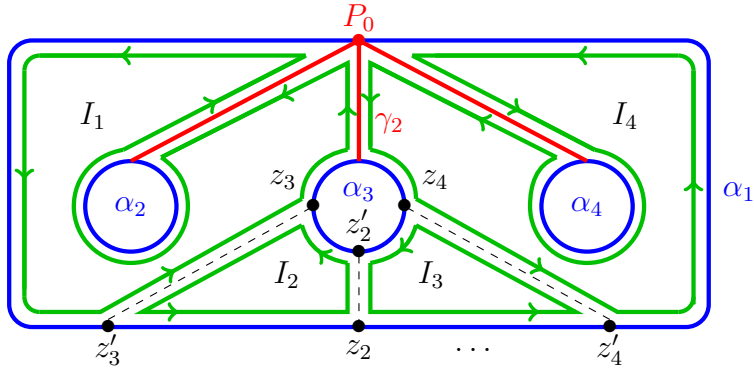
\begin{center}
\begin{tikzpicture}[decoration={
    markings,
    mark=at position 0.5 with {\arrow{>}}}
    ] 
\draw[rounded corners=0.3cm, ultra thick, blue] (-4.6,2) rectangle (4.6,-1.8);
\node at (-3.3,-1.8) {$\bullet$};
\node at (-3.3,-2.1) {$z_3'$};
\node at (0,-1.8) {$\bullet$};
\node at (0,-2.1) {$z_2$};
\node at (1.5,-2.1) {$\ldots$};
\node at (3.3,-1.8) {$\bullet$};
\node at (3.3,-2.1) {$z_4'$};
\draw[ultra thick, blue] (0,-0.2) circle (0.6cm);
\draw[ultra thick, blue] (-3,-0.2) circle (0.6cm);
\draw[ultra thick, blue] (3,-0.2) circle (0.6cm);
\draw[ultra thick, red] (0,2) -- (-3,0.4);
\draw[ultra thick, red] (0,2) -- (0,0.4);
\draw[ultra thick, red] (0,2) -- (3,0.4);
\draw[green!75!black, ultra thick] (0.7,1.8) -- (4.2,1.8); 
\draw[green!75!black, ultra thick] (4.2,1.8) to [out=0, in=90] (4.4,1.6);
\draw[green!75!black, ultra thick] (4.4,1.6) -- (4.4,-1.5);
\draw[green!75!black, ultra thick] (4.4,-1.4) to [out=270, in=0] (4.2,-1.6);
\draw[green!75!black, ultra thick] (-4.2,-1.6) to [out=180, in=270] (-4.4,-1.4);
\draw[green!75!black, ultra thick] (-4.4,-1.4) -- (-4.4,1.6);
\draw[green!75!black, ultra thick] (-4.4,1.6) to [out=90, in=180] (-4.2,1.8);
\draw[green!75!black, ultra thick] (-4.2,1.8) -- (-0.7,1.8);
\draw[green!75!black, ultra thick] (-0.7,1.8) -- (-3.1,0.55);
\draw[green!75!black, ultra thick] (-3.1,0.55) to [out=190, in=90] (-3.75,-0.2) to [out=270, in=180] (-3,-0.95) to [out=0, in=270] (-2.25,-0.2) to [out=90, in=325] (-2.6,0.45);
\draw[green!75!black, ultra thick] (-2.6,0.45) -- (-0.15,1.7) -- (-0.15,0.55);
\draw[green!75!black, ultra thick] (0.15,0.55) -- (0.15,1.7) -- (2.6,0.45);
\draw[green!75!black, ultra thick] (3.1,0.55) to [out=350, in=90] (3.75,-0.2) to [out=270, in=0] (3,-0.95) to [out=180, in=270] (2.25,-0.2) to [out=90, in=200] (2.6,0.45);
\draw[green!75!black, ultra thick] (3.1,0.55) -- (0.7,1.8);
\draw[green!75!black, ultra thick, -<] (4.4,1.6) -- (4.4,0);
\draw[green!75!black, ultra thick, -<] (-4.4,-1.3) -- (-4.4,0.3);
\draw[green!75!black, ultra thick, -<] (-4.2,1.8) -- (-3,1.8);
\draw[green!75!black, ultra thick, -<] (1,1.8) -- (3,1.8);
%
\draw[green!75!black, ultra thick, -<] (-0.7,1.8)-- (-2,1.13);
\draw[green!75!black, ultra thick, -<] (-2.6,0.45)-- (-0.9,1.315);
\draw[green!75!black, ultra thick, -<] (-0.15,1.7)-- (-0.15,1);
\draw[green!75!black, ultra thick, -<] (0.15,0.6)-- (0.15,1.3);
\draw[green!75!black, ultra thick, -<] (0.15,1.7)-- (1.7,0.9);
\draw[green!75!black, ultra thick, -<] (3.1,0.55) -- (2,1.13);
\node[blue] at (5,0) {$\alpha_1$};
\node[blue] at (-3,-0.2) {$\alpha_2$};
\node[blue] at (0,0) {$\alpha_3$};
\node[blue] at (3,-0.2) {$\alpha_4$};
\draw[green!75!black, ultra thick] (-0.15,0.55) to [out=190, in=90] (-0.75,-0.1);
\draw[green!75!black, ultra thick] (0.75,-0.1) to [out=90, in=350] (0.15,0.55);
\draw[dashed] (-3.3,-1.7) -- (-0.6,-0.2);
\node at (-0.6,-0.2) {$\bullet$};
\node at (-1,0.15) {$z_3$};
\draw[dashed] (0,-1.8) -- (0,-0.8);
\node at (0,-0.8) {$\bullet$};
\node at (0,-0.45) {$z'_2$};
\draw[dashed] (3.3,-1.7) -- (0.6,-0.2);
\node at (0.6,-0.2) {$\bullet$};
\node at (1,0.15) {$z_4$};
\draw[green!75!black, ultra thick,postaction={decorate}] (-4.2,-1.6) -- (-3.4,-1.6) -- (-0.75,-0.1);
\draw[green!75!black, ultra thick,postaction={decorate}]  (0.75,-0.1) -- (3.4,-1.6) -- (4.2,-1.6);
\draw[green!75!black, ultra thick,postaction={decorate}] (-0.75,-0.45) -- (-2.85,-1.6) -- (-0.15,-1.6) -- (-0.15,-0.95);
\draw[green!75!black, ultra thick,postaction={decorate}] (-0.15,-0.95) to [out=170,in=285] (-0.75,-0.45);
\draw[green!75!black, ultra thick,postaction={decorate}] (0.15,-0.95) -- (0.15,-1.6) -- (2.85,-1.6) -- (0.75,-0.45);
\draw[green!75!black, ultra thick,postaction={decorate}] (0.75,-0.45) to [out=255,in=10] (0.15,-0.95)  ;
\node[ultra thick] at (-3.5,1) {$I_1$};
\node[ultra thick] at (-0.95,-1.1) {$I_2$};
\node[ultra thick] at (0.95,-1.1) {$I_3$};
\node[ultra thick] at (3.5,1) {$I_4$};
\node[red] at (0,2) {$\bullet$};
\node[red] at (0,2.3) {$P_0$};
\node[red] at (0.4,0.9) {$\gamma_2$};
\end{tikzpicture}
\end{center}
The next step is to find a combination of the small contour integrals $I_1,\ldots, I_4$ (each of which vanishes separately $I_a=0$ for $a=1,\ldots,4$), such that only contributions of $z_1$ being integrated along the boundary components remain: in order to ensure a cancellation for the contributions along the lines $\overline{z'_3 z_3}$, $\overline{z_2 z_2'}$ and $\overline{z'_4 z_4}$ we have to add up the integrals as follows
\begin{align}
I_1+I_2+e^{i\pi s_{12}}I_3+e^{i\pi s_{12}}I_4=0\,.
\end{align}
However, with this assignment of relative phase factors, the two integrals along the path $\gamma_2$ do no longer cancel, thus leading also to contributions in the interior of the contractible region. Notice that these contributions correspond precisely to the terms $B^{(1)}(1,2|3,4)$ and $B^{(1)}(2,1|3,4)$ we found explicitly in the case of the four-point functions on the annulus (see (\ref{Mono2})). The presence of these terms indicates, however, that a priori the monodromy relation is not independent of the choice of the base point $P_0$: indeed, moving the read lines $\gamma_a$ (particular across any of the lines $\overline{z'_3 z_3}$, $\overline{z_2 z_2'}$ and $\overline{z'_4 z_4}$) a priori changes the quantities involved in the monodromy relation\footnote{Notice that this explains also the presence of the two inequivalent orderings $A(1,2|3,4)$ and $A(2,1|3,4)$ in the relation (\ref{Mono2}): while there is only one inequivalent color ordering of the double non--planar four--point amplitude on the annulus, there are two inequivalent orderings of the three points $\{z_1,z_2,P_0\}$, which indeed appear in (\ref{Mono2}).}. 

This indicates that the problems we have encountered at the one-loop level continue also at higher order for the choice of the mirror points $z'_2,z'_3,z'_4$ (\emph{i.e.} the tessellation of the contour integral) we have chosen. A systematic study of possible tessellations and the ensuing integral equations will be presented elsewhere \cite{Progress}.

\section{Concluding remarks}\label{Sect:Conclusions}

In this work we have derived world--sheet string monodromy relations of orientable  one--loop open string scattering amplitudes and discussed their generalization to 
non--orientable world--sheets and higher loops. We have found relations involving different color--ordered subamplitudes, both in the orientable (planar and non--planar cylinder) and non--orientable (M\"obius strip) case. However, these relations also involve a number of new objects, cf. e.g. eqs. \req{GENERIC2} and \req{GENERIC3a}, which contain crucial position dependent phase factors and make a physical interpretation of our results in full--fledged  string theory difficult. We have given a tentative interpretation of these new objects for the simplest relation involving four open string states as scattering amplitudes in the presence of a non--trivial gauge field flux. While we have shown that it is possible to rewrite these quantities in this framework, their physical interpretation is less clear. For the future, it will be important to clarify the source of the gauge  flux and how to generalize it to higher loop orders \cite{Progress}.

We have given a thorough discussion of monodromy relations of one--loop 
open string amplitudes with the results \req{AmplitudeNRelation} and \req{MonoN}.  In particular, we have performed  a careful analysis 
of the branch cut structure leading to more subtle monodromy effects stemming from the position dependent phase factors. When integrating along the various open string source boundaries in the string monodromy relations these branch effects have to be respected and  the integrands have to be rendered single--valued along this integration path by placing respective phases at the right places. An operation, which has been overlooked in the findings \cite{Tourkine:2016bak} leading to incorrect results.
Moreover, monodromy relations involving non--planar amplitudes with at least two open string states on two different boundaries require additional bulk 
terms. These terms, which are important to cancel tadpole contributions in the monodromy relations, are also missing in the results of \cite{Tourkine:2016bak}.
Furthermore, we have demonstrated that BCJ type relations for loop integrands, which are related by integral reduction techniques, 
become trivial identities in the string monodromy relations involving the full fledged integrated string amplitude.

In the tree--level case the monodromy relation \req{TreeAmpRel} and permutations thereof allow to rewrite generic disk amplitudes with $N$ open string insertions in terms of a basis of $(N-3)!$ integrals \cite{Stieberger:2009hq,BjerrumBohr:2009rd}.
The additional bulk terms and the new objects with the position dependent phase factors seem to make it difficult to setup a closed system of 
monodromy relations and apparently prohibit a similar iterative solution of the monodromy relations, which eventually allows to reduce the number of independent subamplitudes appearing in \req{Fullfledged} to a minimal basis.
On the other hand, in the open string channel the position dependent phase factors also involve 
inverse factors of the Schwinger proper time parameter $\Im(\tilde\tau)$, cf. e.g. \req{GENERIC2} and \req{GENERIC3a}.
This dependence signals that the effect of these phase factors is related to 
IR or reducible field theory effects. Indeed, as we have seen for the $N\!=\!4$ case the amplitudes \req{TriAll}  arising from expanding these phase factors 
 can still be decomposed w.r.t. triangle graphs in $D$ dimensions.
It would be desirable to disentangle these IR effects order by order in $\ap$, which eventually may allow for finding closed subamplitude relations for each order in $\ap$.
In fact, we have seen in subsection \ref{transcend} that for certain terms in the $\ap$--expansion such relations can easily be found.

In order to verify our monodromy relations we have performed various checks both in the open and closed string channel.
In particular, we have computed in the closed string channel $\ap$--expansions of both planar and non--planar one--loop amplitudes giving rise to elliptic iterated integrals whose periods amount  to eMZVs. 
While the planar case has already been studied in the literature yielding one system of eMZVs \cite{Broedel:2014vla}, the more involved non--planar case yielding a different system of eMZVs is studied for the first time in this work. Our one--loop open string monodromy relations
 provide identities  between these two different  systems of eMZVs. Eventually,  such relations should give further input in  understanding the algebra of  eMZVs along the lines \cite{Broedel:2015hia}.

Another  generalization of the present work is to include closed string insertions in the bulk of the string world--sheet diagrams giving rise to mixed amplitudes. At the tree--level such amplitudes have been discussed in 
\cite{Stieberger:2009hq,Stieberger:2015vya} and a generalization of the monodromy relations \req{TreeAmpRel} has been recently established in 
\cite{Stieberger:2016lng}. It is interesting  to understand how the corresponding generalization at the loop--level can be obtained. In fact, the additional bulk terms in the monodromy relations might be interpreted as certain amplitudes with additional closed string insertions. Hence, a generalization of one--loop monodromy relations to mixed amplitudes might allow to clarify and handle the bulk terms in our monodromy relations \cite{Progress}.

\section*{Acknowledgements}
We wish to thank Rutger Boels and Henrik Johansson for one helpful email exchange.
SH would like to thank the MPI Munich and the LMU Munich for kind hospitality during several stages of this work. The work of SH was partly supported by the BQR Accueil EC 2015. St.St. would like to thank Henry Tye for kind hospitality and financial support at the Hong Kong University of Science and Technology (HKUST Jockey Club) during completion of this work.
\appendix
\section{Jacobi theta functions}\label{Sect:JacobiTheta}
\label{Sect:JacobiTheta}

A meromorphic function $f(z,\tau)$ of $z$ is defined to be elliptic if it is doubly periodic on a torus, i.e. $f(z+1,\tau)=f(z,\tau)$ and $f(z+\tau,\tau)=f(z,\tau)$.
Elliptic functions having a prescribed set of zeroes and poles in a period parallelogram can be constructed as quotients of Jacobi theta functions.
In this appendix for the latter  we collect useful identities.

In the computation of one--loop string amplitudes we encounter Jacobi theta functions which are defined as 
\begin{align}
&\theta\big[^a_b\big](z,\tau)=\sum_{n=-\infty}^\infty e^{\pi i\left(n-\tfrac{a}{2}\right)^2\tau}\,e^{2\pi i\left(z-\tfrac{b}{2}\right)\left(n-\tfrac{a}{2}\right)}\,,&&\text{for} &&a,b=0,1\,.
\end{align}
Throughout the text we will use the notation
\begin{align}
&\theta_1(z,\tau)=\theta\big[^1_1\big](z,\tau)\,,&&\theta_2(z,\tau)=\theta\big[^1_0\big](z,\tau)\,,&&\theta_3(z,\tau)=\theta\big[^0_0\big](z,\tau)\,,&&\theta_4(z,\tau)=\theta\big[^0_1\big](z,\tau)\,.\nonumber
\end{align}
Under the modular transformation $(z,\tau)\rightarrow \left(\tfrac{z}{\tau},-\tfrac{1}{\tau}\right)$, the theta-function $\theta\big[^a_b\big](z,\tau)$ transforms in the following way
\begin{align}
\theta\big[^a_b\big]\left(\frac{z}{\tau},-\frac{1}{\tau}\right)=\sqrt{-i\tau}\,e^{\frac{i\pi}{2}\,ab+i\pi\,\frac{z^2}{\tau}}\,\theta\big[^{\hspace{0.15cm}b}_{-a}\big](z,\tau)\,.\label{TrafoTheta}
\end{align}
The derivative of the theta function $\theta[^1_1\big](z,\tau)=\theta_1(z,\tau)$ with respect to the first argument can be related to the Dedekind eta function
\begin{align}
\theta'_1(0,it)=2\pi\,\eta^3(it)\,,
\end{align}
where
\begin{align}
&\eta(\tau)=q^{\tfrac{1}{24}}\ \prod_{n=1}^\infty(1-q^n)\,,&&\text{with} &&q=e^{2\pi i\tau}\,,
\end{align}
which transforms as
\begin{align}
\eta(-1/\tau)=\sqrt{-i\tau}\,\eta(\tau)\ .\label{TrafoEta}
\end{align}
Furthermore, under shifts of the first argument, the Jacobi theta functions transform as
\begin{align}
\theta\big[^a_b\big]\left(z+\frac{\epsilon_1}{2}\,\tau+\frac{\epsilon_2}{2},\tau\right)=e^{-\frac{i\pi\tau}{4}\,\epsilon_1^2-\frac{i\pi\epsilon_1}{2}(2z-b)-\frac{i\pi}{2}\,\epsilon_1\epsilon_2}\,\theta\big[^{a-\epsilon_1}_{b-\epsilon_2}\big](z,\tau)\,.\label{IdentityShiftTheta}
\end{align} 
Finally, we borrow the following representations as power series expansions \cite{MOS}
\begin{align}
\ln\theta_1(z,\tau)&=\fc{1}{12}\ln q+\ln\eta(q)+\ln[2\sin(\pi z)]-2\ \sum_{n=1}^\infty
\fc{q^n}{1-q^n}\fc{\cos(2\pi nz)}{n}\ ,\\
\ln\theta_2(z,\tau)&=\fc{1}{12}\ln q+\ln\eta(q)+\ln[2\cos(\pi z)]-2\ \sum_{n=1}^\infty
\fc{(-1)^n\ q^n}{1-q^n}\fc{\cos(2\pi nz)}{n}\ ,\\
\ln\theta_3(z,\tau)&=-\fc{1}{24}\ln q+\ln\eta(q)-2\ \sum_{n=1}^\infty
\fc{(-1)^n\ q^{n/2}}{1-q^n}\fc{\cos(2\pi nz)}{n}\ ,\\
\ln\theta_4(z,\tau)&=-\fc{1}{24}\ln q+\ln\eta(q)-2\ \sum_{n=1}^\infty
\fc{q^{n/2}}{1-q^n}\fc{\cos(2\pi nz)}{n}\ ,
\end{align}
from which we may deduce:
\begin{align}
\ln\lf(\fc{\theta_1(z,\tau)}{\theta_1'(0,\tau)}\ri)&=\ln\lf(\fc{\sin(\pi z)}{\pi}\ri)+4\ \sum_{m\geq 1}\fc{q^m}{1-q^m}\fc{\sin^2(m\pi z)}{m}\ ,\label{Ma1}\\[2mm]
\ln\lf(\fc{\theta_4(z,\tau)}{\theta_4(0,\tau)}\ri)&=4\ \sum_{m\geq 1}\fc{q^{m/2}}{1-q^m}\fc{\sin^2(m\pi z)}{m}\ .\label{Ma2}
\end{align}
With $\theta_1'(0)=2\pi\eta^3=\pi \theta_2(0)\theta_3(0)\theta_4(0)$ the last equation implies
\be
\ln\lf(\fc{\theta_4(z,\tau)}{\theta_1'(0,\tau)}\ri)=-\ln\pi-\ln\theta_2(0)\theta_3(0)+4\ \sum_{m\geq 1}\fc{q^{m/2}}{1-q^m}\fc{\sin^2(m\pi z)}{m}\ ,
\ee
which in turn enters \req{GT}.

\section{Iterated integrals on the torus}
\label{App:IteratedIntegs}

For convenience in this section we write down the iterated integrals on the cylinder necessary. 
For the planar amplitude we need  (with $G(z):=G(z,\tau)$):
\begin{align}
\h\int_0^1dz_4 \int_0^{z_4} dz_3\int_0^{z_3}dz_2\ G(z_2)&=-\fc{1}{6}\ln(2\pi)-\fc{\z_3}{4\pi^2}+\sum_{m\geq 1}\fc{q^m}{1-q^m}\lf(\fc{1}{3m}-\fc{1}{2\pi^2m^3}\ri)\ ,\\
\h\int_0^1dz_4 \int_0^{z_4} dz_3\int_0^{z_3}dz_2\ G(z_3)&=-\fc{1}{6}\ln(2\pi)+\fc{\z_3}{2\pi^2}+\sum_{m\geq 1}\fc{q^m}{1-q^m}\lf(\fc{1}{3m}+\fc{1}{\pi^2m^3}\ri)\ ,\\
\h\int_0^1dz_4 \int_0^{z_4} dz_3\int_0^{z_3}dz_2\ G(z_4)&=-\fc{1}{6}\ln(2\pi)-\fc{\z_3}{4\pi^2}+\sum_{m\geq 1}\fc{q^m}{1-q^m}\lf(\fc{1}{3m}-\fc{1}{2\pi^2m^3}\ri)\ ,\\
\h\int_0^1dz_4 \int_0^{z_4} dz_3\int_0^{z_3}dz_2\ G(z_3-z_2)&=-\fc{1}{6}\ln(2\pi)-\fc{\z_3}{4\pi^2}+\sum_{m\geq 1}\fc{q^m}{1-q^m}\lf(\fc{1}{3m}-\fc{1}{2\pi^2m^3}\ri)\ ,\\
\h\int_0^1dz_4 \int_0^{z_4} dz_3\int_0^{z_3}dz_2\ G(z_4-z_2)&=-\fc{1}{6}\ln(2\pi)+\fc{\z_3}{2\pi^2}+\sum_{m\geq 1}\fc{q^m}{1-q^m}\lf(\fc{1}{3m}+\fc{1}{\pi^2m^3}\ri)\ ,\\
\h\int_0^1dz_4 \int_0^{z_4} dz_3\int_0^{z_3}dz_2\ G(z_4-z_3)&=-\fc{1}{6}\ln(2\pi)-\fc{\z_3}{4\pi^2}+\sum_{m\geq 1}\fc{q^m}{1-q^m}\lf(\fc{1}{3m}-\fc{1}{2\pi^2m^3}\ri)\ .
\end{align}
Furthermore, for the non--planar amplitude we determine (with $G_T(z):=G_T(z,\tau)$):
\begin{align}
\h\int_0^1dz_4 \int_0^{z_4} dz_3\int_0^{z_3}dz_2\ G_T(z_2)&=-\fc{1}{6}\ln(2\pi)-\fc{1}{48}\ln q
+\sum_{m\geq 1}\fc{q^{m/2}}{1-q^m}\lf(\fc{1}{3m}-\fc{1}{2\pi^2m^3}\ri)+\fc{1}{6}Q_3\ ,\\
\h\int_0^1dz_4 \int_0^{z_4} dz_3\int_0^{z_3}dz_2\ G_T(z_3)&=-\fc{1}{6}\ln(2\pi)-\fc{1}{48}\ln q
+\sum_{m\geq 1}\fc{q^{m/2}}{1-q^m}\lf(\fc{1}{3m}+\fc{1}{\pi^2m^3}\ri)+\fc{1}{6}Q_3\ ,\\
\h\int_0^1dz_4 \int_0^{z_4} dz_3\int_0^{z_3}dz_2\ G_T(z_4)&=-\fc{1}{6}\ln(2\pi)-\fc{1}{48}\ln q
+\sum_{m\geq 1}\fc{q^{m/2}}{1-q^m}\lf(\fc{1}{3m}-\fc{1}{2\pi^2m^3}\ri)+\fc{1}{6}Q_3\ ,
\end{align}
with $Q_3$ defined in \req{Q3} 
and:
\begin{align}
\h\int_0^1dz_2 \int_0^1 dz_3 \int_0^{z_3} dz_4\ G(z_2)&=-\fc{1}{2}\ln(2\pi)+\sum_{m\geq 1}\fc{q^m}{1-q^m}\fc{1}{m}\ ,\\
\h\int_0^1dz_2 \int_0^1 dz_3 \int_0^{z_3} dz_4\ G(z_3-z_4)&=-\fc{1}{2}\ln(2\pi)+\sum_{m\geq 1}\fc{q^m}{1-q^m}\fc{1}{m}\ ,\\
\h\int_0^1dz_2 \int_0^1 dz_3 \int_0^{z_3} dz_4\ G_T(z_3)&=-\fc{1}{2}\ln(2\pi)-\fc{1}{16}\ln q+
\sum_{m\geq 1}\fc{q^{m/2}}{1-q^m}\fc{1}{m}+\h\ Q_3\ ,\\
\h\int_0^1dz_2 \int_0^1 dz_3 \int_0^{z_3} dz_4\ G_T(z_4)&=-\fc{1}{2}\ln(2\pi)-\fc{1}{16}\ln q+
\sum_{m\geq 1}\fc{q^{m/2}}{1-q^m}\fc{1}{m}+\h\ Q_3\ ,\\
\h\int_0^1dz_2 \int_0^1 dz_3 \int_0^{z_3} dz_4\ G_T(z_4-z_2)&=-\fc{1}{2}\ln(2\pi)-\fc{1}{16}\ln q+
\sum_{m\geq 1}\fc{q^{m/2}}{1-q^m}\fc{1}{m}+\h\ Q_3\ ,\\
\h\int_0^1dz_2 \int_0^1 dz_3 \int_0^{z_3} dz_4\ G_T(z_3-z_2)&=-\fc{1}{2}\ln(2\pi)-\fc{1}{16}\ln q+
\sum_{m\geq 1}\fc{q^{m/2}}{1-q^m}\fc{1}{m}+\h\ Q_3\ .
\end{align}

\section[Non--planar four--point monodromy relation in the gauge $\bm{z_3=1-\frac{\tau}{2}}$]{Non--planar four--point monodromy relation in the gauge $\bm{z_3=1-\frac{\tau}{2}}$}\label{App:4PtOtherGauge}

The relation (\ref{PredictGaugez1}) derived in section~\ref{Sect:AlphaExpansion} crucially depends on the gauge choice $z_1=0$ such that the contour integral of $z_2$ never comes close to $z_1$. It also simplifies the computation in so far as we never had to deal with the case $\Re(z_2)<\Re(z_1)$. Complications of this type will, however, arise when we pick a different gauge, for example $z_3=1-\frac{\tau}{2}$, as we shall schematically discuss presently. Notice that the very fact that the to gauge choices give inequivalent results is due to a seeming non-reparameterization invariance of the amplitude.

Choosing the gauge $z_3=1-\frac{\tau}{2}$ is schematically shown in the following figure
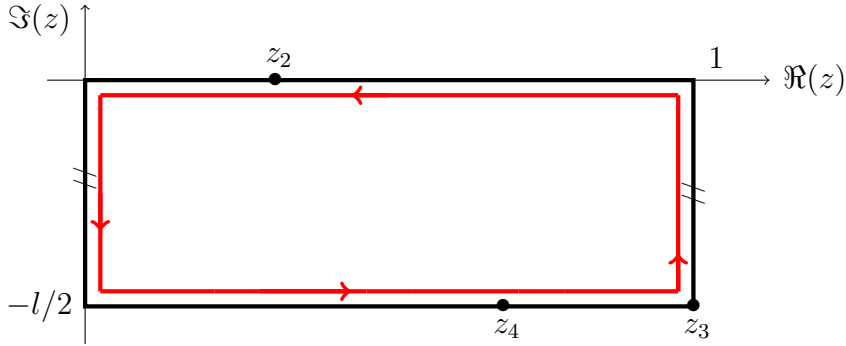
\begin{center}
\begin{tikzpicture}
\draw[->] (-0.5,3) -- (9,3);
\draw[->] (0,-0.5) -- (0,4);
\draw[ultra thick] (0,0) rectangle (8,3); 
\node at (9.6,3) {$\Re (z)$};
\node at (-0.6,3.8) {$\Im (z)$};
\node at (-0.6,0) {$-l /2$};
\node at (8.3,3.3) {1};
\node at (2.5,3) {$\bullet$};
\node at (2.55,3.3) {$z_2$};
\node at (5.5,0) {$\bullet$};
\node at (5.55,-0.3) {$z_4$};
\node at (8,0) {$\bullet$};
\node at (8.05,-0.3) {$z_3$};
\node[rotate=90,scale=0.8] at (0,1.7) {{\small //}};
\node[rotate=90,scale=0.8] at (8,1.5) {{\small //}};
\draw[red, ultra thick,yshift=0.2cm] (0.2,0) -- (1.7,0);
\draw[red, ultra thick,yshift=0.2cm] (2.3,0) -- (3.7,0);
\draw[red, ultra thick,yshift=0.2cm] (4.3,0) -- (5.7,0);
\draw[red, ultra thick,yshift=0.2cm] (6.3,0) -- (7.8,0);
%
\draw[red, ultra thick,yshift=0.2cm] (1.7,0) -- (2.3,0);
\draw[red, ultra thick,yshift=0.2cm] (3.7,0) -- (4.3,0);
\draw[red, ultra thick,yshift=0.2cm] (5.7,0) -- (6.3,0);
\draw[red, ultra thick,->,yshift=0.2cm] (2.3,0) -- (3.5,0);
%
\draw[red, ultra thick,->] (7.8,0.2) -- (7.8,0.7);
\draw[red, ultra thick,->] (0.2,1.5) -- (0.2,1);
\draw[red, ultra thick,->] (4,2.8) -- (3.5,2.8); 
\draw[red, ultra thick] (7.8,0.2) -- (7.8,2.8);
\draw[red, ultra thick] (0.2,0.2) -- (0.2,2.8);
\draw[red, ultra thick] (0.2,2.8) -- (7.8,2.8);%
\end{tikzpicture}
\end{center}
To follow the conventions of section~\ref{Sect:MonodromyOrientable}, we have opted to integrate the point $z_1$ along the contour (instead of $z_2$). As above, we divide this contour integration into three different regions. However, in doing so, we have to distinguish two cases
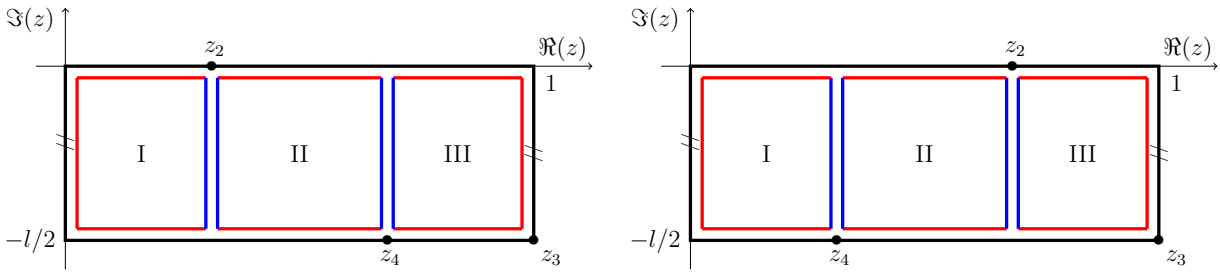
\begin{center}
\scalebox{0.77}{\parbox{10.5cm}{\begin{tikzpicture}
\draw[->] (-0.5,3) -- (9,3);
\draw[->] (0,-0.5) -- (0,4);
\draw[ultra thick] (0,0) rectangle (8,3); 
\node at (8.5,3.3) {$\Re (z)$};
\node at (-0.6,3.8) {$\Im (z)$};
\node at (-0.6,0) {$-l /2$};
\node at (8.3,2.7) {1};
\node at (8,0) {$\bullet$};
\node at (8.3,-0.3) {$z_3$};
\node at (2.5,3) {$\bullet$};
\node at (2.55,3.3) {$z_2$};
\node at (5.5,0) {$\bullet$};
\node at (5.55,-0.3) {$z_4$};
\node[rotate=90,scale=0.8] at (0,1.7) {{\small //}};
\node[rotate=90,scale=0.8] at (8,1.5) {{\small //}};
%
\draw[red, ultra thick,yshift=0.2cm] (0.2,0) -- (2.4,0);
\draw[red, ultra thick,yshift=0.2cm] (2.6,0) -- (5.4,0);
\draw[red, ultra thick,yshift=0.2cm] (5.6,0) -- (7.8,0);
\draw[red, ultra thick,yshift=-0.2cm] (0.2,3) -- (2.4,3);
\draw[red, ultra thick,yshift=-0.2cm] (2.6,3) -- (5.4,3);
\draw[red, ultra thick,yshift=-0.2cm] (5.6,3) -- (7.8,3);
%

%
%
\draw[red, ultra thick] (7.8,0.2) -- (7.8,2.8);
\draw[blue, ultra thick] (5.6,0.2) -- (5.6,2.8);
\draw[blue, ultra thick] (5.4,0.2) -- (5.4,2.8);
\draw[blue, ultra thick] (2.6,0.2) -- (2.6,2.8);
\draw[blue, ultra thick] (2.4,0.2) -- (2.4,2.8);
\draw[red, ultra thick] (0.2,0.2) -- (0.2,2.8);
\node at (1.3,1.5) {I};
\node at (4,1.5) {II};
\node at (6.7,1.5) {III};
%
\end{tikzpicture}}}
\scalebox{0.77}{\parbox{10.5cm}{\begin{tikzpicture}
\draw[->] (-0.5,3) -- (9,3);
\draw[->] (0,-0.5) -- (0,4);
\draw[ultra thick] (0,0) rectangle (8,3); 
\node at (8.5,3.3) {$\Re (z)$};
\node at (-0.6,3.8) {$\Im (z)$};
\node at (-0.6,0) {$-l /2$};
\node at (8.3,2.7) {1};
\node at (8,0) {$\bullet$};
\node at (8.3,-0.3) {$z_3$};
\node at (2.5,0) {$\bullet$};
\node at (2.55,-0.3) {$z_4$};
\node at (5.5,3) {$\bullet$};
\node at (5.55,3.3) {$z_2$};
\node[rotate=90,scale=0.8] at (0,1.7) {{\small //}};
\node[rotate=90,scale=0.8] at (8,1.5) {{\small //}};
%
\draw[red, ultra thick,yshift=0.2cm] (0.2,0) -- (2.4,0);
\draw[red, ultra thick,yshift=0.2cm] (2.6,0) -- (5.4,0);
\draw[red, ultra thick,yshift=0.2cm] (5.6,0) -- (7.8,0);
\draw[red, ultra thick,yshift=-0.2cm] (0.2,3) -- (2.4,3);
\draw[red, ultra thick,yshift=-0.2cm] (2.6,3) -- (5.4,3);
\draw[red, ultra thick,yshift=-0.2cm] (5.6,3) -- (7.8,3);
%

%
%
\draw[red, ultra thick] (7.8,0.2) -- (7.8,2.8);
\draw[blue, ultra thick] (5.6,0.2) -- (5.6,2.8);
\draw[blue, ultra thick] (5.4,0.2) -- (5.4,2.8);
\draw[blue, ultra thick] (2.6,0.2) -- (2.6,2.8);
\draw[blue, ultra thick] (2.4,0.2) -- (2.4,2.8);
\draw[red, ultra thick] (0.2,0.2) -- (0.2,2.8);
\node at (1.3,1.5) {I};
\node at (4,1.5) {II};
\node at (6.7,1.5) {III};
%
\end{tikzpicture}}}

\end{center}
namely $\Re(z_2)<\Re(z_4)$ and $\Re(z_4)<\Re(z_2)$. Thus, now the three regions are 
\begin{align}
&\phantom{x} &&\text{case 1:} &&\text{case 2:}\nonumber\\
&\text{contour I:} &&\Re(z_1)<\Re(z_2)<\Re(z_4)\,,&&\Re(z_1)<\Re(z_4)<\Re(z_2)\,,\nonumber\\
&\text{contour II:} &&\Re(z_2)<\Re(z_1)<\Re(z_4)\,,&&\Re(z_4)<\Re(z_1)<\Re(z_2)\,,\nonumber\\
&\text{contour III:} &&\Re(z_2)<\Re(z_4)<\Re(z_1)\,,&&\Re(z_4)<\Re(z_2)<\Re(z_1)\,.\nonumber
\end{align}
Due to the fact that upon crossing from region I into region II in the first case (and from region II into region III) in the second case, we now have to consider different integrands for the various contributions. To describe each contribution explicitly, we introduce
\begin{align}
F_1(z_1,z_2,z_4)=&\,\text{exp}\bigg[s_{12}\,\mathfrak{g}(z_{21})+s_{34}\,\mathfrak{g}(1-z_{4})+s_{14}\,\mathfrak{g}_T(z_{41})+s_{13}\,\mathfrak{g}_T(1-z_1)\nonumber\\
&\hspace{1cm}+s_{24}\,\mathfrak{g}_T(z_{42})+s_{23}\,\mathfrak{g}_T(1-z_{2})\bigg]\nonumber\\
F_2(z_1,z_2,z_4)=&\,\text{exp}\bigg[s_{12}\,\mathfrak{g}(z_{12})+s_{34}\,\mathfrak{g}(1-z_{4})+s_{14}\,\mathfrak{g}_T(z_{41})+s_{13}\,\mathfrak{g}_T(1-z_1)\nonumber\\
&\hspace{1cm}+s_{24}\,\mathfrak{g}_T(z_{42})+s_{23}\,\mathfrak{g}_T(1-z_{2})\bigg]\,.
\end{align}
The corresponding versions of these integrands for $z_1$ integrated along the upper boundary are obtained by shifting $z_1\to z_1-\frac{\tau}{2}$, resulting in the usual phase-factors. Notice furthermore 
\begin{align}
F_2(z_1,z_2,z_3)=e^{-i\pi s_{12}} F_1(z_1,z_2,z_3)\,,
\end{align} 
We can once again distinguish contributions for the various different regions (see table~\ref{Tab:ContributionsNonPlanar}). Combining the three contour integrals $I_\text{I}$, $I_\text{II}$ and $I_\text{III}$ associated with the three regions in the following way
\begin{align}
&\text{case 1:} &&0=I_{\text{I}}+e^{i\pi s}I_{\text{II}}+e^{i\pi s}I_{\text{III}}\,,\nonumber\\
&\text{case 2:} &&0=I_{\text{I}}+I_{\text{II}}+e^{i\pi s}I_{\text{III}}\,,\nonumber
\end{align}
all the vertical contributions in blue mutually cancel. With the explicit expressions of table~\ref{Tab:ContributionsNonPlanar}\footnote{All quantities in the table are understood to be computed in the gauge $z_3=1-\tfrac{\tau}{2}$.} we therefore find for the sum of the two cases\footnote{W have checked this relation analytically to linear order in $\alpha'$ and numerically to order $\mathcal{O}({\alpha'}^3)$.}
\begin{align}
&a_{1,1}^{(1)}(1,2|3,4)+e^{i\pi s}\left[a_{1,2}^{(1)}(1,2|3,4)+a_{1,3}^{(1)}(1,2|3,4)\right]+a_{2,1}^{(1)}(1,2|3,4)+a_{2,2}^{(1)}(1,2|3,4)+e^{i\pi s}a_{2,3}^{(1)}(1,2|3,4)\nonumber\\
&-\hat{a}^{(1)}_{1,1}(2|3,4,1)-e^{i\pi s}\left[\hat{a}^{(1)}_{1,2}(2|3,4,1)+\hat{a}^{(1)}_{1}(2|3,1,4)\right]-\hat{a}^{(1)}_{2,1}(2|3,4,1)-\hat{a}^{(1)}_{2,2}(2|3,4,1)\nonumber\\
&-e^{i\pi s}\hat{a}^{(1)}_{2}(2|3,1,4)-b_{1,1}^{(1)}(1,2|3,4)-b_{1,2}^{(1)}(1,2|3,4)+e^{i\pi s}\left[b_{1,1}^{(1)}(1,2|3,4)+b_{2,2}^{(1)}(1,2|3,4)\right]=0\,.\label{PredictGaugez4}
\end{align}
This expression is valid for any value of $\tau$ and can therefore be integrated over the world-sheet cylinder modulus. However, the ensuing objects lack an interpretation as color ordered open string sub-amplitudes. Indeed, there are relative phase-factors in (\ref{PredictGaugez4}) between the boundary contributions which a priori prohibit an interpretation in terms of physical objects. For example, in terms of the objects introduced in table~\ref{Tab:ContributionsNonPlanar}, the physical double non-planar subamplitude (in the gauge $z_4=1-\tfrac{\tau}{2}$) is
\begin{align}
A^{(1)}(1,2|3,4)=(s_{12}s_{14})\,A^{(0)}_{YM}(1,2,3,4)\int_0^\infty dl&\bigg[a_{1,1}^{(1)}(1,2|3,4)+a_{1,2}^{(1)}(1,2|3,4)+a_{1,3}^{(1)}(1,2|3,4)\nonumber\\
&+a_{2,1}^{(1)}(1,2|3,4)+a_{2,2}^{(1)}(1,2|3,4)+a_{2,3}^{(1)}(1,2|3,4)\bigg]\,.\nonumber
\end{align}
However, this combination of terms is does not appear in (\ref{PredictGaugez4}), instead the various contributions are multiplied with relative phase factors. The latter are due to the fact that we have to distinguish the cases $\Re(z_1)<\Re(z_2)$ and $\Re(z_2)<\Re(z_1)$, which is in contrast to (\ref{PredictGaugez1}). This also indicates that (at least qualitatively) the relations (\ref{PredictGaugez1}) and (\ref{PredictGaugez4}) are different hinting to an unphysical dependence on the gauge choice. 
\begin{landscape}
\begin{table}
\begin{tabular}{lll}
&{\bf case 1} & {\bf case 2}\\[20pt]
\hspace{-0.6cm}\scalebox{0.9}{\parbox{3cm}{
\begin{tikzpicture}
\draw[red, ultra thick,yshift=0.2cm] (0.2,0) -- (2.4,0);
\node[rotate=90] at (-0.2,1.5) {\footnotesize $b_{i,1}^{(1)}(1,2|3,4)$};
\draw[red, ultra thick] (0.2,0.2) -- (0.2,2.8);
\node at (1.3,-0.2) {\footnotesize $\hat{a}^{(1)}_{i,1}(2|3,4,1)$};
\draw[red, ultra thick,yshift=-0.2cm] (0.2,3) -- (2.4,3);
\node at (1.3,3.2) {\footnotesize $a_{i,1}^{(1)}(1,2|3,4)$};
\draw[blue, ultra thick] (2.4,0.2) -- (2.4,2.8);
\node at (1.3,1.5) {I};
\end{tikzpicture}}}
& 
\parbox{7cm}{\small\begin{align}
&a_{1,1}^{(1)}(1,2|3,4)=\int_0^1dz_4 \int_0^{z_4}dz_2\int_0^{z_2}dz_1\,F_1(z_1,z_2,z_4)\,,\nonumber\\
&\hat{a}^{(1)}_{1,1}(2|3,4,1)=\int_0^1dz_4 \int_0^{z_4}dz_2\int_0^{z_2}dz_1\,F_{T,1}(z_1,z_2,z_4)\,,\nonumber\\
&b_{1,1}^{(1)}(1,2|3,4)=\frac{\tau}{2}\int_0^1dz_4 \int_0^{z_4}dz_2\int^0_{-1}dz_1\,F_1(\tfrac{\tau}{2}\,z_1,z_2,z_4)\,.\nonumber
\end{align}}
&  \parbox{7cm}{\footnotesize\begin{align}
&a_{2,1}^{(1)}(1,2|3,4)=\int_0^1dz_2 \int_0^{z_2}dz_4\int_0^{z_4}dz_1\,F_2(z_1,z_2,z_4)\,,\nonumber\\
&\hat{a}^{(1)}_{2,1}(2|3,4,1)=\int_0^1dz_2 \int_0^{z_2}dz_4\int_0^{z_4}dz_1\,F_{T,2}(z_1,z_2,z_4)\,,\nonumber\\
&b_{2,1}^{(1)}(1,2|3,4)=\frac{\tau}{2}\int_0^1dz_2 \int_0^{z_2}dz_4\int^0_{-1}dz_1\,F_2(\tfrac{\tau}{2}\,z_1,z_2,z_4)\,.\nonumber
\end{align}}
\\[70pt]

\hspace{-0.6cm}\scalebox{0.9}{\parbox{3cm}{\begin{tikzpicture}
\draw[red, ultra thick,yshift=0.2cm] (0.2,0) -- (2.4,0);
\node[rotate=90] at (-0.2,1.5) {\footnotesize $\phantom{b_{i,2}^{(1)}(1,2|3,4)}$};
\draw[blue, ultra thick] (0.2,0.2) -- (0.2,2.8);
\node at (1.3,-0.2) {\footnotesize $\hat{a}_{i,2}^{(1)}(2|3,4,1)$};
\draw[red, ultra thick,yshift=-0.2cm] (0.2,3) -- (2.4,3);
\node at (1.3,3.2) {\footnotesize $a_{i,2}^{(1)}(1,2|3,4)$};
\draw[blue, ultra thick] (2.4,0.2) -- (2.4,2.8);
\node at (1.3,1.5) {II};
\end{tikzpicture}}}
& 
\parbox{7cm}{\small\begin{align}
&a_{1,2}^{(1)}(1,2|3,4)=\int_0^1dz_4 \int_0^{z_4}dz_1\int_0^{z_1}dz_2\,F_1(z_1,z_2,z_4)\,,\nonumber\\
&\hat{a}_{1,2}^{(1)}(2|3,4,1)=\int_0^1dz_4 \int_0^{z_4}dz_1\int_0^{z_1}dz_2\,F_{T,1}(z_1,z_2,z_4)\,,\nonumber
\end{align}
}
&
\parbox{7cm}{\footnotesize\begin{align}
&a_{2,2}^{(1)}(1,2|3,4)=\int_0^1dz_2 \int_0^{z_2}dz_1\int_0^{z_1}dz_4\,F_2(z_1,z_2,z_4)\,,\nonumber\\
&\hat{a}_{2,2}^{(1)}(2|3,4,1)=\int_0^1dz_2 \int_0^{z_2}dz_1\int_0^{z_1}dz_4\,F_{T,2}(z_1,z_2,z_4)\,,\nonumber
\end{align}
}
 \\[70pt]

\scalebox{0.9}{\parbox{3cm}{
\begin{tikzpicture}
\draw[red, ultra thick,yshift=0.2cm] (0.2,0) -- (2.4,0);
\node[rotate=270] at (2.8,1.5) {\footnotesize $b_{i,2}^{(1)}(1,2|3,4)$};
\draw[blue, ultra thick] (0.2,0.2) -- (0.2,2.8);
\node at (1.3,-0.2) {\footnotesize $\hat{a}_i^{(1)}(2|3,1,4)$};
\draw[red, ultra thick,yshift=-0.2cm] (0.2,3) -- (2.4,3);
\node at (1.3,3.2) {\footnotesize $a_{i,3}^{(1)}(1,2|3,4)$};
\draw[red, ultra thick] (2.4,0.2) -- (2.4,2.8);
\node at (1.3,1.5) {III};
\end{tikzpicture}}}
&
\parbox{7cm}{\footnotesize\begin{align}
&a_{1,3}^{(1)}(1,2|3,4)=\int_0^1dz_1 \int_0^{z_1}dz_4\int_0^{z_4}dz_2\,F_1(z_1,z_2,z_4)\,,\nonumber\\
&\hat{a}_1^{(1)}(2|3,1,4)=\int_0^1dz_1 \int_0^{z_1}dz_4\int_0^{z_4}dz_2\,F_{T,1}(z_1,z_2,z_4)\,,\nonumber\\
&b_{1,2}^{(1)}(1,2|3,4)=\frac{\tau}{2}\int_0^1dz_4 \int_0^{z_4}dz_2\int^0_{-1}dz_1\,F_1(\tfrac{\tau}{2}\,z_1+1,z_2,z_4)\,\nonumber
\end{align}}
&
\parbox{7cm}{\small\begin{align}
&a_{2,3}^{(1)}(1,2|3,4)=\int_0^1dz_1 \int_0^{z_1}dz_2\int_0^{z_2}dz_4\,F_2(z_1,z_2,z_4)\,,\nonumber\\
&\hat{a}_2^{(1)}(2|3,1,4)=\int_0^1dz_1 \int_0^{z_1}dz_2\int_0^{z_2}dz_4\,F_{T,2}(z_1,z_2,z_4)\,,\nonumber\\
&b_{2,2}^{(1)}(1,2|3,4)=\frac{\tau}{2}\int_0^1dz_2 \int_0^{z_2}dz_4\int^0_{-1}dz_1\,F_2(\tfrac{\tau}{2}\,z_1+1,z_2,z_3)\,.\nonumber
\end{align}} \\[70pt]
\end{tabular}
\caption{Contributions of the various regions for the non-planar relation in the gauge $z_4=1+\tfrac{\tau}{2}$.}
\label{Tab:ContributionsNonPlanar}
\end{table}
\end{landscape}
\break

\end{document}